\documentclass[12pt]{article}

\usepackage[colorlinks,linkcolor=blue,citecolor=blue,urlcolor=blue,bookmarks,bookmarksnumbered]{hyperref}
\usepackage{amsmath,amssymb,amsfonts,mathrsfs}
\usepackage[paper=letterpaper,margin=1.0in]{geometry}
\usepackage{graphicx,xcolor}
\usepackage{cite}
\RequirePackage{tikz}
 \usetikzlibrary{arrows}
 \def\TikZ#1{\begin{tikzpicture}#1\end{tikzpicture}}
\usepackage{enumitem}

\newcommand{\be}{\begin{equation}}
\newcommand{\ee}{\end{equation}}
\newcommand{\bea}{\begin{eqnarray}}
\newcommand{\eea}{\end{eqnarray}}
\newcommand{\ba}{\begin{array}}
\newcommand{\ea}{\end{array}}
\newcommand{\ben}{\begin{enumerate}}
\newcommand{\een}{\end{enumerate}}
\newcommand{\bi}{\begin{itemize}}
\newcommand{\ei}{\end{itemize}}
\newcommand{\bc}{\begin{center}}
\newcommand{\ec}{\end{center}}
\newcommand{\bfig}{\begin{figure}}
\newcommand{\efig}{\end{figure}}
\newcommand{\bq}{\begin{quotation}}
\newcommand{\eq}{\end{quotation}}
\newcommand{\bt}{\begin{table}}
\newcommand{\et}{\end{table}}
\newcommand{\btab}{\begin{tabular}}
\newcommand{\etab}{\end{tabular}}
\newcommand{\bs}{\begin{slide}}
\newcommand{\es}{\end{slide}}

\def\fm{{\mathfrak{m}}}
\def\te{{\tilde{e}}}
\def\tp{{\mathord{\tilde p\mkern1mu}}}

\def\cH{{\mathcal{H}}}
\def\cL{{\mathcal{L}}}

\def\cD{{\mathcal{D}}}

\newcommand{\pa}{\partial}

\newcommand{\IC}{\mathbb{C}}
 
\newcommand{\IR}{\mathbb{R}}

\def\ZZ{\mathbb{Z}}
\def\sY{\mathscr{Y}}
\def\sW{\mathscr{W}}

\newcommand{\cK}{\mathcal{K}}
\newcommand{\cO}{\mathcal{O}}

\newcommand{\X}{\mathbb{X}}
\newcommand{\cX}{\mathcal{X}}
\newcommand{\sX}{\mathscr{X}}

\newcommand{\vev}[1]{\langle #1 \rangle}

\DeclareMathOperator{\SO}{SO}
\DeclareMathOperator{\SU}{SU}
\newdimen\lft\lft=0pt
\newcommand{\dec}[4][2pt]{\setlength{\lft}{#1}\mathrel{\smash{\underset{\smash{\raisebox{\lft}{#3}}}{\overset{\smash{\raisebox{-5.5\lft}{#2}}}{#4}}}}}

\newcommand{\beq}{\begin{eqnarray}}
\newcommand{\eeq}{\end{eqnarray}}
\newcommand{\beqn}{\begin{eqnarray}}
\newcommand{\eeqn}{\end{eqnarray}}

\def\pa{\partial}
\newcommand{\rd}{\mathrm{d}}

\let\SS=\S 
\def\S{\mathbb{S}}
\def\s{\sigma}

\def\lL{l_\Lambda}

\let\tw=\tilde
\let\Tw=\widetilde
\let\a=\alpha
\def\tA{\tilde{A}}
\let\ba=\overline
\def\tB{\tilde{B}}

\let\b=\beta
\let\d=\delta
\def\const{\hbox{\it const.\/}}
\def\rd{{\rm d}}
\def\define{\buildrel{\rm def}\over=}
\let\f=\phi

\def\e{\epsilon}

\let\g=\gamma
\def\inv#1{{\textstyle{\frac1{#1}}}}

\let\j=\psi

\let\lv=\l 
\let\l=\lambda
\let\L=\Lambda
\def\Lb{\Lambda_b}
\let\q=\theta

\let\p=\pi
\def\Seff{S_{\rm eff}}
\let\t=\tau
\let\vd=\partial
\def\vev#1{\left\langle#1\right\rangle}
\let\z=\zeta

\let\w=\omega \let\om=\omega

\def\frc#1#2{{\textstyle{\frac{#1}{#2}}}}
\def\RR{\relax\leavevmode
       \ifmmode\mathchoice
       {\hbox{\sf R\kern-.4em R}}
       {\hbox{\sf R\kern-.4em R}}
       {\lower.9pt\hbox{\scriptsize\sf R\kern-.36em R}}
       {\lower1.2pt\hbox{\tiny\sf R\kern-.36em R}}
       \else{\sf R\kern-.4em R}\fi}

\def\resetby#1#2{\@addtoreset{#2}{#1}}
\def\seceq{\@addtoreset{equation}{section}
              \def\theequation{\thesection.\arabic{equation}}}

\def\Label#1{\label{#1}%
                \smash{\hbox to0pt{\raise1ex\hbox{\tiny[#1]}\hss}}}
\def\noLabels{\let\Label=\label}
\def\Eq#1{Eq.~(\ref{#1})}

\def\BHM{axilaton}

\DeclareMathOperator{\SL}{\textrm{SL}}
\DeclareMathOperator{\Tr}{\textrm{Tr}}
\DeclareMathOperator{\sech}{\textrm{sech}}
\let\sss=\scriptscriptstyle
\def\dS{\text{dS}}
\def\AdS{\text{AdS}}

\def\tx{{\mathord{\tilde x}}}
\def\hx{{\mathord{\hat x}}}
\def\htx{{\mathord{\hat{\tilde x}}}}

\def\brb#1#2{\boldsymbol{[\mkern-2.5mu[}%
              \mkern1mu#1,\mkern-2mu#2\mkern2mu\boldsymbol{]\mkern-2.5mu]}}

 \allowdisplaybreaks
 \numberwithin{equation}{section}

 \parskip\medskipamount

\begin{document}
\setcounter{page}{-1}
\thispagestyle{empty}

\begin{center}

\vskip 1.0cm
\centerline{\Large \bf On de~Sitter Spacetime and String Theory}
\vskip 1.5cm

\renewcommand{\thefootnote}{\fnsymbol{footnote}}

\centerline{{\bf
Per Berglund${}^{1}$\footnote{\tt per.berglund@unh.edu},
Tristan H\"{u}bsch${}^{2}$\footnote{\tt thubsch@howard.edu}
and
Djordje Minic${}^{3}$\footnote{\tt dminic@vt.edu}
}}

{\it
${}^1$Department of Physics and Astronomy, University of New Hampshire, Durham, NH 03824, U.S.A. \\
${}^2$Department of Physics and Astronomy, Howard University, Washington, D.C.  20059, U.S.A. \\
${}^3$Department  of Physics, Virginia Tech, Blacksburg, VA 24061, U.S.A. \\
${}$ \\
}

\end{center}

\vskip 5mm

\begin{abstract}
We review various aspects of de~Sitter spacetime in string theory: its status as an effective field theory spacetime solution, 
its relation to the vacuum energy problem in string theory,
its (global) holographic definition in terms of two entangled and non-canonical
conformal field theories, 
as well as a realization of a realistic de Sitter universe endowed with the observed visible matter and
the necessary dark sector in order to reproduce the
realistic
cosmological structure.
In particular, based on the new insight regarding the cosmological constant problem in string theory,
we argue that in 
a doubled, $T$-duality-symmetric, phase-space-like and non-commutative generalized-geometric formulation, string theory can naturally lead to a small and positive cosmological constant that is radiatively stable and technically natural.
Such a formulation is fundamentally based on a quantum spacetime, but in an effective spacetime description of this general formulation of string theory,
the curvature of the dual spacetime is the cosmological constant of the observed spacetime, while the size of the dual spacetime is the gravitational constant of the
same observed spacetime. Also, the three scales associated with intrinsic non-commutativity of string theory, the cosmological constant scale and the Planck scale, as well as the Higgs scale, 
can be arranged to satisfy various seesaw-like formulae. 
Along the way, we show that these new features of string theory can be implemented in a particular
deformation of cosmic-string-like models.
\end{abstract}

\vfill
{\it Dedicated to the memory of Joe Polchinski and Steve Weinberg}
\vfill
\clearpage
\begingroup
\baselineskip=10pt \parskip=\smallskipamount
\setcounter{tocdepth}{2}
\tableofcontents
\endgroup
\thispagestyle{empty}

\setcounter{footnote}{0}
\renewcommand{\thefootnote}{\arabic{footnote}}

\section{Introduction} 
\label{s:Intro}
Whether asymptotically de~Sitter spacetime can exist as a solution of string theory has been one of the fundamental conundrums in string theory ever since the dramatic discovery of 
dark energy in the late 1990s\cite{Riess:1998cb, Perlmutter:1998np}.
This question is still considered open\cite{Danielsson:2018ztv,Cicoli:2018kdo}, and the interest in this hard and fundamental issue has been reignited recently\cite{Obied:2018sgi, Agrawal:2018own}. In fact, the details of cosmological evolution are more demanding: This coveted solution of string theory, with all its degrees of freedom, must also accommodate all the observations typically modeled by the inflationary expansion and other particulars of our universe, including the so-called $H_0$-tension\cite{Verde:2019ivm} and other challenges of the Standard $\L$CDM Model\cite{Perivolaropoulos:2021jda}. Fortunately and as we will discuss below, string theory harbors a multitude of degrees of freedom, some of which turn out to be {\em\/hidden\/} ({\em\/dark\/}) to the matter of the particle physics Standard Model, and such that their cumulative effects model the asymptotically de~Sitter spacetime, while their particulars help modeling other details of the cosmological evolution.\footnote{We are encouraged by the resonances with David Gross' lecture at the \href{https://doi.org/10.26081/K6FK95}{KITP Conference: Snowmass Theory Frontier},
\url{https://doi.org/10.26081/K6FK95}.}

We aim to frame ``a discussion of a few key issues''\footnote{We borrow the phrase from Joe Polchinski's report\cite{Polchinski:2006gy} and endeavor to follow his conceptual guidance.} that warrants a particular extension of effective field theory and showcases some of its relevant consequences along the way.
We address the question of the existence of de~Sitter spacetime in string theory, including the crucial problem of
vacuum energy (the cosmological constant problem). We summarize the recent new insight about this crucial problem and we
emphasize the relation between the quantum spacetime (phase-space) aspects of string theory and how they
mesh with the requirements of holography in order to provide a radiatively stable and technically natural cosmological
constant. 
Based on this central insight we discuss a holographic definition of de Sitter spacetime, and argue that there is an important lesson that can be drawn from these considerations, that reinforces a $T$-duality-symmetric, chiral and intrinsically non-commutative
formulation of string theory, called the metastring.
Such a formulation can, in principle, accommodate the observed universe with a small and positive cosmological constant
and the visible and dark sectors needed for the realistic cosmological structure.

We contrast this new view with the more familiar
spacetime-geometric form usually associated with the standard and well studied low-energy effective point-field theory limit of string theory\cite{rCHSW,rGSW1,rGSW2,Polchinski:1998rq,Polchinski:1998rr}.
 We also comment on the small and positive cosmological constant in this context.
 Our central point here is that de~Sitter spacetime is realized in a generalized-geometric and non-commutative, phase-space-like doubled formulation of string theory, for which the familiar local effective field theory is not a generic limit. 

At a coarsest initial approach in this novel framework, we find that the curvature of the dual spacetime serves the cosmological constant of the observed spacetime, whereas the size of the dual spacetime determines the gravitational constant in the
observed spacetime. Also, the three scales associated with intrinsic non-commutativity of string theory, the cosmological constant scale and the Planck scale,
as well as the Higgs scale,
 satisfy seesaw-like relations.
We contrast our viewpoint with the recent discussion in the literature.

This report starts
 firmly rooted in a brief and filtering review of the key approaches to finding de~Sitter space in string theory in Section~\ref{s:Find}; here we present the facets of the core problem of finding de~Sitter vacua in string theory while following the standard approach via Effective Field Theory (EFT).
 A detailed discussion of a particular deformation family of toy models\cite{rBHM5,rBHM6,rBHM7,Berglund:2020qcu} in Section~\ref{s:dSdefo} reveals a seesaw UV/IR correlation and the inherent non-perturbativity of such solutions.
 In the next three sections we address:
  the vacuum energy problem and its various implications (Section~\ref{s:VEST}),
  the holographic formulation of de Sitter spacetime (Section~\ref{s:Define}),
  and the realistic de Sitter universe filled with the observed visible as well as dark matter needed for structure formation (Section~\ref{s:Lessons}).
 Section~\ref{s:Outlook} concludes with an overview of the presented results and outlines some future developments.

\section{Finding de~Sitter Spacetime in String Theory} 
\label{s:Find}
There have been many attempts at finding de~Sitter spacetime (\dS) in string theory, as detailed in the excellent recent 
reviews\cite{Danielsson:2018ztv,Cicoli:2018kdo} and their extensive reference lists; see also\cite{Grana:2005jc, Douglas:2006es} as well as\cite{Polchinski:2006gy} and references therein for earlier reviews.
Here we highlight some quite well known sources (such as\cite{Kachru:2003aw} and\cite{Balasubramanian:2005zx}) and also revisit the lesser known Refs.\cite{rBHM5, rBHM6}.

In the generally accepted and standardly practiced approach\cite{rGSW2,rGSW1,Polchinski:1998rq,Polchinski:1998rr}, by now formulated in an undergraduate textbook\cite{rBZ-StrTh}, string theory leads to a low-energy effective  (point-)field theory (EFT) action for Einstein's gravity (we concentrate on the observed 4-dimensional case, regardless of how this is reached; see however below) in the familiar format:
\begin{equation}
 S_{\text{eff}} = - \int\rd^4 x \sqrt{g} \Big(\frac{1}{8 \pi G} \Lambda + \frac{1}{16 \pi G} R
 +a R_{\mu \nu} R^{\mu \nu} +b R^2 + cR_{\mu \nu \rho \sigma} R^{\mu \nu \rho \sigma} +\dots\Big),
 \label{e:Seffective}
\end{equation}
where the coefficients $a,b,c$ as well as the omitted terms are completely determined by the renormalization of the underlying worldsheet theory\cite{rF79b,rF79a}.
Producing a positive and an uncharacteristically small cosmological constant $\Lambda$ turns out to be a very difficult task, at least by following this usual formulation of string theory. Besides the notorious difficulty of obtaining a $\L$ that is {\em\/uncharacteristically small\/} as compared with usual characteristic energy-scales of the EFT, the additional difficulty of obtaining $\L\,{>}\,0$ is exacerbated by the standard focus on models which at least start out with some supersymmetry: de~Sitter spacetime is well known to admit no supersymmetry\cite{r1001}. So, while the presence of supersymmetry, subsequently broken in some technically manageable fashion, does help both with computability in concrete models and with general features, it also narrows the search by considerations of technical management. This very general ``lamppost'' approach is in no way guaranteed to lead to such a demanding solution.

In what follows, we argue that a more general formulation {\em\/does\/} lead to a positive cosmological constant, $\Lambda$. This more general formulation is in fact more natural from the point of view of the intrinsic structure of string theory, and involves a doubled, non-commutative and generalized geometry. This formulation is in fact also indicated by a more careful reexamination of the EFT solutions such as the discretuum of specific axion-dilaton ({\em\/\BHM\/})\footnote{The admittedly frivolous contraction ``\BHM'' reminds that the family of cosmological brane-world models developed in Refs.\cite{rBHM1, rBHM2, rBHM3, rBHM4, rBHM5, rBHM6} are driven by the background values of the axio-dilaton system and their $\SL(2,\ZZ)$ monodromy in the extra 2-dimensional plane. The succinct name also saves us from repeated circumlocutions that perforce include the above-cited reference sequence.} solutions\cite{rBHM5,rBHM7,Berglund:2020qcu} that deform the Minkowski models of Refs.\cite{rBHM1, rBHM2, rBHM3, rBHM4}. We thus first concentrate on this standard and thoroughly familiar EFT approximation of string theory, its compactifications and brane-world scenarios. See also the very recent comprehensive EFT-focused review\cite{Flauger:2022hie}, which also provides an impressive survey of the related literature.

The general message of various attempts to find de~Sitter spacetime via compactifications in string theory is as follows: It is possible, if na{\"\i}vely, to realize dS in these EFT spacetime constructions. What is not clear is whether these attempts are safe from various perils, including the crucial questions of stability, and how generic they are. (The community seems to be divided on these points.)
Even so, we argue below that this approach misses an important physics point: All these constructions and analyses are conducted within the context of an effective field theory spacetime interpretation of string theory, which is not generic,
whereby the search for dS solution in string theory from this viewpoint cannot possibly be exhaustive and conclusive.

To this end, we re-examine in the next section a particular family of \BHM\ brane-world cosmological models which {\em\/does\/} produce a de~Sitter spacetime. Highlighting its main characteristics and implications turns out to relate to our main point, stated later below.
Suffice it here to note that the iterative development of these \BHM\ models is as follows: 
Starting from a deformed co-dimension 2 solution (motivated, in part by\cite{Cohen:1999ia}) with a natural connection to a logarithmic running of coupling constants,\footnote{This is a unique feature of codimension-2 solutions, which owes to the well-known logarithmic nature of Green's function in 2-dimensional space.} we find a deep connection to a non-supersymmetric phase of $F$-theory, the conjectured non-perturbative formulation of the IIB string, via the stringy cosmic strings. Rather unexpectedly from this EFT point of view, this model does accommodate de~Sitter spacetime and automatically produces a seesaw mechanism generating a small cosmological constant. 
Also, although the \BHM\ models are non-supersymmetric for most of their parameters, they do have a well-defined supersymmetric limit --- located centrally in their parameter space. Thereby, this \BHM\ deformation family of models may be regarded as showcasing cosmological supersymmetry breaking deformations of a generic supersymmetric configuration in $F$-theory.

 Furthermore, this turns out to be connected to a deconstruction of the cosmological constant when extending from 3-dimensional to 4-dimensional spacetime, which pivots on the old observation of Witten\cite{Witten:1994cga} about the peculiar features of supersymmetry in 3 dimensions:  Due to the presence of conical defects in 3-dimensional gravity, the supercharges do not have to be globally defined. Therefore, supersymmetry does not necessarily imply masses of superpartners to be degenerate, as the mass-splitting is controlled by the strength of the conical defects.
 The deconstruction performed in Ref.\cite{Jejjala:2002we} (see also Ref.\cite{Becker:1995sp}) produces a 4-dimensional version of Witten's argument, albeit now controlled by stringy defects. These stringy defects are generically strongly coupled in the 4-dimensional continuum limit, and are easily related to the non-supersymmetric stringy cosmic strings in the \BHM\ deformation family of models. The same holds also for the ensuing seesaw formula for the cosmological constant:
$M_{\text{vac}} \sim M^2/M_P$, where $M_{\text{vac}}$ is the vacuum energy/cosmological constant scale,
$M_P$ the Planck scale, and $M$ the characteristic particle physics scale.\footnote{There is an
intriguing connection between this scenario and the recent suggestion by Heckman et al.\cite{Heckman:2017uxe} regarding a Quantum Hall like definition of the chiral IIB string in which the chiral string lives on the 10-dimensional boundary, and the bulk is controlled by a $(10{+}2$)-dimensional topological theory, the analogue of the Chern Simons 3-dimensional bulk description from the usual Quantum Hall set up, in which one ends up with 2d chiral excitations. 
Now the 2-dimensional part of this $(10{+}2$)-dimensional bulk is the axion-dilaton torus of $F$-theory, which emerges from a cylinder that may well be the cylinder of\cite{rBHM1, rBHM2, rBHM3, rBHM4, rBHM5, rBHM6}.
The relation of the 2-dimensional torus to the cylinder might be of the same nature like the deconstruction of the cosmological constant between 3-dimensional and 4-dimensional spacetime mentioned above.}

Our main point (see Section~\ref{s:VEST}) is that the chirally doubled (phase-space like), non-commutative generalized-geometric formulation of string theory does lead, and naturally, to a positive cosmological constant and asymptotically dS backgrounds.
Essentially and in a coarsest approximation, the curvature of the dual space plays the role of the cosmological constant in the observed spacetime, and the size of the dual space is the gravitational constant in the
same observed spacetime. Also, the three scales associated with intrinsic non-commutativity of string theory, the cosmological constant scale and the gravitational constant (Planck) scale, as well as
the Higgs scale, are naturally arranged via seesaw formulae.
 Our showcasing discretuum of \BHM\ toy models illustrates all these main points, but is not at all isolated: For example, these models bear some striking similarities to the recently analyzed de~Sitter bubbles in unstable AdS backgrounds\cite{Banerjee:2018qey,Banerjee:2019fzz,Dibitetto:2020csn,Petri:2022yhy}.
 In a somewhat similar vein, a recent analysis of hyperbolic compactifications of $M$-theory are found to accommodate a ``relatively simple'' (albeit meta-stable) de~Sitter uplift as well as cosmological inflation within hyperbolic (negative curvature) compactifications of $M$-theory\cite{DeLuca:2021pej}.

\subsection{Some de~Sitter Spacetime Preliminaries}
\label{s:Prelim}
Here we summarize some basic facts about de~Sitter (dS) spacetime following the presentation found in\cite{Balasubramanian:2001nb}.
In $d{+}1$ dimensions, spaces with a positive cosmological constant solve equations
of motions derived from the action
\be
S_{\text{B}} = - \int_M\rd^{d+1} x~ \sqrt{-g} \Big(\frac{1}{8 \pi G} \Lambda + \frac{1}{16 \pi G} R \Big) +
\frac{1}{8 \pi G} \int_{\partial M^-}^{\partial M^+}\!\!\!\rd^d x~ \sqrt{h}\, K.
\ee
Here $M$ is the bulk manifold and the spatial boundaries at early and late times are denoted by $\partial M^{\pm}$,
$g_{\mu \nu}$ is metric in the bulk of spacetime, and $h_{ab}$ and $K$ are the induced metric and the
trace of the extrinsic curvature of the respective boundaries. In de~Sitter spacetime the spacetime
boundaries $I^{\pm}$ are Euclidean surfaces at early and late time infinity. 
The extrinsic curvature
boundary term is necessary to allow a well-defined Euler-Lagrange variation.

It is usually convenient to define a corresponding length scale
\be
\lL = \sqrt{\frac{(d{-}1)(d{-}2)}{2 \Lambda}},
\ee
which itself does depend on the dimension ($d$) of the spacetime, but the order of magnitude of which does not change much.\footnote{For example, for $d\,{=}\,10$ vs.\ $d\,{=}\,4$-dimensional spacetime, the relative ratio is $l_\L^{\sss(10)}\big/l_\L^{\sss(4)}=\sqrt{\frac{9{\cdot}8}{3{\cdot}2}}\approx3.46$.}
Using this notation, the vacuum dS spacetime solution to the equations of motion that follow from the
variation of the above action $\delta S_B =0$, is
\be
\rd s^2 =- \rd t^2 + \lL\!^2 \cosh^2(t/\lL)\, \rd\Omega_{d-1}^2
\ee
where equal time sections are $(d{-}1)$-spheres. The same space admits a coordinate system
where equal time surfaces are flat:
\be
\rd s^2 =- \rd t^2 +\exp(2 \tau/\lL)\, \rd\vec{x}\,^2
     = \frac{\lL\!^2}{\eta^2} (- \rd\eta^2 + \rd\vec{x}\,^2)
 \label{e:dSStat}
\ee
This (flat) patch only covers half of de~Sitter space,
extending from a ``big bang'' at a past horizon to the Euclidean surface at future
infinity. This is also called an inflationary patch.
By replacing $\tau$ by $- \tau$ a patch which covers the other half of de~Sitter (from
past infinity to a future horizon) can be obtained. We will refer to these two patches
as the ``big bang'' and the ``big crunch'' patches. Finally, an inertial observer in dS
spacetime sees a static spacetime with a cosmological horizon:
\be
\rd s^2 = - (1- r^2/\lL\!^2)\, \rd t^2 + (1- r^2/\lL\!^2)^{-1} \rd r^2
+ r^2 \rd\Omega_{d-2}^2
\ee
The relations between these coordinate patches and Penrose diagrams are presented
in the classic book by S. Hawking and G. Ellis on the large scale structure of spacetime and references therein\cite{Hawking:1973uf}.

Note that we can formally ``Wick rotate'' from a positive to a negative cosmological constant
by the analytic continuation $\lL \to i\lL$. This formal transformation (sometimes
accompanied by additional Wick rotations of some of the coordinates) takes patches
of de~Sitter into patches anti-de~Sitter spacetime (AdS). For example, the static patch~\eqref{e:dSStat} rotates
to global AdS. Likewise, by redefining $\exp(\tau/\lL) = r/\lL$ and carrying out some formal
analytic continuations gives the Poincar\'{e} patch of AdS. Tracking these continuations
through the classic computations of properties of AdS spaces gives a powerful method
of inferring some aspects of de~Sitter physics. We will comment more about this in the
section on holography in de~Sitter spacetime.

\subsection{Notes on dS Constructions in String Theory} 
\label{s:Notes}
As summarized in some detail in the review\cite{Danielsson:2018ztv,Cicoli:2018kdo}, the basic assumptions of this standard, EFT-relying approach to string theory include the following:
 {\small({\it\bfseries a\/})}~The string coupling is stabilized at a small value, justifying the EFT approach.
 {\small({\it\bfseries b\/})}~The derivative corrections to~\eqref{e:Seffective} are suppressed.
 {\small({\it\bfseries c\/})}~The stringy winding modes can be neglected.
 {\small({\it\bfseries d\/})}~The cosmological constant is small in the units
of the masses of the Ka{\lv}u{\.z}a-Klein modes in the context of string compactification, i.e., the size of the observable universe and
the size (and curvature\footnote{In the Brane-World cosmological scenarios such as pioneered in\cite{Randall:1999vf} extra dimensions may well have a large extent, but are highly curved, and it is this curvature length-scale that is relevant in this context.}) of extra dimensions have to be clearly and well separated. (This is related to the previous assumption.)

\subsubsection{Difficulties}
The general expectation for the appearance of a cosmological constant in string theory (and thus de~Sitter spacetime) is that it is determined by a minimum of the effective potential $V(\phi)$ where 
$\phi^i$ are the moduli
\begin{equation}
S_{\text{eff}} = \int\!\sqrt{|g|}\;
 \Big( \frac12 R
       - \frac12 g^{\mu\nu}G_{ij}(\phi)\,\partial_\mu \phi^i\, \partial_\nu \phi^j
       - V(\phi) + \dots \Big)
 \label{e:Veff}
\end{equation}
The immediate and obvious concern here is whether one is computing the bare or the actual cosmological constant by minimizing $V(\phi)$. This concern is usually addressed by:
 {\small({\bf1})}~citing the above-enumerated assumptions and a careful arrangement the computations (the ``rules of the game''), and
 {\small({\bf2})}~specifying that one is ``working in a fine-tuned corner of the moduli space, where the weakly coupled space-time solution is argued to be a good approximation to the full string theory solution''\cite{Danielsson:2018ztv}.
 These complementing assumptions are then argued to insure that the resulting value of the cosmological constant in this self-consistent framework of flux compactifications is fairly close to its actual value, and safe from the otherwise typical quantum field-theoretic sensitivity to UV cutoff energies. 

Finding a phenomenologically acceptably small and positive cosmological constant within this general approach seems highly improbable upon surveying numerous attempts in the reviewed literature\cite{Danielsson:2018ztv,Cicoli:2018kdo}.
 In turn, a phenomenologically acceptable cosmological constant determined by the minimum of an effective potential such as in~\eqref{e:Veff} has in fact been argued to be outright incompatible with quantum gravity\cite{Obied:2018sgi, Agrawal:2018own}.
 Without a feasible de~Sitter spacetime solution, this line of reasoning then only allows for modeling the details of the observed cosmology with {\em\/quintessence\/} --- which models then have their own challenges and difficulties.
 Typically, these include a high degree of fine-tuning: For one, the mass of the scalar field does get quadratically divergent corrections (as do masses of all scalars in 4d, such as the Higgs) inducing the standard hierarchy problem. Here however, this problem is much more serious. The mass of the scalar field in quintessence models must be very small (typically ${\sim}\,10^{-33}\,$eV) so as to ``fit'' the observed dark energy scale, exacerbating the standard Planck/Higgs ${\sim}\,10^{17}$ mass-scale hierarchy problem to a Planck/quintessence ${\sim}\,10^{61}$ problem! Furthermore, this scalar will also contribute to the cosmological constant problem, so that the dark energy problem does not really seem to be solved from a quantum perspective. Even on the observational side, quintessence tends to induce too low a value for the Hubble constant ($H_0$)\cite{Lee:2022cyh,Banerjee:2020xcn}; see also\cite{SPT-3G:2022hvq}.

Within this standard EFT-relying approach and fairly generally, one can distinguish three broad classes of options\cite{Danielsson:2018ztv}:

\paragraph{Non-critical string theory:} 
Here, the effective action in the string frame has an extra contribution that for
dimensions larger than critical can be understood as a positive contribution to the effective potential.
The problem with this option is that one does not have a control over the perturbative sector of these models; see\cite{Maloney:2002rr}.

\paragraph{Critical string theory, with no known classical geometry:} 
This so-called non-geometric route relies on the non-geometric ($Q$ and $R$) fluxes, and in that case one has an increase in the number of parameters. In such cases, one may be able to find metastable dS vacua in principle, although such models tend to be very scarce; see\cite{Hertzberg:2007wc}.

\paragraph{Critical string theory, with classical geometry:} 
In this, geometric route, one typically starts with 10d supergravity with fluxes at the two-derivative level.
The effective potential has two terms: one coming from curvature of the extra dimensions and the other
coming from the flux terms.
Schematically, in 10 dimensions and using the notation of Danielsson and Van Riet\cite{Danielsson:2018ztv}, we have:
\begin{equation}
S_{\text{eff}} = \int\!\sqrt{|g|}\;
 \Big( \frac12 R 
     - \frac12 \|\partial \phi\|^2
     - \sum_p c_p\, e^{a_p \phi} F_p\!^2 \Big) + \textit{CS}
\end{equation}
where ``\textit{CS}\,'' denotes the Chern Simons terms and $F_p$ are the appropriate rank-$p$ field strengths of 10d supergravity
($p$ is odd for IIB and even for IIA supergravity).
Then the effective potential is found to be composed of a term coming from the curvature of 6 extra dimensions and
the flux term:
\begin{equation}
V = {-}\int\!\sqrt{g_6}\; R_6 + \int\!\sqrt{g_6}\; c_p\, e^{a_p \phi} F_p\!^2
\end{equation}
Localized brane sources include an extra term to the effective potential
\begin{equation}
V_{\rm src} = \mu \int\!\sqrt{|g_n|}
\end{equation}
where $\mu$ denotes the brane's tension and $g_n$ is the determinant of the 10d metric pulled-back onto the
internal submanifold wrapped by the brane.

The no-go theorem of Maldacena and Nunez\cite{Maldacena:2000mw} asserts that it is impossible to obtain a dS background 
if the space is static, compact and without singularities.
While adding D-branes does not help, orientifolds (which provide negative tension) can help, but there are further no-go theorems in this case as well\cite{Grana:2005jc,Bena:2009xk}, which are in general expected to be resolved by stringy UV completion\cite{Michel:2014lva,Cohen-Maldonado:2015ssa,Armas:2018rsy}.
 Indeed, classical solutions that include appropriate back-reactions have been constructed\cite{Cordova:2019cvf,Cordova:2018dbb}.

Most of the approaches to dS space in string theory start with classical flux compactifications with 
orientifold sources and by looking at various quantum effects that stabilize the moduli.
Perhaps the best understood are type~IIB string theory models\cite{Giddings:2001yu}; see also\cite{Dasgupta:1999ss}.
This analysis leads to the following main groups of ``uplift-vacuum'' scenarios:
 Kachru-Kallosh-Linde-Trivedi (KKLT)\cite{Kachru:2003aw}, as well as the
 ``(racetrack) K{\"a}hler uplift''
  (RKU)\cite{Balasubramanian:2004uy,Westphal:2006tn,Rummel:2011cd} and the related
 ``large volume scenario''
  (LVS)\cite{Balasubramanian:2005zx,Conlon:2005ki,Crino:2020qwk}.

Let us provide here but a telegraphic summary of a few salient features:
 ({\small\bf1})~The KKLT and LVS scenarios both rely on the inclusion of a $\overline{\text{D3}}$-brane (anti-D3-brane) to break supersymmetry spontaneously by lifting the overall energy, following the over two decades old exploration of ``brane supersymmetry breaking.'' For a comprehensive description of the role of the  $\overline{\text{D3}}$-brane in producing a de~Sitter spacetime in the KKLT framework, see\cite{Bergshoeff:2015jxa}; for the fully (linearly and nonlinearly realized) supersymmetric analysis see\cite{Cribiori:2019hod}. Issues with loss of parametric control owing to strong warping have been discussed in\cite{Gao:2020xqh}.
 ({\small\bf2})~The LVS formalism was, besides breaking supersymmetry spontaneously and generating the $\AdS\to\dS$ uplift with a $\overline{\text{D3}}$-brane, also found to induce a logarithmic distribution (in the string landscape) of the supersymmetry breaking scale\cite{broeckel2021moduli}. Warping corrections are here also found to likely induce a loss of parametric control\cite{Junghans:2022exo,Junghans:2022kxg,Gao:2022fdi}.
 ({\small\bf3})~The RKU scenario avoids using $\overline{\text{D3}}$-branes to break supersymmetry and generate the $\AdS\to\dS$ uplift, but finds it possible to generate a {\em\/metastable\/} dS vacuum by balancing 3-form fluxes and D7-branes against leading order perturbative $O(\a'\,^3)$-corrections to the K\"ahler potential.
 These three scenarios are in fact closely related, and several variations and combinations have been considered in the literature\cite{Rummel:2011cd}
 ({\small\bf4})~In turn, recent study\cite{Bento:2021nbb} shows that a careful balancing of various contributions ($\overline{\text{D3}}$-branes, Whitney branes\cite{Collinucci:2008pf,Collinucci:2008sq}) near a (warped deformed) conifold locus in the moduli space permits a metastable $3{+}1$-dimensional de~Sitter spacetime; see \SS\,\ref{s:TT}.
 ({\small\bf5})~The KKLT scenario was also rather quickly extended to also provide for cosmic inflation\cite{Kachru:2003sx}.

Given the variety in details of these approaches and the so-called {\em\/totalitarian principle,}\footnote{``Everything not forbidden is compulsory''\cite{rTHW-tot} has been heavily popularized by M.~Gell-Mann as the chief overarching tenet of quantum physics, and is often misattributed to him.} one should expect that at least certain aspects of {\em\/all\/} of these scenarios ultimately contribute to the realization of \dS\ spacetime in string theory. It is then, naturally, the intricacies of the interplay between the various details of the involved effects (rather than the isolated characteristics of any particular one of them) that will decide on the viability of this program. In turn, such careful balancing tends to depend on the technical details of the concrete models considered, and the ultimate outcome may well, {\em\/a priori,} be sensitive to next-order and even higher perturbative corrections.

This analysis can be extended also to the type~IIA and heterotic models, and the results generally turn out rather similar to the canonical type~IIB case.
 Furthermore, it is of course phenomenologically relevant to distinguish models in these categories by the energy where supersymmetry appears effectively broken, i.e., whether the supersymmetry-breaking mass-scale is as high as the Ka{\lv}u{\.z}a-Klein scale or below it, and how close it is to experimentally attainable energies.
 Finally, let us summarize several of the key concerns arising in all of these approaches (see also\cite{Danielsson:2018ztv}):

\paragraph{Classical concerns:}
 There are no known classical 4d de~Sitter solutions that are free of tachyons. One simple argument for this is as follows: Of all  terms contributing to the scalar potential, the orientifold tension is negative, which at some critical point of the potential and generically, signals the growth of a tachyonic direction. Most probably then, the 4d (broken) supergravity effective description of dS is problematic.

\paragraph{Geometric vs.\ non-geometric concerns:}
 One distinction between geometric and non-geometric is that
the non-geometric vacua are not $T$-duals of geometric ones. Also, there is an issue whether the non-geometric solutions are locally geometric. From all known metastable dS vacua of the non-geometric (double field theory like) approach, none are locally geometric, and thus they are hard to interpret as the observed dS vacua.

\paragraph{Quantum concerns:}
 These include
 {\small({\it\bfseries\/a\/})}~anti-brane back-reaction inside extra dimensions,
 {\small({\it\bfseries\/b\/})}~anti-brane back-reaction on the moduli
(see\cite{Bena:2009xk}),
and perhaps the most severe,
 {\small({\it\bfseries\/c\/})}~generic issues with non-supersymmetric Giddings-Kachru-Polchinski (GKP) solutions\cite{Giddings:2001yu}.
 Given the various string-string dualities connecting the type~II and the heterotic theories, it is reasonable to inquire how do the various constraints on heterotic models reflect on type~II models\cite{Sethi:2017phn}. This detailed study employs several different dualities and finds that while de~Sitter solutions are not necessarily ruled out, they invariably ``must involve an unconventional, or at least hard-to-compute ingredient.'' In particular, inclusion of anti-branes and orientifolds does not pass this threshold of unconventionality. This detailed study is however rather carefully framed in exhibiting the key reliance on EFT, and finds that there indeed is a universal issue with the effective field theory constructions.
 Similarly, a recent analysis of the quantum back-reaction of thermal matter in a dS spacetime suggests the need for a phase-transition around the string-scale, $M_s\,{<}\,M_P$, and a ``still mysterious, presumably topological, high-temperature regime of string theory''\cite{Blumenhagen:2020doa}; see also\cite{Agrawal:2020xek}.

\subsubsection{Evading the Difficulties}
 The above, rather general conclusions strongly depend on several of the above-listed assumptions of the standardly adopted EFT-relying framework. Even so, specially tuned and somewhat exceptionally ``balanced'' models have been found that do produce a phenomenologically acceptable de~Sitter spacetime even within this self-consistent framework; for example, see\cite{Bento:2021nbb}.
 
 More to the point however, we argue below against relying on several of the above-listed assumptions, and especially against neglecting the stringy winding modes. Indeed, as presented below and in the next section, a discretuum of toy axilaton models with dS spacetime\cite{rBHM5,rBHM6,rBHM7,Berglund:2020qcu} initially developed following this standard approach to string theory turns out to invalidate several of the above assumptions, and so motivates our subsequent survey of a more general approach. With Section~\ref{s:dSdefo} reviewing the details of the construction, here we summarize several of the key aspects by which these models evade the above-mentioned difficulties and indicate that an approach beyond EFT is needed.

\paragraph{The \BHM\ toy models:}
{\em First}, these models\cite{rBHM5,rBHM6,rBHM7,Berglund:2020qcu} stem from a non-perturbative configuration of IIB string theory --- namely the D7-brane\cite{Cohen:1999ia}, which is a stringy cosmic string like configuration\cite{Greene:1989ya,rCYCY}: a codimensionsion 2 solution in 10 dimensions.
Now, this non-perturbative configuration is deformed in our toy models\cite{rBHM5,rBHM6}, which turn out to be non-perturbative\cite{rBHM7}, and so immediately violate one of the assumptions of the canonical approach
to dS via string compactifications. The other important aspect is that the model mixes UV and IR scales (precisely because of its codimensionsion 2 nature) and thus it violates the assumption regarding the separation of scales in the canonical approaches. Also, the model can be understood as an $S$-fold in the context of a doubled formulation\cite{Berglund:2020qcu}, in which, generically the ``would be winding modes'' are kept as coordinates of a dual spacetime, which, in general, does not commute with the observed
spacetime. 

{\em Second}, the cosmological constant is stabilized via an effective non-commutative field theory\cite{Berglund:2020qcu}. To the lowest order in the non-commutativity parameter (the size of the string), this can be treated as a sequestered effective field theory\cite{Kaloper:2013zca,Kaloper:2014dqa}. Thus the cosmological constant, or vacuum energy, is natural, by the seesaw formula of the toy model\cite{rBHM5,rBHM6,rBHM7}, and radiatively stable.
 Analogous arguments apply to the stabilization of the moduli of the K3 space in these toy models, which involve a co-dimension two solution in six dimensional spacetime, where the extra four dimensions are compactified into a K3.
 These moduli are also subject to a non-commutative effective field theory, and thus to a sequestered effective field theory treatment. 

{\em Third}, the \BHM\ toy models are non-holomorphic deformations of the stringy cosmic string, and thereby break supersymmetry --- but again in a way that can be understood from a general formulation of string theory. In fact, the \BHM\ models interpolate non-holomorphically between two distinct holomorphic superstring solutions\cite{rBHM7}.
 This supersymmetry breaking should be treated as the stabilization of vacuum energy, via the non-commutative field theory involved, and its sequestered effective field theory limit.
 This supersymmetry breaking is thereby ``cosmological'' and directly related to the
seesaw formula for the cosmological constant of the effective four dimensional dS space.

Furthermore, the \BHM\ solution matches the two fundamental backgrounds of string theory, the axion and dilaton, with the two dimensional transverse space, and in this sense is somewhat similar to what happens in non-critical string theory, where the Liouville field, originating from the conformal anomaly, leads to one extra dimension. Axilatons appear in other situations as effective moduli, so that this discretuum of models can be translated also into that context.

Alternatively, it is also possible to treat the dS 4d spacetime as (the Lorentzian Wick-rotation of) the exceptional set of a resolution of a singularity in 
a Calabi-Yau (complex) 5-fold, which in turn is itself a Euclidean Wick-rotation of the 10d spacetime of type~IIB strings\cite{rBHM7}. Via this Wick-rotation, this relates the resolution of a singularity in the (auxiliary) Euclidean picture to the emergence of the cosmological constant in the (actual) Lorentzian geometry of spacetime.

{\em In turn}, these various non-standard features of the \BHM\ models fit naturally within a recent reformulation of string theory\cite{Freidel:2015pka,Freidel:2017xsi} and its particle-like limit\cite{Freidel:2018apz}.
The crucial feature here is that string theory is revealed to have 
a non-commutative and chirally doubled target spacetime with a phase-space like structure. Therein, the familiar, ``visible'' target spacetime ($x$) of string theory has a non-commuting dual $(\tx$): $[x^\mu,\tx_\nu]\,{=}\,\l^2\d^\mu_\nu$.
 This affords a continuum of distinct polarizations, i.e., ``representations'' or ``pictures.''
The usual string theory (including double field theory) is a commutative limit of this non-commutative formulation. This is analogous to the so-called Seiberg-Witten limit of non-commutative field theories\cite{Seiberg:1999vs} found in open string theory in $B$-backgrounds. This commutative limit is quite useful in practical calculations, but it is a singular limit: In it the general non-commutative observables appear as degenerate, and this limit must therefore be interpreted with care.
Our formulation is closer to the Grosse-Wulkenhaar formulation\cite{Grosse:2004yu}, which is intrinsically non-commutative and is formulated with a doubled UV/IR covariant regulator. As such this formulation does not suffer from UV/IR mixing, and it does have a well defined and non-trivial continuum limit.

{\em Finally}, the \BHM\ toy models are not supersymmetric, and their results require additional care and justification.
 The underlying logic of the usual approach is that supersymmetric constructions are controllable and reliable in calculations. This of course always brings up the question of the inherently non-supersymmetric nature of dS spacetime, as well as the fact that both general relativity and the Standard Model of particle physics are non-supersymmetric. The basic point here is that if the \BHM\ toy models are embedded within this non-commutative, chirally doubled, phase-space like general formulation of string theory, then this provides the technical stability instead of supersymmetry.
 To wit: In the case of double field theory\cite{Tseytlin:1990nb,Tseytlin:1990hn,Tseytlin:1990va}, it is found that the uniqueness of certain crucial geometric data (more precisely, the uniqueness of the connection in the context of generalized geometry) is predicated upon supersymmetry.
 On the other hand, the non-commutatively and chirally doubled target spacetime with a phase-space like structure exhibits the so-called Born geometry, which features a symplectic structure (but no supersymmetry), and which also guarantees the uniqueness of the connection\cite{Freidel:2017yuv}.
 In this technical sense, the symplectic structure replaces supersymmetry as a powerful underlying structure in the context of Born geometry and the general non-commutative phase-space like bosonic formulation of string theory, the metastring. It is natural to conjecture that metastring theory does possess a {\em\/complete\/} structural (and utilitarian) analogue of supersymmetry\footnote{In particular, it is tempting to suppose that the structure of Born geometry, perhaps amended by some natural restrictions such as anomaly cancellation, can provide for some non-renormalization theorems akin to supersymmetry, and consequently insure the existence of ``quasi-topological'' sectors. As discussed in \SS\,\ref{s:GS} and definitions~\eqref{e:IJK} in particular, Born geometry features a hallmark triple of mutually anticommuting complex structures, which may well provide a rigidity comparable to supersymmetry.} that is based on the robustness of Born geometry,
viewed as the geometry of general quantum theory. We will return to this point below.

\subsection{Swampland, Landscape and All That}
\label{s:Swamp}
One possible implication of the approach to dS via flux compactifications is that 
string theory leads to the concept of landscape of vacua\cite{Susskind:2003kw, Polchinski:2006gy}, and that as such it
represents a complex system\cite{Denef:2004ze, Denef:2006ad}. This then should be studied using techniques from complex systems, such as spin glasses, protein folding etc., ultimately leading to the implementation of machine learning and AI technology\cite{He:2017aed}.
The key question then is the evaluation of measures on the landscape\cite{Linde:2006nw}, and there exists a holographic measure for the cosmological constant\cite{Horava:2000tb}.

The string landscape is sometimes put together with the multiverse picture\cite{Vilenkin:2006xv}
of eternal inflation, and also with the anthropic principle, the latter of which is used to argue for a small and positive cosmological constant, following Weinberg's classic work\cite{Weinberg:1987dv,Weinberg:1988cp, Weinberg:2000yb}.
All this raises fundamental questions regarding the predictability of string theory
(and eternal inflation). Note that the problem of finding dS space in string theory
is also related to finding inflation in string theory with natural slow-roll parameters.

Another possibility considered in the recent literature is that there is no dS in canonical
string theory compactifications, which forces us to model the cosmological evolution of the universe with a quintessence model of dark energy.
This idea is compatible with the swampland approach, which seeks to find general
constraints on the coupling of effective field theories to {\em\/quantum gravity,} which is incompletely understood and so is typically modeled by standard approaches to string theory. While other approaches to quantizing gravity do exist, it has been argued that those are in fact special limits of string theory\cite{Dijkgraaf:2004te}. Although this contention does require considerable further analysis for an ultimate confirmation (or correction), it behoves to recall the well known fact that the ``quantum-to-classical'' limit mapping is ``many-to-one'': There almost always exist multiple {\em\/different\/} quantum theories that have the same given classical (and even semi-classical) limit.
 
Therefore, aiming for an effective Standard Model-like quantum field theory coupled to Einstein gravity in the classical limit, including the ``semiclassical gravity'' approach, cannot possibly suffice in singling out the complete underlying quantum theory. This is where we use the underlying string theory for guidance, in that --- whatever the ultimate theory of quantum gravity may turn out to be, we assume it will share the self-consistent characteristics that can be gleaned from string theory. This is sometimes referred to as the {\em\/String Universality\/} or the {\em\/String Lamppost Principle\/}\cite{vanBeest:2021lhn}.
 Furthermore, we argue herein that this coveted goal requires a string theory framework\cite{Freidel:2015pka,Freidel:2017xsi} that is more refined than the standard description\cite{rGSW1,rGSW2,Polchinski:1998rq,Polchinski:1998rr}.

\paragraph{A Baker's Dozen of Swampland Criteria and Conjectures:}
A notion developed over the last two decades, that the vast {\em\/landscape\/} of string theory models is in fact immersed within an even larger collection, dubbed {\em\/swampland,} of self-consistent semiclassical effective field theories, which are however inconsistent as quantum theories when gravity is incorporated\cite{Vafa:2005ui}. Subsequent work refined this notion and resulted in a collection of increasingly better refined statements, starting with the so-called {\em\/Swampland Criteria\/}\cite{Agrawal:2018own,Andriot:2018wzk}; see also the lectures\cite{Agmon:2022thq}. These preemptive requirements imply that the envisioned effective field theory description of models coupled to quantum gravity {\em\/breaks down\/} if either of the following two circumstances occurs:
\begin{enumerate}\itemsep=-1pt\vspace*{-1mm}

 \item\label{i:smallD}
  The scalar quantum fields $\phi^i$ in the model (such as the moduli) vary much more than $O(1)$ in natural units\cite{Ooguri:2006in,Klaewer:2016kiy,Grimm:2018ohb,Heidenreich:2018kpg,Blumenhagen:2018hsh}.

 \item\label{i:slowV}
  The logarithmic gradient of the potential of those scalar fields ($|\nabla_{\!\!\phi^i}V|/V$) varies much more than $O(1)$ in natural units
\cite{Obied:2018sgi,Agrawal:2018own,Andriot:2018wzk,Denef:2018etk}; see also\cite{Garg:2018reu,Ooguri:2018wrx}.
\vspace*{-1mm}
\end{enumerate}
The first of these criteria is {\em\/complementary\/} to the so-called {\em\/swampland distance conjecture\/} (``SDC'')\cite{Ooguri:2006in}, in that they pertain to the ultimately opposite extremes in field variations:
\begin{enumerate}[resume]\itemsep=-1pt\vspace*{-1mm}
\item\label{i:SDC}
 Given two infinitely distant values of a moduli field, $\vev{\f_1}$ and $\vev{\f_2}$, the infinite tower of massive states at $\vev{\f_1}$, with the characteristic mass-scale $M_{\vev{\f_1}}$, collapses to a heap of massless states at $\vev{\f_2}$: $M_{\vev{\f_2}}\sim M_{\vev{\f_1}}\,e^{-\a\,d(\vev{\f_1},\vev{\f_2})}$, $\a>0$; see also\cite{Klaewer:2016kiy,Blumenhagen:2018nts,Lee:2019wij}.
 
  \item\label{i:newPh}
  As an immediate corollary of~\ref{i:SDC}, extending an EFT to an infinitely distant value of some moduli field collapses infinitely many states to become massless --- which therefore must stem from a new massless {\em\/extended\/} object, i.e., new physics,\footnote{Notice that, e.g., the ``conifold'' points are in the physically relevant (Weil-Petersson-Zamolodchikov) metric at finite\cite{Candelas:1988di}, in fact $O(1)$ distance\cite{Blumenhagen:2018nts} in higher-dimensional moduli spaces; they presumably host ``new physics,'' but are undetected by just the SDC, listed as item~\ref{i:newPh} above. In turn, the $\psi\,{\to}\,\infty$ limit in the Dwork family of quintics\cite{Candelas:1990qd,Candelas:1990rm} corresponds to a highly singular model, quite likely inducing the string compactification to include ``new'' extended objects, suggesting that SDC may not not be a sufficient predictor of ``new physics''; see also the novel results on conifold transitions\cite{Anderson:2022bpo}.} such as in\cite{Baume:2019sry,Marchesano:2019ifh,Font:2019cxq}; see also\cite{Montero:2022prj}.
\end{enumerate}

We continue with the key conjectured properties of consistently gravitational EFTs following the comprehensive, detailed and much recommended surveys in\cite{Brennan:2017rbf,Palti:2019pca,vanBeest:2021lhn,Grana:2021zvf}; for a recent, astrophysically minded survey, see\cite{Silk:2022xbg}. In most statements, we also provide {\em\/some\/} pointers to additional literature, and we number the properties continuing the swampland criteria, since these statements are closely interrelated.
\begin{enumerate}[resume]\itemsep=-1pt\vspace*{-1mm}

 \item\label{i:noGlobalS}
  No symmetry is global and exact in a gravitational EFT; it is either gauged or broken, or was approximate to begin with\cite{Banks:1988yz,Banks:2010zn}.

 \item \label{i:finiteN}
  There can only be a finite number of massless fields in a gravitational EFT.
  (By its nature, this is an observation about {\em\/most\/} and {\em\/well-understood\/} models; in fact, the emergence of an infinite number of massless fields is accepted as an indication of a massless/tensionless extended object or decompactification, i.e., some new physics.\footnote{On the face of it, this conjecture excludes some {\em\/more exotic\/} proposals with infinitely many massless fields. Such is the example of so-called Vasiliev gravity\cite{Fradkin:1987ks}, which however {\em\/can\/} be reinterpreted in terms of tensionless strings\cite{Anninos:2011ui}, and thereby as corroborating this conjecture.})
  
 \item\label{i:noParams}
  There are no free/independent parameters in a gravitational EFT; all parameters are vacuum expectation values of corresponding quantum fields. 
  (By its nature, this conjecture is both a {\em\/credo\/} of ultimate unification in
   physics, as well as an observation of circumstances of string theory.)
  
 \item\label{i:Cobordism}
  A $d$-dimensional theory of quantum gravity can be compactified only on $k$-dimensional compact spaces of trivial cobordism class\cite{McNamara:2019rup}. 
  (This conjectured complete connectivity is the ultimate generalization of the early
   indications of `rolling connections' among string
    vacua\cite{Green:1988uw,Green:1988wk,Candelas:1989ug}.)
 
 \item\label{i:allChargss}
  A gravitational EFT with a gauge symmetry group $G$, states of all possible $G$-charges consistent with Dirac quantization must exist\cite{Polchinski:2003bq,Heidenreich:2021xpr}. 
 
 \item\label{i:WGC}
  A $d$-dimensional gravitational EFT with a $p$-form $U(1)$ gauge symmetry and gauge coupling $e_p$ has the following {\em\/weak gravitational conjecture\/} (``WGC'') bounds\cite{Arkani-Hamed:2006emk,Heidenreich:2015nta}:
  \begin{enumerate}\itemsep=-1pt\vspace*{-1mm}
   \item Electric WGC:
   At least one $(p{-}1)$-brane must have its tension, $T_p$ ($T_1\,{=}\,m$), and charge, $q_p$, satisfy the inequality $\sqrt{\frac{p(d{-}p{-}2)}{d{-}2}}\,T_p\,{\leqslant}\,q_p\,e_p\,(M_P^{\sss(d)})^{\frac{d-2}2}$.
   \item Magnetic WGC: The UV cutoff scale of the EFT is bounded by the relation $\L_{UV} \lesssim\big(e_p\,(M_P^{\sss(d)})^{d-2}\big)^{1/2p}$.
  \vspace*{-.5\baselineskip}
  \end{enumerate}
The generalization to multiple $U(1)$s results in the {\em\/Convex Hull Condition,\/} which is more restrictive than the conjecture applied to each $U(1)$ gauge symmetry separately.

 \item\label{i:nonSySyAdS}
  Non-supersymmetric AdS vacua must be unstable in quantum gravity\cite{Ooguri:2018wrx}; in fact, this is argued to be the case for all non-supersymmetric vacua\cite{vanBeest:2021lhn}.
 
 \item\label{i:specScale}
  A $d$-dimensional EFT with $N_s$ particle states is weakly coupled to gravity below the energy scale $\L_s=M_P^{\sss(d)}/\sqrt[d-2]{N_s}$.

 \item\label{i:FestinaLente}
  In a gravitational EFT in a 4-dimensional dS background, the mass of every charged particle must satisfy $m^2 \gtrsim eqM_P^{\sss(4)}H$, where $H$ is the Hubble constant\cite{Montero:2019ekk,Montero:2021otb}.
 
\end{enumerate}

\paragraph{Commentary:}
The criteria~\ref{i:smallD} and~\ref{i:slowV} are restated so as to remind of the standard expectations in the classical treatment of small oscillations:
small field variations within a regime of a slowly varying potential, respectively. Here however the inclusion of quantum gravity provides the natural, Planckian units for the field space.\footnote{Logically, these ``Planckian units'' ought to pertain to the ultimate high-energy regime of string theory, and this oft-cited as 10d Planck energy may well be exponentially lower than its effective 4d value of $10^{19}$\,GeV; see, e.g., the toy models described in Section~\ref{s:dSdefo}.} Indeed, even the formulation of the second of these (list item~\ref{i:slowV} above) is predicated on the existence of a semi-classically behaving effective potential for the considered fields.
 This then in fact corroborates one of the main claims herein: that the effective field theory description of fundamental physics cannot be complete.

Furthermore, conjecture~\ref{i:SDC} implies that an EFT with quantum gravity formulated at any particular point in the moduli space cannot possibly be used  consistently to describe the situation at an infinitely distant point in the moduli space; see also the statements~\ref{i:newPh} and~\ref{i:finiteN} in the listing above.

Statements~\ref{i:finiteN} and~\ref{i:noParams} in the above listing are generalized from direct observations in all known string theory models, and those explicitly exclude ostensibly esoteric field theory models, such as those with infinitely many massless fields\cite{Fradkin:1987ks}.

Also, a closer analysis of several of the above statements indicate a key role of back-reaction, which present issues that cannot seem to be resolved within EFT, such as the ``Weak Gravity Conjecture'' (list item~\ref{i:WGC} above) in systems that include extended objects\cite{Lanza:2020qmt}. Furthermore, so-called axionic strings indicate an ``effective breakdown'' of EFT\cite{Lanza:2021udy}. In turn, WGC has recently also been used to argue that a phenomenologically acceptable value of the cosmological constant can be induced by loop corrections\cite{Danielsson:2022lsl}.

All in all, rather than interpreting these statements as condemning various logical options (including dS spacetime in particular) to swampland or worse, they highlight the range and span of applicability of gravitational EFTs. Among them is also then the conjectured conclusion: ``dS space does not exist as [{\em\/a gravitational EFT description of\;}] a consistent quantum theory of gravity''\cite{Brennan:2017rbf}, where we explicitly qualified the quoted statement by inserting (in square brackets) the tacit assumption.

Indeed, the abundant reports supporting the above-listed conjectures may be seen as corroborating evidence indicating the need to go beyond gravitational EFT modeling, such as adopting a more detailed picture in which the chirally doubled, non-commutative generalized-geometric formulation of string theory can lead to a positive cosmological constant.
Therein, the curvature scalar of the dual spacetime is the cosmological constant in the observed spacetime, and the size of the dual space is the gravitational constant in the same observed spacetime. These are lowest-order results\cite{Berglund:2020qcu}. Also, the three scales associated with
 {\small({\bf\itshape a\/})}~intrinsic non-commutativity of string theory,
 {\small({\bf\itshape b\/})}~the cosmological constant scale and
 {\small({\bf\itshape c\/})}~the Planck scale,
naturally satisfy a seesaw-like relation by $T$-duality. We also discuss various implications for
dark matter, hierarchy problem as well as a new non-perturbative approach to string theory and quantum gravity.
Along the way, we show that these novel features of string theory can be implemented in a particular deformation of cosmic-string like models.

Regarding the question ``what physics is responsible for the masses and coupling of elementary particles
and their interactions?'' an alternative approach that fits the narrative reviewed in this article is the so called
``Universe as an attractor idea'' in which the fundamental interactions vertex of string field theory,
or the above-mentioned non-perturbative formulation of metastring theory in terms of an abstract, chirally doubled, non-commutative
and non-associative matrix quantum theory, introduces an effective ``horizontal transfer of information''
that has been used in biological physics to argue for the universality of the genetic code\cite{Argyriadis:2019fwb}.
We will address this in the penultimate section of this review.

\subsection{A Sorting Survey of Assumptions}
\label{s:Class}
It is worth taking a step back, to survey some of the general assumptions that have over the past four decades been built into seeking string vacua that admit a physically realistic interpretation.
 Foremost, the foregoing discussion relied on considerations of presumed Ans{\"a}tze for the target spacetime structures (topology and geometry), which are complemented by
 ({\small\bf1})~insights and results from the worldsheet (superconformal) quantum field theory point of view (which also incorporate models for which no known target space geometry has yet been identified), and
 ({\small\bf2})~the continually developing network of dualities, mirror symmetry and other relations among the resulting models.
 Although immensely varied and rich, this palette of possibilities still hides previously unexpected ``pockets of possibilities'' (see for example\cite{Bento:2021nbb}), and as we will argue below, a more detailed analysis of the underlying worldsheet level of string theory harbors structural details\cite{Freidel:2015pka,Freidel:2017xsi}, which have far-reaching consequences that still remain to be explored\cite{Berglund:2020qcu,Berglund:2021xlm,Berglund:2021hbo}.

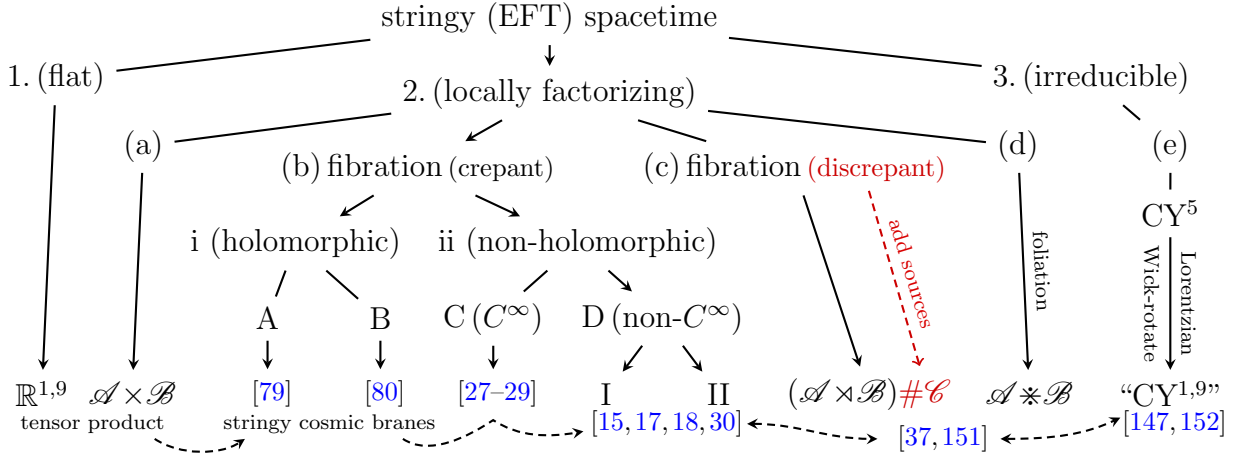
\begin{figure}[htb]
 \begin{center}
   \begin{tikzpicture}[xscale=1.5,every node/.style={inner sep=1mm,outer sep=.5mm}]
     \path[use as bounding box](-.2,.4)--(10.2,-5.2);
     \node(S) at(4.5,0) {stringy (EFT) spacetime};
      \node(1) at(.125,-.8) {1.\,(flat)};
       \node(F) at(0,-5) {$\IR^{1,9}$};
      \node(2) at(4.5,-1) {2.\,(locally factorizing)};
       \node(2a) at(.9,-1.7) {(a)};
        \node(T) at(.8,-5) {$\mathscr{A}{\times}\mathscr{B}$};
       \node(2b) at(3.33,-2) {(b)\,fibration\,\footnotesize(crepant)};
        \node(2bi) at(2.25,-3) {i (holomorphic)};
         \node(2biA) at(2,-4) {A};
          \node(SCS) at(2,-5) {\footnotesize\cite{Greene:1989ya}};
         \node(2biB) at(3,-4) {B};
          \node(CYCY) at(3,-5) {\footnotesize\cite{rCYCY}};
        \node(2bii) at(4.75,-3) {ii (non-holomorphic)};
         \node(2biiA) at(4,-4) {C\,($C^\infty$)};
          \node(BHM1-3) at(4,-5) {\footnotesize\cite{rBHM1,rBHM2,rBHM3}};
         \node(2biiB) at(5.5,-4) {D\,(non-$C^\infty$)};
          \node(I) at(5,-5) {I};
          \node(II) at(6,-5) {II};
       \node(2c) at(6.67,-2) {(c)\,fibration\,\color{red!80!black}\footnotesize(discrepant)};
        \node(fd) at(7.33,-5)
                    {$(\mathscr{A}{\rtimes}\mathscr{B})\color{red!80!black}{\#}\mathscr{C}$};
        \draw[red!80!black, densely dashed, ->, thick](7.33,-2.3) to node
                [above, rotate=-73]{\scriptsize\color{red!85!black}add sources}++(.45,-2.33);
       \node(2d) at(8.65,-1.7) {(d)};
        \node(fo) at(8.75,-5) {$\mathscr{A}{\divideontimes}\mathscr{B}$};
      \node(3) at(9.3,-.8) {3.\,(irreducible)};
       \node(3e) at(10,-1.7) {(e)};
       \node(5) at(10,-2.6) {CY$^5$};
       \node(HH) at(10,-5) {``CY$^{1,9}$''};
       \node(A-dS) at(7.95,-5.6) {\footnotesize\cite{Banerjee:2018qey,Blaback:2019zig}};
     \draw[-stealth, thick](S)--(1)--(F);
     \draw[-stealth, thick](S)--(2); \draw[-stealth,thick](2)--(2a)--(T);
      \draw[-stealth, thick](2)--(2b);
       \draw[-stealth, thick](2b)--(2bi);
        \draw[-stealth, thick](2bi)--(2biA)--(SCS);
        \draw[-stealth, thick](2bi)--(2biB)--(CYCY);
       \draw[-stealth, thick](2b)--(2bii);
        \draw[-stealth, thick](2bii)--(2biiA)--(BHM1-3);
        \draw[-stealth, thick](2bii)--(2biiB);
         \draw[-stealth, thick](2biiB)--(I);
         \draw[-stealth, thick](2biiB)--(II);
      \draw[-stealth, thick](2)--(2c)--(fd);
      \draw[-stealth, thick](2)--(2d) to node[above, rotate=-87]{\scriptsize foliation}(fo);
     \draw[-stealth, thick](S)--(3)--(3e)--(5) to node
                [above, rotate=-90]{\scriptsize Lorentzian} node
                [swap, below, rotate=-90] {\scriptsize Wick-rotate} (HH);
     \path(.45,-5.4) node {\scriptsize tensor product};
     \path(2.55,-5.4) node {\scriptsize stringy cosmic branes};
     \path(5.5,-5.4) node {\footnotesize\cite{rBHM4,rBHM5,rBHM7,Berglund:2020qcu}};
     \path(10,-5.4) node {\footnotesize\cite{rHitch,Berglund:2021xlm}};
     \
     \draw[densely dashed,-stealth,thick]
             (1,-5.6)to[out=-45,in=-135](1.8,-5.6);
     \draw[densely dashed,-stealth,thick]
             (3.2,-5.6)to[out=-45,in=-135](4,-5.35)to[out=-45,in=-165](4.8,-5.5);
     \draw[densely dashed,stealth-stealth,thick]
             (6.25,-5.4)to[out=0,in=-180](A-dS);
     \draw[densely dashed,stealth-stealth,thick]
             (A-dS)to[out=0,in=-150](HH);
   \end{tikzpicture}
 \end{center}
 \caption{A (partial) family tree of various constructions mentioned herein, lined up here from simplest at the far left to the tentatively highest complexity at the far right}
 \label{f:FTree}
\end{figure}

The notion of spacetime geometry that is admissible and realizable in string theory has evolved over the past half a century, since\cite{Scherk:1974ca,Scherk:1974mc}, and has brought about many key models and their variations, as (partially and coarsely) sketched in Figure~\ref{f:FTree}; see also\cite{Berglund:2021xlm}.
The evolutionary/family branch 2.(b).ii.D.II (where the de~Sitter \BHM\ models live) may well converge with 3.(e), since the hallmark non-analyticity in the former provides in the latter for the ($\d$-function like) isolation of dS-like sub-spacetimes.
In the 2.(c)-branch, ``discrepant''\footnote{This terminology stems from the technical terms in algebraic geometry, where the smoothing of singularities introduces a {\em\/discrepancy,} a change in the 1st Chern class, i.e., the Ricci tensor (modulo total derivatives). Independently, higher-rank curvature and more general geometric characteristics may also change\cite{Anderson:2022bpo}. 
 Furthermore, some of the ``source locations'' ($\mathscr{C}$) of such changes were even identified as {\em\/exo-curves\/}\cite{rMR-MM3,rMR-YPG,rAGM04,rHR01}, whereby the total spacetime becomes a {\em\/stratified pseudomanifold\/}: 
 A spacetime in this branch may well have lower-dimensional components that connect to the top-dimensional bulk via a yet lower-dimensional intersection; see also the ``hybrid models'' of\cite{rPhases}.} means that (stringy) Ricci-flatness (anomaly cancellation) is restored by including suitable additional {\em\/sources,} i.e., geometrical/topological {\em\/defects,} $\mathscr{C}$. Also, the 2.(d) branch features {\em\/foliations,} $\mathscr{A}{\divideontimes}\mathscr{B}$, which are only locally, but not globally, product spaces; this type of geometry will re-emerge below, in the metastring description of spacetime\cite{Freidel:2018tkj,Berglund:2021hbo}.

\paragraph{Backtracking:}
Inspired by Refs.\cite{rFrGaZu86,rBowRaj87,rBowRaj87a,rBowRaj87b,Oh:1987sq,rHHRR-sDiffS1,Pilch:1987eb,rBowRaj88,Bowick:1988nj,Bowick:1990wt}, we adopt the Ricci-flatness of the configuration space as a necessary (and possibly sufficient) condition for quantum consistency. The the free loop-space, $\mathrm{Ric}(\cL\sX)$, of the stringy spacetime\footnote{By ``stringy spacetime'' we mean a topological space equipped with a metric that may in certain locations degenerate/singularize but in ways that ultimately prove to be innocuous to (super)string dynamics. Such include spacetimes with local finite quotient (orbifold) singularities, conifolds (``$A$-$D$-$E$ singularities''), ``normal stratified pseudovarieties'' (where strata of different dimensions are connected in real codimension-2 loci, thus featuring external, ``exo-branes'' first appearing, {\em\/tacitly\/} in Ref.\cite{rAGM04}, then explicitly so-identified in Ref.\cite{rHR01}), and possibly worse singular spaces\cite{rSingS}. Without a nonperturbatively complete (super)string theory, this specification must necessarily remain open-ended.} $\sX$ in this original analysis would have to be generalized to include all collective degrees of freedom (branes, defects, etc.) and should also be double-covered to account for left- and right-movers, i.e., loop orientations. The need for this latter modification was foreshadowed by the original computations of vacuum energy by Polchinski, as discussed in \SS\;\ref{s:VEmeta}; see p.\;\pageref{p:Joe's2}, below~\eqref{e:mGreen}.
 The homotopy and (co)homology of $\cL\sX$ is fully determined by the analogous data on $\sX$\cite{rBeast,rBottTu}, so that $\mathrm{Ric}(\cL\sX)=0$ implies $R_{\mu\nu}(\sX)=0$ up to total derivatives.\footnote{On spacetimes with boundaries or singularities, this requirement is, in general, corrected by contributions localized to those special subspaces; the analogous follows for all other types of {\em\/topological defects.\/}} These statements presume a metric, $g_{\mu\nu}$, of Lorentzian signature on $\sX$, and of course imply that $R(\sX)=g^{\mu\nu}R_{\mu\nu}(\sX)=0$, but this latter (scalar) condition is significantly weaker:
 ({\small\bf1})~the metrics in the 2.(b).ii branches varies non-analytically over spacetime, rendering $R_{\mu\nu}$ a {\em\/distribution\/} (with step-function, $\d$-function and other discontinuities and defects), while
 ({\small\bf2})~the 2.(c) branch involves {\em\/stratified pseudomanifolds,} which admit only piece-wise defined metrics.
 These technical details both help evading various dS no-go theorems (which presume compactness, sufficient differentiability, \dots), but also render a comprehensive and conclusive analysis much more difficult.

 Ricci-flatness is also related to supersymmetry: The topological space(time) $\sX$ is K{\"a}hler if it admits a metric $\mathfrak{g}_{\mu\nu}$ with Euclidean signature and a complex structure with respect to which $\mathfrak{g}_{\mu\nu}$ is Hermitian, everywhere positive definite, so that  $\mathfrak{g}_{\mu\bar\nu}\,{=}\,\pa_\mu\pa_{\bar\nu}K$ for a (K{\"a}hler) potential $K$. Then, the existence of a Ricci-flat K{\"a}hler metric on $\sX$ implies the existence of the covariantly-constant ``holomorphic volume form'' (the holomorphic anticanonical bundle, $\cK_\sX^*\!\define\!\det(T_\sX^{(1,0)})\!\approx_\IC\!\cO_\sX$ is trivial), which on spin manifolds implies the existence of a covariantly constant spinor, and so the preservation of a globally well-defined supersymmetry. For a priori Lorentzian spacetimes $\sX$, one usually presumes to be able to factor (at least in some fibration or foliation sense) a flat $\IR^{1,d}$, leaving a nontrivial $(9{-}d)$-dimensional factor with a Euclidean metric. The real world is of course not supersymmetric, and we seek more general geometries, which include --- as the {\em\/generic\/} case --- the irreducible (non-factorizable, non-product) CY$^5\to\text{``CY}^{1,9}$,'' with a Lorentzian metric of the dS type (and broken supersymmetry) at least within a well isolated/separated $(1,3)$-dimensional subspace.
 
Finally, it bears emphasizing that all these and related Ans{\"a}tze presume stringy spacetime (as in Figure~\ref{f:FTree} and more) to be describable by some EFT in the target space of vev's of the worldsheet quantum fields --- which in turn have been standardly identified over the past 15 years\cite{Polchinski:1998rq,Polchinski:1998rr}.
{\em\/This\/} is the key assumption, which has been questioned recently\cite{Freidel:2015pka,Freidel:2017xsi}, which we argue lacks sufficient justification, and which opens an avenue for novel insights.
This need is in fact indicated by the through studies of ``swampland criteria'' as we discussed in Section~\ref{s:Swamp}, and some systemic modification has even been explicitly called for from very different aspects\cite{Sethi:2017phn,Blumenhagen:2020doa,Lanza:2021udy,Anderson:2022bpo}.

\paragraph{Conventional Approach:}
One seeks to identify {\em\/a part\/}\footnote{This {\em\/part\/} may well be a simple subspace of the total spacetime $\sX$, a quotient of $\sX$, an exceptional subset within $\sX$, an external stratum attached to $\sX$, \dots\ and any combination thereof.} of the target spacetime, $\sX$, in superstring models as a candidate for the observed $3{+}1$-dimensional asymptotically de~Sitter spacetime, $\sW^{3,1}_{\sss\rm dS}$. This may be (re)organized according to how the various superstring degrees of freedom couple to each other, and in particular to those spanning $\sW^{3,1}_{\sss\rm dS}$ and the matter localized within it; see Figure~\ref{f:FTree}. In cases where a geometric description is known,\footnote{This pool increases over time, both by introducing genuinely new constructions, but also as many constructions previously deemed ``non-geometric'' have since acquired a geometric interpretation, and if by generalizing the notion of geometry itself.} we may organize this information by assigning the target spacetime a topological space, over which certain {\em\/structures\/} are defined: In superstring models, supersymmetry is typically associated with a 
 ({\small\bf1})~{\em\/complex structure,} at least in some directions of the target spacetime, $\sX$. Mirror symmetry, which is closely related to $T$-duality, implies the analogous relevance of a 
 ({\small\bf2})~{\em\/symplectic structure\/}\cite{Strominger:1996it}.
The loop-space approach\cite{rFrGaZu86,rBowRaj87,rBowRaj87a,rBowRaj87b,rHHRR-sDiffS1,Pilch:1987eb} in turn indicates the existence of an overall 
 ({\small\bf3})~{\em\/K{\"a}hler structure,} i.e., choices of (Ricci-flat) K{\"a}hler metrics, also to be a very general and inherent characteristic --- which is well defined both from the complex-analytic and from the symplectic ``side.''
 Finally, any one of these structures may well {\em\/fail to extend\/} to certain suitably small subsets (``singularities'') and in suitably ``innocuous'' ways. This was first discovered to be the case with ``orbifold singularities''\cite{Dixon:1985jw,Dixon:1986jc}, but is a considerably more general feature in string theory\cite{Green:1988wk,Green:1988uw,Candelas:1989ug,Candelas:1988di,Horowitz:1989bv,Hubsch:1990ua}; see also\cite{McNamara:2019rup}.
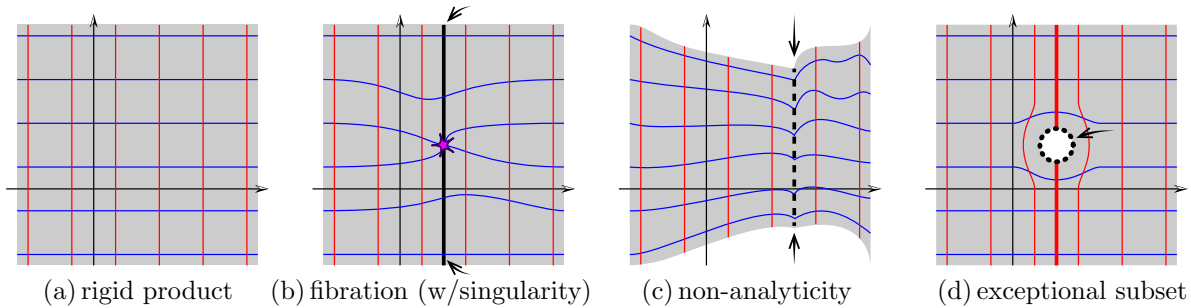
\begin{figure}[htbp]
 \begin{center}
  \begin{picture}(160,35)
   \put(0,0){\includegraphics[width=160mm]{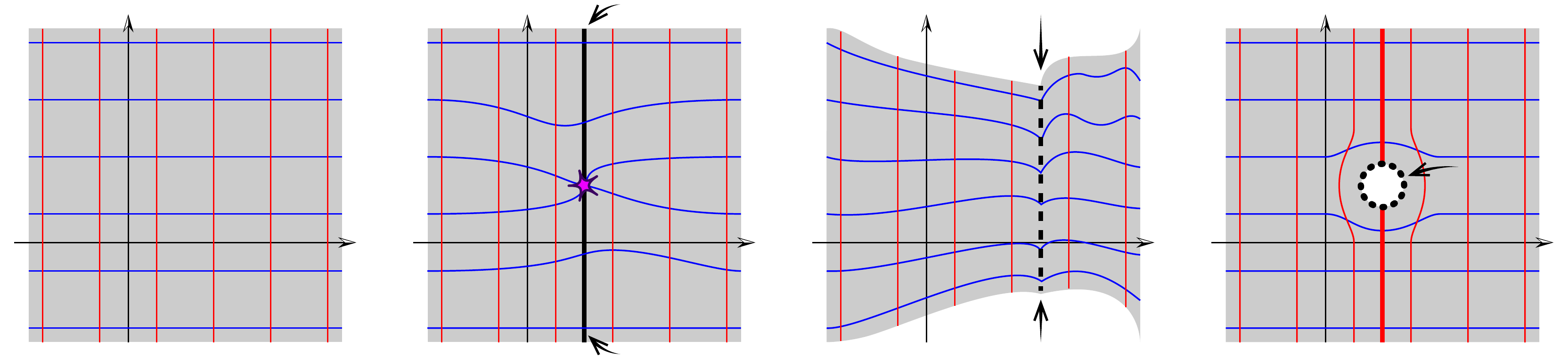}}
   \put(19,-3){\makebox[0pt][c]{\footnotesize(a)\,rigid product}}
   \put(58,-3){\makebox[0pt][c]{\footnotesize(b)\,fibration (w/singularity)}}
   \put(100,-3){\makebox[0pt][c]{\footnotesize(c)\,non-analyticity}}
   \put(142,-3){\makebox[0pt][c]{\footnotesize(d)\,exceptional subset}}
  \end{picture}
 \end{center}
 \caption{A cartoonish sketch of some possible geometries in target spacetime hints at the behavior of any of the relevant structures, which may combine in various ways}
 \label{f:4Cases}
\end{figure}

\noindent~~$\triangleright$\;{\bf\itshape Rigid product:}
The simplest of course is flat target spacetime, $\IR^{1,9}$: all relevant structures conform to the rigid tensor product of $\IR^1$-like degrees of freedom;
 see branch 1.\ in Figure~\ref{f:FTree} and Figure~\ref{f:4Cases}\,(a).
The same is true also of various (``static'') compactification frameworks, where some global, rigid factor in spacetime is different from $\IR^n$, such as the Freund-Rubin\cite{Freund:1980xh} or one of the myriads of Calabi-Yau compactification Ans{\"a}tze\cite{rCHSW}, in branch 2.(a) in Figure~\ref{f:FTree}; for a comprehensive listing see\cite{Kreuzer:2000xy,Avram:1997rs,Altman:2014bfa}.

\noindent~~$\triangleright$\;{\bf\itshape Fibration:}
In ``stringy cosmic strings''\cite{Greene:1989ya,rCYCY}, the direct product space is ``warped,'' so that the ``small,'' compact factor, $\sY^{\sss(6)}$, varies over the ``large,'' non-compact spacetime $\sW^{3,1}$. This variation may be smooth (as in a vector bundle), or occasionally singular (in various ways and to various degrees);
 see branches 2.(b) and 2.(c) in Figure~\ref{f:FTree} and
     Figure~\ref{f:4Cases}\,(b), where $\sY^{\sss(6)}$ (vertical) may singularize at some special locations of $\sW^{3,1}$ (horizontal). In the latter case, the structures (smoothness, complex, symplectic, K{\"a}hler, etc.) of the total spacetime may or may not remain continuous or analytic.

\noindent~~$\triangleright$\;{\bf\itshape Defect:}
 For over two decades by now, a non-analytic dependence of the metric\cite{Randall:1999ee,Randall:1999vf} has been known to enable novel geometries, as showcased by the \BHM\ models reviewed in Section~\ref{s:dSdefo}; see Figure~\ref{f:4Cases}\,(c). The total spacetime remains a direct product, but non-analyticity (or other structural non-uniformity) in ``warping'' singles out some of the instances of one factor, which then is a ``defect'' within the total spacetime; see branch 2.(b).ii.D in Figure~\ref{f:FTree}.

\noindent~~$\triangleright$\;{\bf\itshape Exceptional sets:}
 Finally, the various types of algebro-geometric ``surgery'' (e.g., ``blow-ups'' and ``small resolutions''\cite{rBeast}) proffer yet another possibility where the  ``exceptional sets'' are in no way a factor in the total spacetime, but an isolated subset, possibly of key interest and sometimes describable in terms of gravitational instantons\cite{Eguchi:1978xp,Eguchi:1978gw,Eguchi:1979yx}; see Figure~\ref{f:4Cases}\,(d). In Figure~\ref{f:FTree}, branches 2.(b,c,d) and 3.(e) all include this. By virtue of exoflops, exceptional (but {\em\/internal\/}) subsets are also related to connected but {\em\/external\/} strata of stratified pseudomanifolds\cite{Aspinwall:1993nu,Hubsch:2002st} and ``hybrid'' models\cite{rPhases}.
 
Features such as those outlined in the above (illustrative and by no means exhaustive) listing may well be combined in concrete models: For example, non-analyticity of a structure relevant to stringy spacetime dynamics at certain locations within the spacetime makes them {\em\/isolated\/} and {\em\/exceptional.} At certain locations, more than one such exception or singularization can occur while remaining admissible to string dynamics.

The main point of this very telegraphic and incomplete survey of possible geometries is that judicious combinations of even just some of the features illustrated in Figure~\ref{f:4Cases} easily evade the known ``no-go'' results about de~Sitter spacetime in string theory by invalidating at least one of their assumptions; see\cite{Sethi:2017phn} for fairly general results indicating this, as well as Section~\ref{s:dSdefo} for a show-casing toy example.
 In particular, direct product geometries (even with warped but not too singular metrics) result in their Ricci curvature tensors (and so also scalars) being additive. This however fails in general, when the spacetime does not factor.\footnote{This is an obvious corollary of the rigorously proven Tian-Yau theorem\cite{rTY1,rTY2}: Compact complex Fano ($R_{\mu\nu}\!>\!0$) varieties are unions of their compact Calabi-Yau ($R_{\mu\nu}\!=\!0$) hypersurfaces and their non-compact Calabi-Yau ($R_{\mu\nu}\!=\!0$) complements. The simplest example of this is $S^2$ (of positive curvature), which is the union of a {\em\/non-compact\/} (Ricci-flat) cylinder, and two points that form the {\em\/compact\/} (Ricci-flat) 0-fold. The key point is that these are not {\em\/reducible\/}: in no way do they involve a direct product of spaces.} Even more intricate is the case of ``exceptional sets,'' which are subspaces that are isolated by some lack of uniformity in some of the relevant structures (complex, symplectic, K{\"a}hler,\dots), and so imply exceptional behavior of at least some geodesics in their vicinity. Ultimately for a successful phenomenological application, it suffices for a given stringy model total target spacetime to have but a single $3{+}1$-dimensional de~Sitter subset, even if it is some sort of a structural ``defect,'' and is {\em\/sufficiently\/} isolated by any means. That is, the subset of the stringy degrees of freedom spanning the observable Standard Model ought to couple to {\em\/the rest\/} of stringy degrees of freedom only below current experimental bounds.

 A complete classification and a corresponding precise extent (and loopholes!) of all the various ``no-go'' results is far from known, but is clearly desirable for a definitive resolution.

Another point is that the detailed metric structure is important in determining properties such as geodesic (in)completeness, and for various distinct types of geodesics. For example, whereas Wick-rotating to Euclidean geometry can be helpful in certain aspects of analysis as it maps de~Sitter spacetime into an ordinary sphere, it also obscures certain key structures such as horizons and other geodesic features.

The foregoing repeatedly confirms the notion that the various string theory models are members of a (possibly multiply) connected\footnote{Since the early results about Calabi-Yau compactification\cite{Green:1988wk,Green:1988uw,Candelas:1989ug,rCHS-ZWP}, this notion has been extended over all other stringy models\cite{McNamara:2019rup}, affording even a collective study of certain subsets of this entire ensemble; see for example\cite{Perez:2020klz}.} ``theory space,'' giving a meaning to the physically relevant distance between them. It then seems reasonable to expect models that admit some form of a de~Sitter sub-spacetime (including all options, from a global factor to an isolated, exceptional ``bubble,'' and however isolated a ``structural defect'') to form a possibly very high codimension (``special'') subspace in this stringy theory space, somewhat akin to how conifolds (and their ``small resolutions'') are located within the discriminant locus in the moduli space of Calabi-Yau manifolds. Similarly then, at least some of the stringy models admitting a de~Sitter sub-spacetime may well be at finite distance from other (presumably more generic and non-de~Sitter) models. To this line of reasoning, the determination of the correct physically relevant metric is of key importance, and has been studied recently: for example, Ref.\cite{Stout:2021ubb} examines the physical reasons for such distances to diverge.
 This relates the issue to the ``Swampland Distance Conjecture''
\cite{Ooguri:2006in,Heidenreich:2018kpg,Grimm:2018ohb,Brennan:2017rbf,Palti:2019pca,vanBeest:2021lhn}, its CFT analogue\cite{Baume:2020dqd,Perlmutter:2020buo}, and even the quantum information theoretic metric and distance\cite{Provost:1980nc,Wootters:1981ki,Shapere:1989kp,Zanardi:2006wn,Gu:2010vv,Campos-Venuti:2007vv,Carollo:2019ygj}.
 Thus, not only is it relevant to determine if stringy models can admit a de~Sitter sub-spacetime (and if as an isolated ``structural defect,'' such as\cite{rBHM7,Berglund:2020qcu,rHitch} and\cite{Bento:2021nbb}), but the physical distance of such models from others would indicate if such de~Sitter sub-spacetimes can turn up via a phase transition, and whether it can be ``rolling'' or requires tunneling. Such inquiries are clearly a ``tall order,'' and remain wide open for now.

\paragraph{Lamppost Effect:}
At heart is the {\em\/embarras de richesse\/} of string theory, the configuration space of which is transfinitely richer than that in ordinary quantum field theory. Within this abundance, we seek to identify a {\em\/sector\/} that could approximate the observed asymptotically de~Sitter spacetime, $\sW^{3,1}_{\scriptscriptstyle\text{dS}}$, with the Standard Model of elementary particles and other observed details in it. The historically followed simple-to-complex progression of the explored geometries (left-to-right in Figure~\ref{f:FTree}) is then only reasonable, from a practitioner's point of view. It is also by no means guaranteed to find a solution.

\subsection{de~Sitter Bubbles}
\label{s:TT}
The early discovery that certain singular features in spacetime are important but innocuous to string dynamics\cite{Dixon:1985jw,Dixon:1986jc} shows that the choice of available geometries is, in this respect, considerably more general than in conventional Field Theory. A large class of such singular geometries affords topology change\cite{rGHC,Green:1988uw,Candelas:1989ug,Partouche:2000uq}, in a milder sense also\cite{Aspinwall:1993nu,Aspinwall:1994zd}, and so involve some form of a phase transition. This tends to harbor radical effects both in physics and in geometry: massless black holes\cite{Strominger:1995cz}, exoflops\cite{Aspinwall:1993nu,Hubsch:2002st}, $D$-branes\cite{Polchinski:1996fm}, and orientifolds\cite{Sagnotti:1987tw,Horava:1989vt,Bianchi:1990tb}, --- which frequently involve strong coupling regimes.

Models that are {\em\/near\/} being singular then invariably involve multiple competing contributions to the effective action, some of which are often hard to estimate. Nevertheless, at least in some cases, suitable approximations have been well argued, as is the case of warped deformations of conifold singularities\cite{Klebanov:2000hb,DeWolfe:2002nn,Douglas:2007tu}; see also\cite{Douglas:2009zn,Crino:2020qwk,Marchesano:2021gyv}. Here, the deformation of the conifold singularity and the relative size of the ``vanishing cycles,'' the $\overline{\text{D3}}$-brane, and the metric warping all affect the target spacetime physics. Detailed examinations of such nearly singular warped direct product compactification scenarios (Figure~\ref{f:4Cases}.b) find regimes in which the multiple competing contributions do induce the proper uplift for a metastable de~Sitter metric in the non-compact spacetime factor\cite{Bento:2021nbb,Marchesano:2021gyv}. Taken as case-in-point counter-examples to general conjectures\cite{Bena:2020xrh,Bena:2021wyr}, these indicate that models with stringy de~Sitter spacetime may well be very far from generic, and may well require a highly specialized (``goldilocks'') syzygy of circumstances and qualities; see also footnote~\ref{fn:DDVvsOV+FK} below.

By mirror symmetry, the same phenomenon will also occur when the K{\"a}hler (or symplectic) structure of the model is nearly singular, as is the case with very small relative sizes of the exceptional sets (``$\cO(-1,-1)$-curves'') in a Calabi-Yau 3-fold compactification. Computationally, worldsheet instanton contributions are expected to make the analysis of this ``mirror image'' more demanding, but one should again expect a detailed balance of various contributing factors as with the nearly singular (complex structure) deformations\cite{Bento:2021nbb}. In turn however, Calabi-Yau 3-folds are known to generically have very many such exceptional sets (and conifold singularizations)\cite{rReidK0}, which implies that the pool of such nearly singular models is very numerous, even if requiring highly specialized arrangements.

The direct product and the exceptional subset geometries (Figure~\ref{f:4Cases}, a--d) may well combine such as in\cite{Randall:1999vf,Randall:1999ee}, which may be interpreted as two copies of $\AdS_5$ glued together non-analytically across a 3-brane-World, $\sW^{1,3}_{z=0}$\cite{Banerjee:2018qey}; see also\cite{Hawking_2000, Karch_2001,Danielsson:2021tyb}. The remaining 5 dimensions presumably compactified on a suitable space of positive curvature, such as $S^5$, complete a direct product 10-dimensional spacetime, $(\AdS_5\#\AdS_5)\,{\times}\,S^5$. This combination of geometries reminds of the \BHM\ toy model discussed in \SS\,\ref{s:dSdefo}, where the ``cylinder'' $\sW^{1,3}_{z=0}\,{\times}\,S^1$ is the non-analytic ``interface'' between the ``inside,'' $\sW^{1,3}_-\,{\rtimes}\,\sY^2_{\rm in}$, and ``outside'' region, $\sW^{1,3}_+\,{\rtimes}\,\sY^2_{\rm out}$. 
Here, the total spacetime is
$\big((\sW^{1,3}_-{\rtimes}\,\sY^2_{\rm in})\#
      (\sW^{1,3}_+{\rtimes}\,\sY^2_{\rm out})\big)\,{\times}\,\text{K3}$.
In both of these types of models, the metric depends on a local coordinate as $|z|$ near $\sW^{1,3}_{z=0}$, which induces a $\d(z)$-like singularity in the Ricci tensor.  This hallmark non-analyticity of the $9{+}1$-dimensional metric at $\sW^{1,3}_{z=0}$ has been shown to insure both gravity and matter to exhibit properly localized modes at $\sW^{1,3}_{z=0}$, which thereby may serve as a candidate for ``our'' world.
 Just as in \SS\,\ref{s:dSdefo} herein, the model of\cite{Banerjee:2018qey} shows the metric within the so embedded $\sW^{1,3}_{z=0}$ to be of the de~Sitter type, with the effective, $3{+}1$-dimensional cosmological constant in the phenomenologically relevant ballpark.
 
In the model of Ref.\cite{Banerjee:2018qey}, $\sW^{1,3}_{z=0}$ emerges as a ``bubble,'' a domain-wall like interface between a metastable false $\AdS_5$ and the nucleated, expanding true $\AdS_5$. Such vacuum decay phenomena have been estimated to be generic in string theory\cite{Ooguri:2016pdq,Freivogel:2016qwc},\footnote{\label{fn:DDVvsOV+FK}Refs.\cite{Ooguri:2016pdq,Freivogel:2016qwc} in fact conjecture that de~Sitter vacua with small cosmological constant {\em\/generally\/} decay faster than their horizon size; the latter of these two notes the contradiction with the {\em\/concrete example\/} results of Ref.\cite{Danielsson:2016rmq}.} which would then imply the same also for nucleation, a subset of which resulting in de~Sitter ``bubble''-worlds. These phenomena quite explicitly involve phase transitions, which are generally known to be inherently non-perturbative and the description of which is certain to become improved in time. Both in the claimed ubiquity and in this inherently non-perturbative nature, this reminds of the superstring realization\cite{Green:1988uw,Candelas:1989ug} of ``Reid's conjecture''\cite{rReidK0}, and also of the much more general cobordism connection\cite{McNamara:2019rup} --- however, with\cite{Banerjee:2018qey} focusing on the dynamically enfolding aftermath of such a phase transition.

It behooves us then to give an old whimsy\cite{rHitch} a second look.
Overall, string dynamics is generally understood to require the overall target spacetime to be Ricci-flat\cite{rFrGaZu86,rBowRaj87,rBowRaj87a,rBowRaj87b,rHHRR-sDiffS1,Pilch:1987eb}. Its Euclidean Wick-rotation in Type-II theories is by the $N\!=\!2$ supersymmetry guaranteed to admit both a complex structure and a K{\"a}hler metric, making it a Calabi-Yau 5-fold; various dualities should extend this also to other models. Just as Calabi-Yau 3-folds generically contain numerous isolated ``$\cO(-1,-1)$ curves''\cite{rReidK0}, Calabi-Yau 5-folds generically contain numerous isolated Fano ($R_{\mu\nu}>0$) compact complex surfaces $\mathcal{S}$\cite{rHitch}. In the Lorentzian original, at least some of these surfaces should acquire a $(1,3)$-signature and a de~Sitter metric (owing to positive curvature), with $\Lb>0$ parametrizing the relative size of $\mathcal{S}\!\mapsto\!\sW^{1,3}$, and very much akin to the de~Sitter desingularizing deformation discussed in Section~\ref{s:BHM}.
 Most candidate complex surfaces, $\mathcal{S}$, easily contain real codimension-2 subspaces to serve as $I^\pm$ (past and future ``infinity,'' see Sections~\ref{s:Prelim} and~\ref{s:Define}) and a natural ``flow'' between them to serve as an ``arrow of time''~\cite{Banks:1984cw,Hawking:1985af}.
 Being the exceptional sets of ``small resolutions'' of conifold singularities the dynamical evolution of the Lorentzian (and so de~Sitter) of these $3{+}1$-dimensional subspaces $\mathcal{S}\!\mapsto\!\sW^{1,3}$ may well enfold not unlike the de~Sitter bubbles of Ref.\cite{Banerjee:2018qey,Banerjee:2019fzz,Dibitetto:2020csn,Petri:2022yhy}.

\section{de~Sitter Deformations of Cosmic Branes}
\label{s:dSdefo}
The foregoing detailed review of the many various approaches to identifying a de~Sitter $3{+}1$-dimensional spacetime sector, $\sW^{1,3}$, within concrete string theory models demonstrates both
 the systematic and often technical difficulties in doing so,
 but also the general conceptual conundrum:
 Owing to the various non-renormalization {\em\/stability\/} guarantees, it is the supersymmetric models that are most predictive, controlled and reliable.
 However, the ultimate practical goal is of course {\em\/real-world physics\/}: a Standard-like model of particle physics in a de~Sitter-like spacetime --- which is most definitely not supersymmetric.

The encountered and highlighted technical difficulties also indicate several assumptions, sometimes tacit, which were pivotal in deducing the non-viability of de~Sitter spacetime in string theory models.
 To summarize, these include, sometimes several of the following, and in various combinations:
 ({\small\bf1})~non-singularity of phenomenologically relevant structures within the full string theory, such as the spacetime;
 ({\small\bf2})~compactness of extra spacetime dimensions;
 ({\small\bf3})~an essentially static approximation to our dynamically evolving universe;
 ({\small\bf4})~reliability and definiteness of the effective, low-energy point-field limit of the full string theory.
Some of the attempts in the literature do present exceptions to this, as we discuss below.

To showcase a possible avenue in evading these assumptions, we review a particular illustrative discretuum of axion-dilaton (\BHM) toy models\cite{rBHM5,rBHM7,Berglund:2020qcu}, which correlates supersymmetry breaking and emergence of a de~Sitter sub-spacetime. We focus on the universal subsystem of all superstring models: the combined axion-dilaton field, $\t=\a + i\exp(-\phi)$, familiar from $F$-theory\cite{rFTh}, coupled to gravity. Ignoring all other degrees of freedom permits embedding this analysis within most any superstring model. For simplicity and definiteness however, one may focus on the type~IIB string theory. This strategy derives from the original ``stringy cosmic strings''\cite{Greene:1989ya,rCYCY} and was generalized to a specific type of ``stringy cosmic branes'', wherein exponential mass-hierarchy, localized gravity and a realistic cosmological constant all emerge within the same cosmic 3-brane candidate Universe\cite{rBHM1,rBHM4,rBHM5,rHTdS03,Berglund:2021xlm}.

To this end, compactification on a {\it fixed\/} $(10{-}D)$-dimensional supersymmetry-preserving space, $\sY^{10-D}$, reduces the uncompactified ``large'' spacetime to $D$-dimensional, wherein the axion-dilaton field, $\t$, varies over a ``transverse'' $\sY^2_\perp=(x_{D-2},x_{D-1})$-plane.
The relevant part of the low-energy effective $D$-dimensional action for this simple but universal subsystem, coupled to gravity, reads
\begin{equation}
                \Seff
                = \frac{1}{2\kappa^2}\int\rd^D x \sqrt{-g} ( R
                   - {\cal G}_{\t \bar{\t}}g^{\mu \nu}
                     \vd_{\mu} \t \vd_{\nu} \bar{\t}+\ldots)~.
\label{e:effaction}
\end{equation}
Here $\mu,\nu=0,{\cdots},D-1$, $2\kappa^2=16\p G^{(D)}_N$, where
$G_N^{(D)}$ is the $D$-dimensional Newton constant, and ${\cal
G}_{\t\bar\t}=-(\t{-}\bar \t)^{-2}$ is the metric on 
the complex structure moduli space of a torus.\footnote{Because of its $\SL(2;\ZZ)$-variance, the axion-dilaton, $\tau$, parametrizes the complex structure of a $T^2$\cite{Vafa:1996xn}.}
 While the $D=5{+}1$-dimensional case is of immediate interest, on occasion we exhibit the explicit $D$-dependence of the emergent structure.

\subsection{Spacetime Road-map}
\label{s:Roadmap}
Consider the 9+1-dimensional spacetime of the overall form
 $\sW^{1,3}\!\rtimes\sY^2_\perp\times\sY^4$\cite{rBHM7}:
\begin{enumerate}\itemsep=-1pt\vspace*{-1mm}

 \item $\sY^4$ is fixed, compact, small and supersymmetry-preserving (K3 or a 4-torus). 

 \item $\sY^2_\perp$ is annular, and either
 ({\it a})~non-compact, $\sY^2_{\perp,nc}\approx\IC^*$, or
 ({\it b})~compact by including its circular boundaries, $\sY^2_{\perp,bc}\approx S^1\times I$.
 In both cases, proper distance in $\sY^2_\perp$ extends radially to infinity, while the surface area of $\sY^2_\perp$ is finite. 

 \item $\sW^{3,1}\approx\mathbb{R}^{3,1}$ is fibered (``$\rtimes$'') over $\sY^2_\perp$, the $\sW^{3,1}$-metric depending on $|z\!:=\!\log(r/\ell)|$. The central copy, $\sW^{1,3}_{\!z=0}$, is the candidate $3{+}1$-dimensional spacetime of interest.
 \vspace*{-.5\baselineskip}
\end{enumerate}
The exceptional geometric properties of ``our'' candidate spacetime, $\sW^{1,3}_{z=0}$, stem from being a geometrical/dynamical {\em\/defect\/} within the ``bigger'' spacetime left after {\em\/compactifying\/} on $\sY^4$. The curvature of $\sY^2_\perp$ near the non-analyticity at $z\!=\!0$ then competes with effects that depend on the volume-based length-scales of $\sY^2_\perp$ and $\sY^4$. Even the relative length-scales of the three factors $\sW^{1,3}\!\rtimes\sY^2_\perp\times\sY^4$ are expected to evolve anisotropically, reminding of the Kasner universe, and so present a dynamical toy model.

Alternatively, and so as to compare with the recent ideas\cite{Banerjee:2018qey,Danielsson:2021tyb,Karch_2001,Hawking_2000}, we may cut $\sY^2_\perp$ along the radial location of $\sW^{1,3}_{z=0}$. The ``cylinder'' $\sW^{1,3}_{z=0}\times S^1$ then is the non-analytic ``interface'' between the ``inside,'' $\sW^{1,3}_{z<0}\rtimes\sY^2_{\rm in}$, and ``outside'' region, $\sW^{1,3}_{z>0}\rtimes\sY^2_{\rm out}$.

Compactifying~\eqref{e:effaction} on $\sY^4$ reduces to $D$-dimensional spacetime, and we focus on:
\begin{subequations}
\label{e:adE}
\begin{alignat}9
 \Tw{T}_{\mu\nu}\!\define\!T_{\mu\nu}\!-\!\frc1{D{-}2}g_{\mu\nu}T,\quad\text{so}\quad
 R_{\mu\nu} &=\kappa^2\Tw{T}_{\mu\nu}(\t,\bar\t),\label{e:EinsteinFE}\\
  g^{\mu\nu}\big[(\nabla\!_\mu\nabla\!_\nu\t)
       +\Gamma_{\t\t}^\t(\nabla\!_\mu\t)(\nabla\!_\nu\t)\big] &=0, \label{e:axilatonFE}
\end{alignat}
\end{subequations}
and seek a codimension-2 solution with the warped metric Ansatz\cite{rBHM5}:
\begin{subequations}
 \label{e:metric}
\begin{alignat}9
           \rd s^2 &= A^2(z)\, \bar{g}_{ab}\,\rd x^a \rd x^b
                     + \ell^2 B^2(z)\,(\rd z^2 + \rd\q^2), \label{e:Metric} \\[1mm]
                \bar{g}_{ab}\,\rd x^a \rd x^b &= - \rd x_0^2 +
                e^{2\sqrt{\Lb}\,x_0}\,(\rd x_1^2 + \ldots + \rd x_{D-3}^2). \label{e:gbar}
\end{alignat}
\end{subequations}
Here, the log-radial coordinate $z\!:=\!\log(r/\ell)$ and the angle $\q\in[-\p,+\p]$ parametrize the ``transverse 2-plane,'' $\sY^2_\perp$, and $\bar{g}_{ab}$ is a metric on the $\sW^{D-3,1}_z$ factor with a positive cosmological constant over $\sW^{D-3,1}_{z=0}$.
 Assuming axial symmetry of the metric~\eqref{e:metric} and a complementary radial independence of $\t\!=\!\t(\q)$ separates the system~\eqref{e:adE}, results in $\Tw{T}_{\mu\nu}\!=\!\const$,\footnote{\label{f:avg}Restricted by $\SL(2;\ZZ)$ covariance, the spectrum of $\Tw{T}_{\mu\nu}$ specifies a discrete {\em\/ensemble\/} of $\t$-solutions!} and specifies this explicitly non-holomorphic deformation\cite{rBHM1,rBHM4,rBHM5,rBHM7,Berglund:2020qcu} of the routinely complex-analytic framework of ``stringy cosmic strings''\cite{Greene:1989ya,rCYCY}. This explicit ``holomorphy breaking'' also breaks supersymmetry, as we discuss below.

\subsection{The \BHM}
\label{s:BHM}
The equation of motion~\eqref{e:axilatonFE} for the \BHM\ $\t$ is solved by\cite{rBHM1,rBHM2,rBHM5}:
\begin{subequations}
 \label{e:moduli}
\begin{alignat}9
  \t_I(\q) &=b_0+i\,g_s^{-1}\,e^{\w\q},\\
 \t_{I\!I}(\q) &=\big(b_0\pm g_s^{-1}\tanh(\w\q)\big)
               \pm i\,g_s^{-1}\sech(\w\q).
\end{alignat}
\end{subequations}
Anisotropic and aperiodic, these solutions exhibit a hallmark non-trivial, {\em\/stringy\,} $\SL(2;\ZZ)$ monodromy\footnote{The quantity ${\cal G}_{\t \bar{\t}}^{-1}|\vd\tau|^2$, which appears in the
action~(\ref{e:effaction}), is $\SL(2;\ZZ)$-invariant.} for specific choices in the effective parameter space $(b_0,\w,g_s)$. Since $\SL(2;\ZZ)$ is 3-dimensional, the $\SL(2;\ZZ)$-equivalence classes of the $(b_0,\w,g_s)$-parametrized solutions~\eqref{e:moduli} form a {\em\/discretuum\/}: Although classical and non-supersymmetric, these stringy (rather than merely supergravity) solutions, $\t_I(\q)$ and $\t_{I\!I}(\q)$, have no continuous deformations, and so harbor no continuous instability.
 It is worth noting that these stringy $\SL(2;\ZZ)$-invariant solutions~\eqref{e:moduli} are {\em\/dispersed\/} amongst the analogous $\SL(2;\IR)$-solutions (of just supergravity) somewhat akin to rational numbers amongst the real numbers. It is tantalizing to then expect at least some of the {\em\/landscape\/} to be similarly {\em\/dispersed\/} through (regions of) {\em\/swampland\/}; contrast this with~\cite[Fig.~2]{vanBeest:2021lhn}.

The manifestly non-holomorphic solutions~\eqref{e:moduli} are related to their holomorphic analogues, $\t_I(\q{-}iz)$ and $\t_{II}(\q{-}iz)$\cite{rBHM7}, and their corresponding supersymmetric stringy backgrounds. Thereby,~\eqref{e:moduli} interpolate between $\t_{I,\,I\!I}(\q{-}iz)$ and the well understood ($\w\,{\to}\,0$) constant configurations\cite{rFTh,rS-ForFld}:
\begin{equation}
    \vcenter{\hbox{\begin{tikzpicture}[every node/.style={inner sep=0,outer sep=.75mm}]
     \path[use as bounding box](-.2,.3)--(10.2,-2.3);
     \filldraw[rounded corners=1mm, fill=yellow!20, draw=red!80!black]
      (2.5,-.3) rectangle ++(4.05,.6);
     \node[red!75!black](BHM) at(4.5,0) {\footnotesize\BHM{s}};
     \node(tIh) at(0,0) {\fbox{$\t_I(\q{-}iz)$}};
     \node[red!75!black](tI)  at(3,0) {$\t_I(\q)$};
     \node[red!75!black](tII) at(6,0) {$\t_{I\!I}(\q)$};
     \node(tIIh)at(9,0) {\fbox{$\t_{I\!I}(\q{-}iz)$}};
     \node(tIo) at(0,-1.5) {$\big( [b_0{+}\t_0\w z] , \t_0[1{+}\w\q] \big)$};
     \node(00)  at(4.5,-1.5) {\fbox{$\!\!$\cite{rFTh,rS-ForFld}}};
     \node(tIIo)at(9,-1.5) {$ \big( [b_0{+}\t_0\w\q] , \t_0[1{-}\w z] \big)_{\!\!\text{\cite{Einhorn:2000ct}}}$};
     \draw[thick, ->](tIh) to node[right] {$\scriptstyle O(\w)$} (tIo);
      \draw[thick, ->](tIIh) to node[left] {$\scriptstyle O(\w)$} (tIIo);
     \draw[thick, ->](tIh) to node[above] {$\scriptstyle z\to0$} (tI);
      \draw[thick, ->](tIIh) to node[above] {$\scriptstyle z\leftarrow0$} (tII);
     \draw[thick, ->](tI) to node[above,rotate=-45] {$\scriptstyle\w\to0$} (00);
      \draw[thick, ->](tII) to node[above,rotate=45] {$\scriptstyle0\leftarrow\w$} (00);
     \draw[thick, ->](tIo) to node[above] {$\scriptstyle\w\to0$} (00);
      \draw[thick, ->](tIIo) to node[above] {$\scriptstyle0\leftarrow\w$} (00);
     \draw[blue, densely dotted, thick, <-](tIo) to[out=-30,in=180]++(2,-.7)--++(4,0)
                                                 to[out=0,in=-150]++(2,.4);
      \path[blue](4.5,-2.05) node {\scriptsize$(-\q,z)\leftarrow(z,\q)$};
   \end{tikzpicture}}}
 \label{e:bowtie}
\end{equation}
the latter of which has the (``helicoidal axion'') D7 instanton\cite{Einhorn:2000ct} as the $O(\w)$-linearizing complementary interpolation\cite{rBHM7} in the right-hand half of the network~\eqref{e:bowtie}. The analogous $O(\w)$-linearizing interpolation in the left-hand half of the network~\eqref{e:bowtie}, is the $(z,\q)\to(-\q,z)$ ``rotation'' of the D7 instanton\cite{Einhorn:2000ct}.

These interpolation processes turn out to involve non-weekly coupled strings:
 The $\SL(2;\ZZ)$ symmetry of~\eqref{e:moduli} implies that $\vev{g^D_s}_{\sY^2_\perp}\sim O(1)$. String theory is thus not weakly coupled throughout the whole $D$-dimensional spacetime, and is therefore only approximated in this target-space effective field theory description. Moreover, $\t_I(\q)$ has a $\SL(2;\ZZ)$-controlled discontinuity along $\q\!=\!\pm\p$:
\begin{equation}
   g_s^{\sss\text{(eff)}}[\t_I(\p{-}\e)]\ll1
    \qquad\text{whereas}\qquad
   g_s^{\sss\text{(eff)}}[\t_I(\p{+}\e)]\gg1.
\end{equation}
This forces patching weakly- to ($S$-dually\footnote{$S$-duality patching requires re-incorporating winding modes and ``momentum'' space into the description of target spacetime, implying a ``mixing'' that appears rather natural within the stringy non-commutative and chirally doubled spacetime geometry\cite{Freidel:2015pka,Freidel:2017xsi,Freidel:2018apz} discussed below.}) strongly-coupled string theory across the $\q=\p$ direction in $\sY^2_\perp$.
 Akin to ``$T$-folds''\cite{Dabholkar:2002sy,Flournoy:2004vn}, such inherently nonperturbative constructions have been dubbed ``$S$-folds''\cite{rHIS-nGeoCY}.

 Nevertheless, within ``our'' $D{-}2$-dimensional brane-world,
\begin{equation}
  \sW^{D-3,1}_{z=0}:\quad g_s^{D-2}=g_s^D\sqrt{\a'/V_\perp}\ll1,
 \label{e:weakST}
\end{equation}
since the surface area, $V_\perp$, of the transversal space $\sY^2_\perp$ is exponentially large; see~\eqref{GN} below. Within the brane-world, $\sW^{D-3,1}_{z=0}$, string theory is weakly coupled and there this approximation is very well justified.
 As $R_{\mu\nu}\propto\Tw{T}_{\mu\nu}(\t,\bar\t)=\hbox{diag}[0,{\cdots},0,\inv4\w^2\ell^{-2}]$,
the so-sourced spacetimes~\eqref{e:EinsteinFE} are Ricci-flat over $\sW^{1,3}_z\rtimes\sY^2_\perp$, except along the angular direction in $\sY^2_\perp$: The $\SL(2;\ZZ)$-discrete anisotropy, $\w\,{\neq}\,0$, of the \BHM\ field~\eqref{e:moduli} ``drives'' supersymmetry breaking\cite{rBHM1}.

Finally, Friedan's classic result\cite{rF79b,rF79a} identifies Einstein's field equations~\eqref{e:EinsteinFE} as the lowest-order condition for the renormalization flow fixed point. Restricting to the 2-dimensio\-nal $\sY^2_\perp$, the logarithmic nature of the (leading order) Gell-Mann-Low and Green's functions necessitates the ``dimensional transmutation,'' whereby the length scale ($\ell$, see below) is introduced and set (determined) dynamically --- very much akin to $\L_{\sss\rm QCD}$ in strong interactions. This leaves none of the parameters of the toy model continuously variable, and so leaves no room for continuous instability in this sector.
 A recently considered class of models involving ``defect 7-branes'' as companions of the dynamical cobordism of a domain wall\cite{Blumenhagen:2022mqw} exhibit several of the key technical features of the \BHM\ models --- most prominently, also this logarithmic geometry in $\sY^2_\perp$. This then opens the possibility of reverse-engineering an analogous cobordism-related reinterpretation of the \BHM\ models.

\subsection{The Metric}
\label{s:Metric}
The $\w\neq0$ anisotropic solutions~\eqref{e:moduli} drive a
perturbative, analytic solution to Einstein's field equations~\eqref{e:EinsteinFE}\footnote{This solution is of the same form as that discussed by Gregory\cite{Gregory:1996dd,Gregory:1999gv} for the $U(1)$ vortex solution.}:
\begin{subequations}
 \label{e:newAB}
\begin{align}
    A(z) &= Z(z) \Big(1- \frac{\w^2 z_0^2(D-3)}{24(D-1)(D-2)} Z(z)^2 + O(\w^4)\Big)~,
\label{e:newA}\\
   B(z) &= \frac{1}{\ell z_0\sqrt{\Lb}}\Big(1 -  \frac{\w^2z_0^2}{8(D-1)} Z(z)^2 + O(\w^4)\Big)~, \label{e:newB}\\*[2pt]
   Z(z)&=Z_\pm(z):=1\pm|z|/z_0,~~z_0>0,\quad
   \text{and}\quad
   \L_b:=\frac{\w^2}{2(D{-}2)\ell^2}\geqslant0. \label{e:Z+Lb}
\end{align}
\end{subequations}
For $Z_+(z)$, $|z|<\infty$ and $\sY^2_{\perp,nc}\approx S^1_\q\times\IR^1_z$ is non-compact,
 while
for $Z_-(z)$, $|z|\leqslant z_0$ includes the boundary circles and $\sY^2_{\perp,bc}\approx S^1_\q\times I_z$.
As was shown in\cite{rBHM5}, close to the putative horizon, $z\!=\!z_0$, spacetime is asymptotically flat in agreement with the behavior of Rindler space\cite{Kaloper:1999sm}.
 Thus, the geometry~\eqref{e:metric} with~\eqref{e:newAB} effectively {\em\/desingularizes\/}\cite{rBHM5} the total spacetime of the radically different Minkowski solution\cite{rBHM1} with the warp factors:
\begin{equation}
           \tA(z) = Z(z)^{\frac{1}{(D-2)}}\quad\text{and}\quad
           \tB(z) = Z(z)^{\frac{-(D-3)}{2(D-2)}}
                   e^{2\xi(1-Z(z)^2)/z_0},\qquad
           \text{when}~~\Lb\!=\!0.
\label{e:oldAB}
\end{equation}
This Minkowski limit exhibits a naked singularity at $z\!=\!z_0$: the metric with~\eqref{e:oldAB} is non-analytic, and becomes complex for $z\!>\!z_0$!
 In turn, the de~Sitter solution~\eqref{e:newAB} keeps the metric~\eqref{e:metric} analytic and real,\footnote{The metric factor $A^2(z)$ merely {\em\/bounces\/} at $z\!=\!z_0$, reminding of $\rho^2\rd\q^2$ in standard cylindrical coordinates when passing through the $z$-axis.}
 and the cosmological constant, $\L_b\!>\!0$, parametrizes this de~Sitter desingularization.
 The location $|z|\!=\!z_0$ however does specify a horizon\footnote{The $|z|\!=\!z_0$ horizon bounds the physical regions ``transversally'' within $\sW^{1,3}_z\!\rtimes\sY^2_\perp$, and is distinct and independent from the \dS\ horizons {\em\/within\/} $\sW^{1,3}_z$.} and so restricts the physical region.
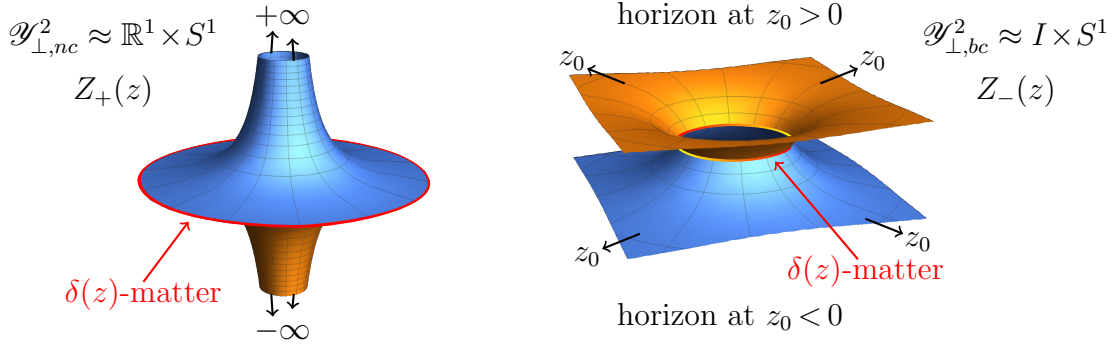
\begin{figure}[ht]
 \begin{center}
  \begin{picture}(110,40)(0,2)
   \put(0,0){\includegraphics[width=50mm]{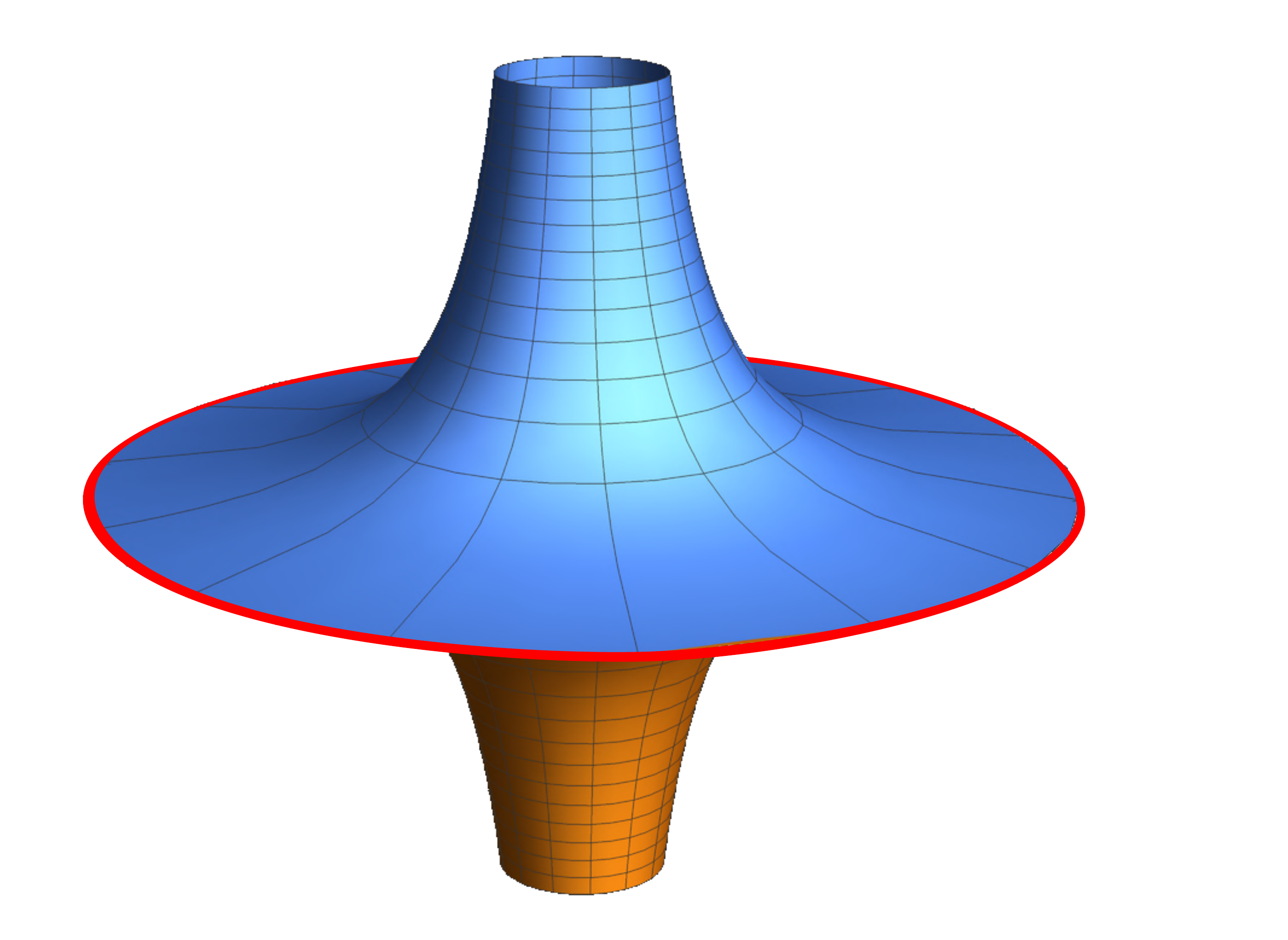}}
    \put(0,0){\TikZ{[scale=2]\path[use as bounding box](0,0);
               \path[red](.2,.1)node{$\d(z)$-matter};
               \draw[thick,red,->](.2,.18)--++(.3,.4);
               \draw[thick,->](1.05,1.7)--++(.01,.15);
               \draw[thick,->](1.2,1.65)--++(-.01,.15);
                \path(1.125,1.95)node{$+\infty$};
               \draw[thick,->](1.05,.1)--++(.01,-.15);
               \draw[thick,->](1.2,.1)--++(-.01,-.1);
                \path(1.125,-.15)node{$-\infty$};
               \path(0,1.45)node{$Z_+(z)$};
               \path(0,1.8)node{$\sY^2_{\perp,nc}\approx\IR^1\!\times\!S^1$};
                    }}
   \put(60,5){\includegraphics[width=50mm]{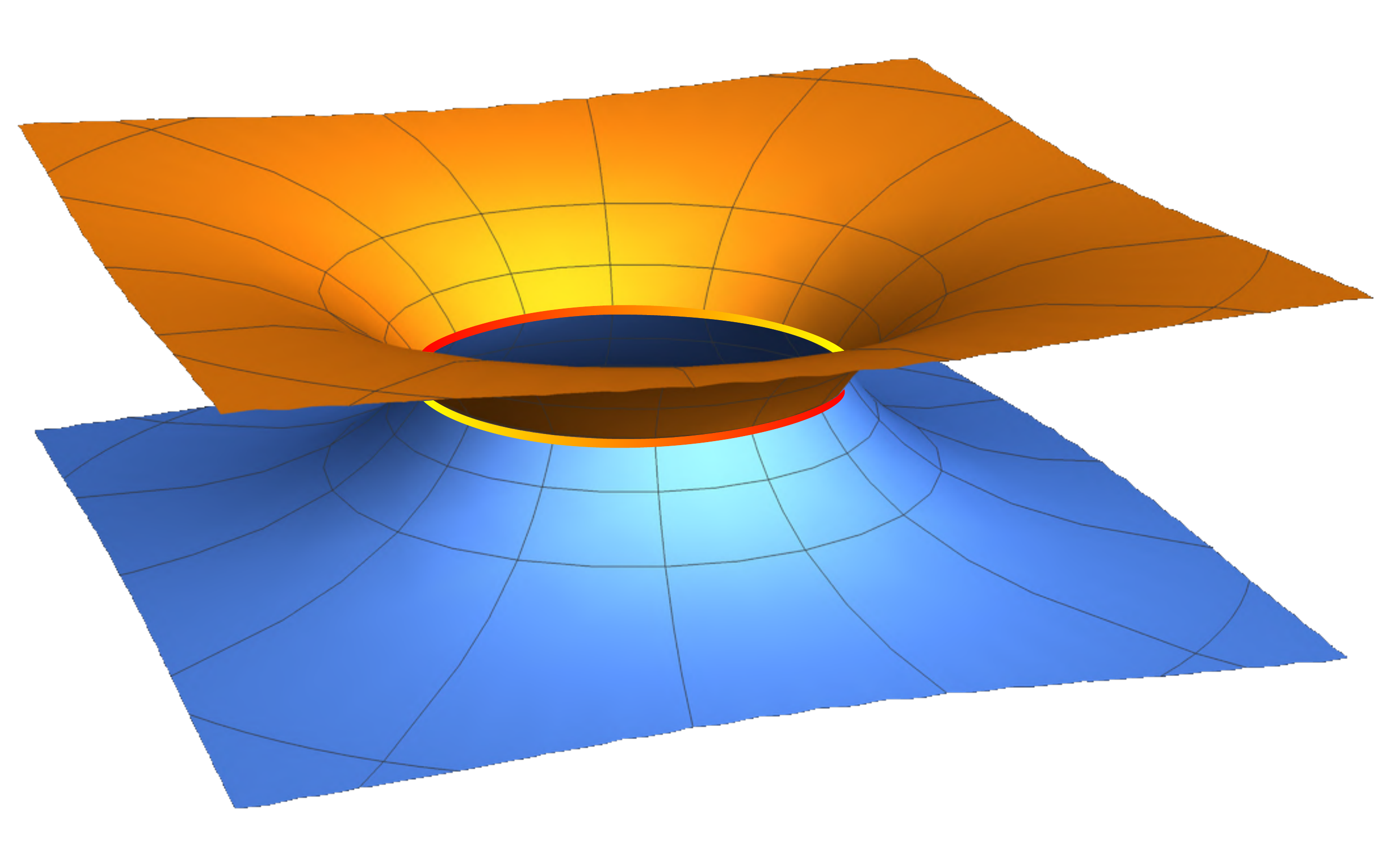}}
    \put(60,5){\TikZ{[scale=2]\path[use as bounding box](0,0);
                 \path[red](2,0)node{$\d(z)$-matter};
                 \draw[thick,red,->](1.95,.08)--++(-.55,.65);
                 \draw[thick,->](1.7,1.25)--++(.25,.1);
                  \path(2.05,1.4)node{$z_0$};
                 \draw[thick,->](.4,1.25)--++(-.25,.1);
                  \path(.05,1.4)node{$z_0$};
                 \draw[thick,->](2,.35)--++(.25,-.1);
                  \path(2.35,.2)node{$z_0$};
                 \draw[thick,->](.5,.25)--++(-.25,-.1);
                  \path(.15,.1)node{$z_0$};
                 \path(1.1,1.7)node{horizon at $z_0\,{>}\,0$};
                 \path(1.1,-.3)node{horizon at $z_0\,{<}\,0$};
               \path(3,1.55)node{$\sY^2_{\perp,bc}\approx I\!\times\!S^1$};
               \path(3,1.2)node{$Z_-(z)$};
                      }}
  \end{picture}
 \end{center}
 \caption{Proper (radial) distance plotted (vertically) against the circumference in $\sY^2_\perp$}
 \label{f:charts}
\end{figure}

This splits the total spacetime into
 the ``inside'' and ``outside'' regions, respectively:
 $[\sW^{1,3}_z\!\rtimes\sY^2_\perp]_{z<0}$ and
 $[\sW^{1,3}_z\!\rtimes\sY^2_\perp]_{z>0}$.
 The interfacing brane-world $\sW^{1,3}_{z=0}$ is thereby a metric {\em\/defect,} owing to the $|z|$-dependence of~\eqref{e:newAB}. There, the Ricci tensor has $\d(z)$-localized contributions, which require both matter and gravity to be localized at $\sW^{1,3}_{z=0}$\cite{rBHM4,rBHM5}; see Figure~\ref{f:charts}.
 Overall, this reinterpretation of the spacetime geometry seems tantalizingly similar to the situation in Refs.\cite{Banerjee:2018qey,Danielsson:2021tyb,Karch_2001,Hawking_2000}.

\subsection{Phenomenology}
\label{s:Pheno}
Whereas $\Lb>0$ (i.e., $\w^2>0$) desingularizes the metric~\eqref{e:metric} with~\eqref{e:newAB},
it was first shown by Gregory\cite{Gregory:1996dd,Gregory:1999gv} (see also\cite{rBHM5}) that the global cosmic brane solution~(\ref{e:oldAB}) is still a good approximation to \Eq{e:newAB} away from the horizon.
In particular, by comparing near $z\!=\!0$ Eqs.~\eqref{e:newAB} and~\eqref{e:oldAB}:
\begin{equation}
  z_0=-\frac{h}{ h'}\Big|_{z=0},\qquad
  \xi:= \Big(\frac{h''}{2 h'} - \frac{\w^2 h}{ 8 h'}\Big)\Big|_{z=0},\qquad
  \ell=\L_b^{-1/2}\sqrt{\frac{h''\,h^{-(D-4)/(D-2)}}{(D-2)(D-3)}}\bigg|_{z=0},
\label{e:z0xiell}
\end{equation}
where $h(z):=A(z)^{D-2}$.
 Close to $z=0$, the smooth solution defined by~(\ref{e:newAB}) and parameterized in terms of $(z_0,\w,\L_b)$ describes a global cosmic brane solution with parameters $(z_0,\xi,\ell)$~\eqref{e:z0xiell}.
 Alternatively, we can solve for $\L_b$,
\begin{equation}
\L_b=\frac{\big(\w^2 - \w^2_{GCB} A^2|_{z=0}\big)}{4 \ell^{2} (D-2)(D-3)}
\define\frac{\Delta \w^2}{4 \ell^{2} (D-2)(D-3)}~,
\label{e:dS=omega}
\end{equation}
where $\w^2_{GCB}\define8\xi/z_0$\cite{rBHM1} is the stress tensor associated to the global cosmic brane to which the solution asymptotes as $z\to 0$; $\w^2$ is the stress tensor for the $\L_b>0$ solution. 
The cosmological constant is thus directly proportional to the anisotropy variance of the \BHM\ $\t$ --- and thereby the string coupling constant!
 This gives a very non-trivial relation between the stringy \BHM\ field,
and hence string theory itself, and a positive $\Lb$.
Furthermore, $\L_b>0$ implies that $\w^2 > \w_{GCB}^2$,\footnote{That $\Lb\geqslant0$ also follows directly from the defining equation for the warp factor $A(z)$\cite{rBHM5}.} so $\w^2\to0$ forces $\w^2_{GCB}\to0$. The $\Lb\to0$ limit is a necessary condition for obtaining a supersymmetric configuration, providing the important relation between a positive cosmological constant and this ({\em\/cosmological\/}) supersymmetry breaking.

Finally, the $(D{-}2)$-dimensional Newton constant, $G^{(D-2)}_N=(M_P^{\sss(D-2)})^{-(D-4)}$, 
 and the surface area, $V_\perp$, of the transversal space are\cite{rBHM5}:
\begin{equation}
\begin{aligned}
  G^{(D-2)}_N &=(M_P^{\sss(D)})^{-(D-2)} V_\perp^{-1},\\
  V_\perp
  &\approx\frac{\p\ell^2}{2z_0}\Big(\frac{2z_0}\xi\Big)^{\frac{D-3}{2(D-2)}}e^{\xi/2z_0}\,
    \g\Big(\frac{D{-}3}{4(D{-}2)};\frac\xi{2z_0}\Big)
  \sim \frac{\pi}{D{-}3}\frac{\ell}{\sqrt{\Lb}}.
\end{aligned}
\label{GN}
\end{equation}
This surface area is large\cite{rBHM5}, driven by the exponential factor, $e^{\xi/2z_0}$, which then also drives the large $M_P^{\sss(D-2)}/M_P^{\sss(D)}$ hierarchy of the effective Planck mass-scales in $D{-}2$ and $D$ dimensions.
 This is in full agreement with the statistical argument of Ref.\cite{Horava:2000tb}, whereby the discrete ensemble of $\SL(2;\ZZ)$-allowed values of $\w$ and so $\L_b$ (see footnote~\ref{f:avg}) strongly favors small but positive values of $\Lb$. In fact, this peculiar result and the consequent seesaw relation~\eqref{e:LongCC} below presage the key results discussed in Section~\ref{s:VEST}.

\subsection{A Seesaw Relation}
\label{s:Seesaw}
For $z_0>0$ and the positive sign-choice in~\eqref{e:Z+Lb}, the volume of $\sY^2_\perp$~\eqref{GN} implies the exponential hierarchy between the Planck mass-scales in $D$ and $D{-}2$ dimensions\cite{Berglund:2020qcu}:
\begin{equation}
 (M_P^{\sss(D-2)})^{D-4}
 = (M_P^{\sss(D)})^{D-2}\,2\p\ell^2\,z_0^{-\frac{D-1}{2(D-2)}}\,e^{z_0}\,
   \gamma\Big(\frc{D-3}{2(D-2)};\frc1{|z_0|}\Big),
   \quad\text{for}~~ Z(z)\!=\!1{\pm}|z|/z_0.
\label{e:normalization}
\end{equation}
Here $|z_0|$ denotes the log-radial distance (in units of $\ell$) from the $(D{-}2)$-dimensional brane-World to the boundary of $\mathscr{Y}_\bot^2$, and $\gamma({\cdots})$ denotes the ``little'' incomplete gamma function. In the phenomenologically relevant $D=6$ case:
\begin{equation}
  M_P^{\sss(4)} \,{=}\, \sqrt{\z_0}\,|z_0|^{-\frac5{16}}\,e^{z_0/2}~
                          \frac{(M_P^{\sss(6)})^2}{M_\ell},
  \qquad M_\ell\define1/\ell,
 \label{e:expHM4}
\end{equation}
where $0\leqslant\z_0\define2\p\Gamma\big(\frc38;\frc1{|z_0|}\big)
    \leqslant2\p\Gamma\big(\frc38\big)\approx14.89$, and the {\em\/exponential hierarchy,} $M_P^{\sss(4)}\,{\gg}\,M_P^{\sss(6)}$, is driven by the auxiliary parameter $z_0$. (The proper distance to the boundary of $\mathscr{Y}_\bot^2$ is infinite.)

Using that $\L_{D-2}=\Lb/G^{(D-2)}_N$ is the $(D{-}2)$-dimensional energy density, the volume result~\eqref{GN} implies the relation
\begin{equation}
\L_{D-2}\sim \Big(\frac{\pi}{D{-}3}\Big)^{\!2}
(M_P^{\sss(D-2)})^{D-2}~(\ell\,M_P^{\sss(D-2)})^2~
\Big(\frac{M_P^{\sss(D)}}{M_P^{\sss(D-2)}}\Big)^{\!2D-4}~.
   \label{e:energydensity}
\end{equation}
In the phenomenologically relevant case of $D=6$ this produces the seesaw-like relation:
\begin{equation}
   \L_{4}
   \sim (M_P^{\sss(4)})^4\>(\ell\,M_P^{\sss(4)})^2~
    \Big(\frac{M_P^{\sss(6)}}{ M_P^{\sss(4)}}\Big)^{\!8}
   \sim (M_P^{\sss(4)})^4\>\Big(\frac{M_P^{\sss(6)}}{ M_P^{\sss(4)}}\Big)^{\!8},
   \label{e:LongCC}
\end{equation}
where the final estimate follows on setting $\ell\sim(M_P^{\sss(4)})^{-1}$.
Using~\eqref{e:expHM4} to eliminate $M_P^{\sss(6)}$ instead, the mass-scale, $M_\L$, corresponding to the cosmological constant~\eqref{e:energydensity} becomes:
\begin{equation}
  M_\L\sim\frac{\pi^2}{9\z_0}|z_0|^{5/4}e^{-2z_0}~\frac{(M_P^{\sss(4)})^2}{\ell^2}
        =\frac{\pi^2}{9\z_0}|z_0|^{5/4}e^{-2z_0}~(M_P^{\sss(4)})^2M_\ell^{~2},
 \label{e:expHL}
\end{equation}
exhibiting the $z_0$-driven {\em\/exponential suppression\/} of $M_\L$ in the \BHM\ models\cite{rBHM5,rBHM6,rBHM7,Berglund:2020qcu}.

 The length-scale $\ell$ is characteristic of the transverse space, $\sY^2_\perp$, and so also for the transverse size of the cosmic brane. For $D=6$, a $\sW^{1,3}$-longitudinal physics mechanism for stabilizing $\ell$ then justifies\footnote{There exist both field and string theory arguments of this type\cite{Antoniadis:2002tr,Dvali:2000xg,Kiritsis:2001bc}.} setting $\ell\sim(M_P^{\sss(4)})^{-1}$ up to factors of $O(1)$.
 In turn, recall that the Ricci form of the Einstein field equations~\eqref{e:EinsteinFE} is in fact the lowest order condition for quantum stability (non-renormalization) of the underlying world-sheet field theory\cite{rF79a,rF79b}.

The original scenario of Ref.\cite{rBHM5} then applies, where the
10-dimensional spacetime of superstring theory is
compactified on a 4-dimensional supersymmetry preserving
space\footnote{All remaining
supersymmetry will be broken by the cosmic brane
solution\cite{rBHM1}.} of characteristic size
$(M_P^{\sss(10)})^{-1}=(M_P^{\sss(6)})^{-1}\sim(10\,\text{TeV})^{-1}\sim10^{-19}$\,m. The
cosmic brane of Ref.\cite{rBHM5} then describes a $3{+}1$-dimensional
de~Sitter
world-brane, with the characteristic scale
$M_P^{\sss(4)}\sim10^{19}\,$\,GeV. 
Furthermore, 
$L\define\L_b^{-1/2}\sim 10^{41}\,\text{GeV}^{-1}\sim 10^{25}$\,m, provides a natural scale that coincides with the Hubble radius.

The relations~\eqref{e:energydensity} and~\eqref{e:LongCC} may well at this stage appear to be a numerical coincidence, but we return to dispel this inference below, in the discussion of the hallmark result~\eqref{e:IRUVseesaw}.

\section{Vacuum Energy and String Theory}
\label{s:VEST}

\subsection{Introduction: Quantum Gravity vs.\ Vacuum Energy}

In the preceding sections we have reviewed the problem of de Sitter space in string theory.
In this section we start from the observation that the problem of de Sitter spacetime in string theory is closely related to the problem of
vacuum energy in string theory. Thus the essential question behind finding de Sitter space in string theory is:
can string theory solve the vacuum energy problem?

The vacuum energy, or the cosmological constant, problem is one of the outstanding fundamental puzzles in physics, as reviewed by Weinberg\cite{Weinberg:1987dv,Weinberg:1988cp, Weinberg:2000yb} and also Polchinski\cite{Polchinski:2006gy}.
This problem has been made even more acute after 
the discovery of dark energy\cite{Riess:1998cb, Perlmutter:1998np}.
These famous observations have established the value of the vacuum energy to be
$\rho_0 \sim (10^{-3}\,\text{eV})^4$, which is somewhere between 50 and 120 orders of magnitude away from the
theoretically predicted values, depending on the  assumptions made\cite{Weinberg:1988cp, Weinberg:2000yb, Polchinski:2006gy}; see also\cite{Martin:2012vx} for a more detailed estimation.

In this section we consider the vacuum energy problem in the context of the 
metastring approach to quantum gravity (as well as its metaparticle limit), the metastring approach being
a duality symmetric and intrinsically non-commutative formulation of string theory.
{ The central result of this section is based on the recent work by Freidel, Kowalski-Glikman, Leigh and Minic\cite{Freidel:2022ryr}. In that paper a UV/IR feedback for the vacuum energy was explored
as arising by defining a notion of quantum (modular) space-time vacua. A regularized microscopic count of vacuum states 
was defined via a process called modular regularization in order to compute the vacuum energy. The calculation of vacuum energy was then tied to a ground state degeneracy. By equating this microscopic ground state degeneracy with macroscopic gravitational entropy (that is, by taking into account holography) a prediction for the vacuum energy 
was deduced that can easily be consistent with observations. In this procedure the smallness of the vacuum energy was tied to the large size of the Universe. In this section we provide some background to this recent insight and comment of its various physics
applications in the context of string theory in de~Sitter spacetime.}

Our central point in this section is that the new insight on 
the vacuum energy problem is naturally offered by
an effective doubled and non-commutative structure of quantum spacetime, the habitat for metastrings and 
metaparticles that was developed in the previous work by 
Freidel, Kowalski-Glikman, Leigh and Minic\cite{Freidel:2013zga, Freidel:2014qna, Freidel:2015pka, Freidel:2015uug, Freidel:2016pls, Freidel:2017xsi, Freidel:2017wst, Freidel:2017nhg, Freidel:2018apz}. 
A generically small cosmological constant
is realized in string theory via the interplay between the phase space structure of the metastring and
the holographic principle of quantum gravity. The same set-up implies a generic seesaw formula
between the vacuum energy scale and the Planck scale and it leads to the natural resolution of the
hierarchy problem in string theory. We spell out explicitly where the usual reasoning based 
on effective field theory breaks down in our approach.

This section is organized as follows: First, we review Polchinski's calculation of the vacuum energy for particles and strings.
Then, we briefly review the basic concepts associated with metaparticles and metastrings.
In Section~\ref{s:VEmeta} we compute the vacuum energy
for metaparticles and metastrings and point out the crucial differences from Polchinski's seminal calculation from Section~\ref{s:VE}.
Then, in Section~\ref{s:resolveCC} we point out how an interplay of the phase space and modular spacetime properties of the metastring
taken together with the holographic nature of low energy gravity in a positive cosmological constant background
that is implied by metastring theory, one obtains a naturally small cosmological constant. 
We also discuss the limit of effective field theory as well as very precise quantization conditions implied
by our argument that might have observable consequences.
Next we discuss how the usual effective field theory results are recovered and
the reasons for the breakdown of effective field theory in the calculation of the cosmological constant
and finally, we point out
the relation between the cosmological constant problem and the gauge hierarchy problem in the context of metastring theory.
We conclude with  an inviting question of how to use this new understanding of the string vacuum in order to illuminate
the problem of masses of elementary particles.

\subsection{Vacuum Energy of Particles and Strings}
\label{s:VE}

Now we turn to the cosmological constant/vacuum energy problem in the context of particle physics and string theory.

A very clear treatment of the vacuum energy problem in string theory and field theory, is given
in\cite[Chapter~7]{Polchinski:1998rq}, pp.~220--222 of Section~7.3; see also\cite{Polchinski:1985zf}. 
Let us follow Polchinski's presentation in the context of a massive scalar particles with mass $m$.
In that case the particle action is ($N$ being the Lagrange multiplier)
\begin{equation}\label{1-Joe}
S \define \int_0^1 \rd\tau \Big[p\cdot \dot x 
- \inv2\,N\left(p^2 + {m}^2\right) 
\Big]\,.
\end{equation}
The vacuum energy is computed from the vacuum particle paths consisting of any number of disconnecting circles. 
By including the correct combinatoric permutation factor and by summing over the paths, this leads to the 
following vacuum partition function\cite{Polchinski:1998rq}
\begin{equation}
Z_{\text{vac}} (m^2) = \exp[Z_{S^1} (m^2)] .
\end{equation}
In the context of a scalar field theory in $D$ spacetime dimensions, whose quanta are represented by 
particles of mass $m$ we have
\begin{equation}
Z_{\text{vac}} (m^2) = \langle 0| \exp(- i H T) |0 \rangle = \exp(- i \rho_0 V_D)
\end{equation}
where the vacuum energy density is given as
\begin{equation}
\rho_0 = \frac{i}{V_D} Z_{S^1} (m^2) .
\end{equation}
Now, the partition function $Z_{S^1} (m^2)$ is the sum over all particle paths with a topology of $S^{1}$\cite{Polchinski:1998rq}
\begin{equation}
 Z_{S^1} (m^2)
  = V_D \int \frac{\rd^D p}{(2\pi)^D} \int_0^{\infty}\frac{\rd l}{2l}\, e^{-(p^2+m^2)l/2}
  = i V_D \int_0^{\infty} \frac{\rd l}{2l}~ (2 \pi l)^{-D/2}\, e^{- m^2 l/2}.
\end{equation}
As emphasized by Polchinski, this expression can be understood intuitively: the modulus of the circle $S^{1}$ is
denoted by $l$, the world line Hamiltonian is $\frac{1}{2}(p^2 +m^2)$, and the $2l$ factor in the denominator
removes the overcounting from translation and reversal of the world line parameter.

We will see later that this particle partition function is a particular limit of the analogous expression for the metaparticle partition function given 
in\cite{Freidel:2018apz}.
Note that $Z_{S^1} (m^2)$ diverges as $l \to 0$. By cutting off the integral at short distance and dropping the
divergent terms one gets\cite{Polchinski:1998rq}
\begin{equation}
\int_0^{\infty} \frac{\rd l}{2l}~ e^{-(p^2 +m^2) l/2} \to - \frac{1}{2} \log(p^2{+}m^2) .
\end{equation}
Also, with the same prescription
\begin{equation}
i \int_0^{\infty} \frac{\rd l}{2l} \frac{\rd p^0}{(2 \pi)}~ e^{-(p^2 +m^2) l/2}
 \to  \frac{1}{2} E_{p}
\end{equation}
where $E^2 = p^2{+}m^2$. 

Thus, in the context of effective field theory of particles one computes the vacuum energy density 
as expected on intuitive physical grounds
\begin{equation}
\rho_0 = \int \rd^3 p~ \frac{1}{2} E_{p}, 
\end{equation}
where $E_{p}^2 = p^2{+}m^2$.
Let us take the effective massless case, because the high momenta dominate the integral. Then this integral
is quartically divergent and it has to be cut-off
$\rho_0 = \int^{\Lambda} \rd^3 p\; \frac{1}{2} E_p$, 
\begin{equation}
\rho_0 \sim \Lambda^{4} .
\end{equation}
A naive Planckian cutoff gives the huge vacuum energy density 120 orders of magnitude off from 
the observed one\cite{Polchinski:1998rq,Weinberg:1988cp}.
Invoking supersymmetry after its breaking (in order to fit the observed world) does not help either\cite{Polchinski:1998rq}.
In that case, by assuming the supersymmetry breaking scale is set by the effective Higgs scale, one gets 52 orders of
magnitude of discrepancy with the observed value.

The corresponding computation in string theory resolves the UV divergence, but
leaves the IR divergence\cite{Polchinski:1998rq}.
We remark that the use of cutoffs in the computation of the cosmological
constant in the context of effective field theory has been recently
carefully discussed by Donoghue in\cite{Donoghue:2020hoh}.

Next we review Polchinski's string theory computation, which generalizes his particle computation of the vacuum energy density.
Polchinski takes the point particle result and generalizes as follows\cite{Polchinski:1998rq}:
\begin{equation}
Z_{S^1} (m^2)   
\define i V_D \int_0^{\infty} \frac{\rd l}{2l}~ (2 \pi l)^{-D/2}\, e^{-m^2 l/2} 
= V_D \int \frac{\rd^D p}{(2 \pi)^D} \int_0^{\infty} \frac{\rd l}{2l}~ e^{-(p^2+m^2)l/2} 
\end{equation}
where mass $m^2$ is given in string theory as 
\begin{equation}
m^2 = \frac{2}{\alpha'} (h + \tilde{h} -2)
\end{equation}
and where the levels are matched $h= \tilde{h}$ as
\begin{equation}
\delta_{h, \tilde{h}} = \int_{- \pi}^{\pi} \frac{\rd \theta}{2 \pi}\, e^{i(h - \tilde{h}) \theta} .
\end{equation}
Thus Polchinski obtains the string analogue of the point particle result (the string partition function on the torus) 
after summing over all physical states (labeled by $i$)
\begin{equation}
\sum_i  Z_{S^1} (m_i^2)
= i V_D \int_0^{\infty} \frac{\rd l}{2l} \int_{- \pi}^{\pi} \frac{\rd \theta}{2 \pi}~
 (2 \pi l)^{-D/2} \sum_i e^{-i (h_i + \tilde{h}_i -2) +i(h_i - \tilde{h}_i) \theta } .
\end{equation}

Returning to Polchinski's calculation, the claim is that after
one introduces the modulus of the torus
\begin{equation}
2 \pi \tau \define \theta + \frac{i l}{\alpha'}\define \tau_1 + i \tau_2, \quad q \define \exp(2\pi i \tau)
\end{equation}
the above string partition function reads
\begin{equation}
Z_{\text{string}}=\sum_i  Z_{S^1} (m_i^2)
= i V_D \int_R \frac{\rd\tau\,\rd\bar{\tau}}{4 \tau_2}~ (4 \pi^2 \alpha' \tau_2)^{-D/2}
\sum_i q^{h_i -1} {\bar{q}}^{\tilde{h}_i -1}
\end{equation}
where the region of integration $R$ is
\begin{equation}
R: \quad \tau_2 >0, \quad |\tau_1|<\frac{1}{2}.
\end{equation}

Polchinski notes that the integrands of $Z_{\text{string}}$ and the toroidal string partition function $Z_{T^2}$ 
are identical but with a completely different region of integration.
In the context of the string partition function on the torus, $Z_{T^2}$, which has the same expression as the above formula
for $Z_{\text{string}}$, the fundamental region of integration is
\begin{equation}
F: \quad |\tau| >1, \quad |\tau_1|<\frac{1}{2}
\end{equation}
and in the string case the UV divergent region is absent.
Thus the string answer is UV finite (but there is still an IR divergence due to the tachyon of the bosonic string).

{ However, as Polchinski points out, even though the string answer is UV finite, the string vacuum energy density still
scales as in the particle case $\rho_{\text{string}} \sim \Lambda^D$, which can be seen by looking explicitly at the momentum
integral in the partition function $Z_{\text{string}}$.}

The main message here is that the vacuum energy problem persists in string theory (as a natural extrapolation of the
vacuum energy problem in particle physics/effective quantum field theory).
Thus, a new insight into the vacuum energy problem is needed. Therefore,
we turn to the concept of metaparticles, metastrings and modular spacetime, and
then we repeat Polchinski computation in that context in order to resolve the vacuum energy problem.

\subsection{Metaparticles, Metastrings and Modular Spacetime}
\label{s:MP+MS}
Here we give a short background on metaparticles\cite{Freidel:2018apz, Freidel:2021wpl}
followed by a brief discussion of metastrings and modular spacetime
and the underlying physics behind these concepts.
The idea of metaparticles (the zero modes of the metastring) is implied by the generic quantum spacetime formulation of quantum theory and
its associated geometry described in\cite{Freidel:2016pls}.
In\cite{Freidel:2016pls}, Freidel, Leigh and Minic have argued that any quantum theory is endowed with
a generic quantum polarization associated with a new concept of
quantum spacetime that we call modular spacetime\cite{Freidel:2015uug}.
Essentially, if one considers a complete basis of unitary observables on a latticized spacetime, one
has to consider the dual lattice as well.  According to the modular spacetime
formulation of quantum theory, quite generically, quantum theory can be formulated using
a quantum spacetime, which can be very roughly understood as a covariant quantum superposition of the spacetime
and its dual. More precisely, modular spacetime is given by a selfdual lattice of (covariant) phase
space lifted to the non-commutative, Heisenberg-Weyl algebra\cite{Weyl:2008aa}, or even more mathematically, the modular spacetime
describes the space of all commuting subalgebras of the Heisenberg-Weyl algebra.
(This is based on the famous theorem of Mackey\cite{Mackey:1949aa} regarding the geometry of all commutative subalgebras of
the Heisenberg algebras. The commutative nature of modular variables of Aharonov\cite{Aharonov:2005uc}
is responsible for
the name ``modular spacetime''.)
This quantum, modular spacetime polarization manifestly realizes quantum non-locality, associated with 
quantum superposition principle, and is consistent with relativistic causality.
It also resolves an apparent contradiction between non-locality (fundamental length and time)
and relativistic covariance.
This generic modular polarization comes equipped with a new geometric structure called Born geometry\cite{Freidel:2013zga, Freidel:2014qna}.
Born geometry unifies symplectic, orthogonal and conformal geometries
and is found to be fundamental in a new formulation of quantum gravity in the guise of
string theory, called metastring theory\cite{Freidel:2015pka}, reviewed briefly in what follows.
Metastring theory lives in modular spacetime and it
represents an intrinsically non-commutative,
chiral phase space-like, and $T$-duality covariant formulation
of string theory\cite{Freidel:2015pka,  Freidel:2017xsi, Freidel:2017wst, Freidel:2017nhg}.
The zero modes of the metastring define a new concept, called metaparticle, that explicitly realizes the
geometry of modular spacetime\cite{Freidel:2018apz}, and that could be considered as a
prediction of the modular representation of quantum theory. The theory of metaparticles can be also understood as a particle model for Born geometry of modular spacetime\cite{Freidel:2013zga, Freidel:2014qna}.

The series of papers\cite{Freidel:2013zga, Freidel:2014qna, Freidel:2015pka, Freidel:2015uug, Freidel:2016pls, Freidel:2017xsi, Freidel:2017wst, Freidel:2017nhg, Freidel:2018apz} (for reviews consult\cite{Freidel:2019jor, Minic:2020oho})
offer a  view on
quantum gravity in terms of ``gravitization of the quantum''.
More precisely, quantum gravity is defined as a theory of
a dynamical Born geometry, as realized in metastring theory. (A mathematical elaboration of Born geometry can be found in\cite{Freidel:2017yuv}).
This development can be understood as a concrete
implementation of the physical concept of Born reciprocity (the fundamental reciprocity between spacetime and
energy-momentum space)\cite{Born:1935dbe,Born:1938zve,Born:1949yva}
as well as the new idea of relative (or observer dependent) locality\cite{Amelino-Camelia:2011hjg,Amelino-Camelia:2011lvm}.
Thus, the path from quantum theory to quantum theory of gravity is closely
analogous to the
familiar path that leads from the physical and geometric underpinnings of special relativity to the general theory of relativity.
Metaparticles could be considered as a generic prediction of this new view on the problem of quantum gravity.
As we will discuss in the penultimate section of this paper, each observed (canonical) particle should have a dual particle, and such dual particles appear as
good candidates for dark matter\cite{Freidel:2019jor, Minic:2020oho} that is correlated to the observed, visible matter\cite{Ho:2010ca, Edmonds:2017zhg, Edmonds:2020iji}. Also, this approach predicts a dual gravity sector (associated with a dynamical dual spacetime)\cite{Freidel:2019jor, Minic:2020oho},
which can be used to model dark energy as a cosmological constant in the observed spacetime\cite{Berglund:2019ctg, Berglund:2020qcu}.
We will discuss this later as well.

It will be sufficient for our purposes to consider free metaparticles, whose dynamics are given by a world-line action involving a doubling of the usual phase space coordinates. The action is of the form\cite{Freidel:2017wst, Freidel:2017nhg, Freidel:2018apz}
\begin{equation}\label{1}
S_{mp} \define \int_0^1\!\rd\tau \Big[\, p\cdot \dot x +\tp\cdot \dot{\tx}+ \alpha'\, p \cdot\dot{\tp}
+ \inv2\,N\left(p^2 +{\tp}^2 + \fm^2\right) +{\tilde N}\left(p\cdot \tp - \mu \right)\Big]\,.
\end{equation}
Here the signature is $(-,+,\ldots,+)$ and the contraction of indices is denoted by ``$\cdot$''.
At the classical level, the theory has world-line reparametrization invariance and two additional features\cite{Freidel:2018apz}.
The first new feature of the model is the presence of an additional local symmetry, which from the string point of view corresponds to the completion of worldsheet diffeomorphism invariance. From the particle world-line point of view,
this symmetry is associated with an additional local constraint. The second new feature is the presence of a non-trivial symplectic form on the metaparticle phase space, the non-zero brackets being
\beq
\{p_\mu, x^\nu\}=\delta_\mu^\nu,\,\,\,
\{\tp_\mu, \tx^\nu\}=\delta_\mu^\nu,\,\, \,
\{ \tx_\mu,x^\nu \}= \pi\alpha' \delta_\mu^\nu.
\label{xtxcomm}
\eeqn
with $\mu,\nu=0,1,\cdots\!,d-1$.
Because of its interpretation as a particle model on Born geometry, associated with the modular representation of quantum theory,
the space-time on which the metaparticle propagates is ambiguous, with different choices related by what in string theory we would call $T$-duality.
The attractive features of this model include world-line causality and unitarity, as well as an explicit mixing of widely separated energy-momentum scales.

The metaparticle propagator follows from the world-line path integral involving the above action
and has the following form\cite{Freidel:2018apz}:
$G(p,\tp; p_i,\tp_i) \sim \delta^{(d)}(p{-}p_i)\delta^{(d)}(\tp{-}\tp_i)
\frac{\delta(p\cdot\tp-\mu)}{p^2+\tp^2+\fm^2-i\varepsilon}
$.\footnote{One can also deduce the corresponding scalar potential from this propagator, which is of
the Friedel form $\frac{e^{-\alpha r}}{r^{\beta}}\cos(\gamma r)$, where $\alpha, \beta, \gamma$ are
functions of $m, \mu, d$ (unpublished work of Freidel, Kowalski-Glikman, Leigh and Minic).}
Although we will not consider metaparticle interactions directly in this paper, such interactions preserve both $p$ and $\tp$.
We see that although $p,\tp$ might resemble a doubling of momentum space, the propagator contains a simple pole along with a $\delta$-function constraint. We note that the usual string spectrum on flat space-time corresponds to the rather singular limit (actually a super-selection sector) in which all states have $\mu=0$ and $\tp=0$; flat compactifications correspond to including states with non-zero $\tp$, at least along space-like directions.

It follows from the metaparticle action that the metaparticle kinematics is defined by two constraints
\begin{alignat}9
\cH &\define p^2 +{\tp}^2 + \fm^2 &&=0, \label{1const}\\*
\cD &\define p\cdot \tp - \mu &&=0. \label{2const}
\end{alignat}
These constraints together are invariant under $O(1,d{-}1)_1\ltimes O(1,d{-}1)_2$.
The group $O(1,d{-}1)_1$ acts as $(\tp,p)\mapsto (\Lambda\tp,p\Lambda^{-1})$, and so can be identified with the usual Lorentz symmetry; $p^2$, $\tp^2$ and $p\cdot\tp$ are all $O(1,d-1)_1$-invariants; two combinations of these are fixed by the constraints, while $\delta=\frac12(\tp^2-p^2)$ is not, and this is {\it not} an invariant of $O(1,d{-}1)_2$. 
Thus, we can interpret the states as elements of  an orbit of $O(1,d{-}1)_2$, with $-p^2=m^2=\delta+\frac12\fm^2$. In addition, for each state, $\tp$ is constrained by \eqref{2const}. Depending on the values of $\mu,\fm^2$, $\delta$ will occupy a certain range of allowed values, and for physical reasons we will need to restrict those values.  Thus we can think of a metaparticle state as being labeled by momentum $p$ and ``internal'' quantum numbers $\tp$ which transform under Lorentz. This might seem to be in conflict with the Coleman-Mandula theorem\cite{Coleman:1967ad}; however, within the sector of fixed non-zero $\tp$, $O(1,d{-}1)_2$ is effectively spontaneously broken.   Here the novel feature is that the internal symmetry is non-compact, and we should consider separately the cases in which $\tp^2$ is time-like, light-like and space-like\cite{Freidel:2021wpl}.


Also, given the fact that the $\mu=0$ superselection sector belongs to usual particles,
the metaparticle extension ($\mu \neq 0$) should be explored in connection with phenomenology of dark matter.\footnote{We also note the similarity between our new parameter $\tp$ and the continuous spin representations of the canonical Poincare group\cite{Schuster:2013pxj}.}
Thus the usual detection analysis of particulate dark matter should be revisited in the
context of metaparticle kinematics and representation theory.
We emphasize that the value of $\mu$ differs for different particles.

We remark that there exists an interesting observation concerning the metaparticle dispersion relation in 
a cosmological context\cite{Freidel:2021wpl}. In curved spacetime with metric $g$ we have 
(as a metaparticle limit\cite{Freidel:2018apz} of the general Hamiltonian constraint for the metastring\cite{Freidel:2015pka, Freidel:2017xsi})
\begin{equation}\label{curvdisp}
g^{\mu\nu}\, p_\mu p_\nu + g_{\mu\nu }\, \tp^\mu \tp^\nu =-\fm^2
\end{equation}
Now, in the case of a cosmological background (such as the space with a positive cosmological constant --- asymptotic de Sitter space --- including
its excitations, such as black holes)
\begin{equation}\label{metr}
  ds^2 = - dt^2 + a^2 d\mathbf{x}^2
\end{equation}
we can deduce\cite{Freidel:2021wpl}
that in the early (small) universe the dual (``dark'') degrees of freedom dominate, and that
in the old (large) universe the visible degrees of freedom are dominant. Also, we can go to a
frame in which the dual 3-momenta are frozen $\vec{\tp}=0$, and thus $E \tilde E =\mu$, leading to an
effective IR deformation of the standard dispersion relation for a scalar particle
$E^2 +\mu^2/E^2 = \vec{p}\,^2 + m^2$, which could be  experimentally tested. The relation $E \tilde{E} = \mu$ will be
important in our discussion of vacuum energy in the following sections.


Now we give a very brief review of metastrings with emphasis on the phase space (non-commutative) structure and extensification of
spacetime.
As already stated, metaparticles are zero modes of the metastring.  Metastring theory\cite{Freidel:2013zga, Freidel:2014qna, Freidel:2015pka, Freidel:2015uug, Freidel:2016pls, Freidel:2017xsi, Freidel:2017wst, Freidel:2017nhg, Freidel:2018apz, Freidel:2019jor} can be thought of as a formulation of string theory in which the target space is doubled in such a way that $T$-duality acts linearly on the coordinates. This doubling means that  momentum and winding modes appear on an equal footing.  We refer to the target space as a phase space since the metastring action requires the presence of a background symplectic form $\omega$. The metastring formulation also requires the presence of geometrical structures that generalize to phase space the spacetime metric and the $B$-field (where the $B$-field originates from the symplectic structure $\omega$). In fact, in the metastring  we have not one but {\it two} notions of a metric.
The first metric $\eta$  is a neutral metric that defines a bi-Lagrangian structure and allows to define the classical spacetime as a Lagrangian sub-manifold\footnote{We remind the reader that in symplectic geometry, a Lagrangian subspace is a half-dimensional submanifold upon which the symplectic form pulls back to zero. In more familiar terms, a Lagrangian submanifold might be the subspace coordinatized by the $q$'s within the phase space coordinatized by $q$'s and $p$'s.} --- more precisely, the classical spacetime is defined as a null subspace for $\eta$ which is also Lagrangian for $\omega$. 
The second metric $H$ is a metric of signature $(2,2(D{-}1))$ that encodes the geometry along the classical spacetime (of dimension $D$) as well as the transverse energy-momentum space geometry.
In this formulation, $T$-duality exchanges the Lagrangian sub-manifold with its image under $J=\eta^{-1}H$.
Classical metastring theory is defined by the following action\cite{Freidel:2015pka} 
which realizes the above comments about quantum field theory in the modular polarization for the 
special case of a two-dimensional world-sheet 
quantum field theory\footnote{See also\cite{Tseytlin:1990va} and the double field formalism\cite{Hull:2009mi}.}
\begin{equation}\label{e:MSAction}
S^{\text{ch}}_{\text{str}}=\frac{1}{4\pi}\int_{\Sigma}\rd^2\sigma~
    \Big[\pa_{\tau}{\X}^{A} \big(\eta_{AB}(\X)+\w_{AB}(\X)\big)
    -\pa_\s\X^A H_{\!AB}(\X)\Big] \pa_\s\X^B, 
\end{equation}
where $\X^A\define (x^a/\lambda ,\tx_a/\lambda )^{T}$ (with $\alpha' = \lambda^2$ and $a=0,\cdots\!,D{-}1$) are dimensionless coordinates on phase-space like (doubled) target spacetime and the fields $\eta, H,\omega$ are all dynamical (i.e., generally $\X$-dependent) target spacetime fields. In the context of a flat metastring we have 
 constant $\eta_{AB}$,  $H_{AB}$
 and $\omega_{AB}$
\be\label{etaH0}
	\eta_{AB} \define \left( \begin{array}{cc} 0 & \delta \\ \delta^{T}& 0  \end{array} \right),\quad
H_{AB} \define  \left( \begin{array}{cc} h & 0 \\ 0 &  h^{-1}  \end{array} \right),
\quad \om_{AB} = \left( \begin{array}{cc} 0 & \delta \\ -\delta^{T}& 0  \end{array} \right) ,
\ee
where $\delta^{\mu}_{\nu}$  
 is the $d$-dimensional identity matrix and $h_{\mu\nu}$ is the $d$-dimensional Lorentzian metric, $T$ denoting transpose.
The Hamiltonian and diffeomorphism constraints of the metastring read as follows:
\be
\cH = H_{AB}\,\partial_{\sigma} \X^A \partial_{\sigma} \X^A, \quad
\cD =\eta_{AB}\,\partial_{\sigma} \X^A \partial_{\sigma} \X^A
\ee
These constraints reduce to the Hamiltonian and diffeomorphism constraints for the metaparticles.
In the standard string notation the metaparticle parameters $m^2$ and $\mu$ become
$\alpha' m^2 = N + \tilde{N} -2$ and $\alpha' \mu = N - \tilde{N}$.
Note the fundamental non-commutativity of the metastring\cite{Freidel:2017wst, Freidel:2017nhg}:
\be
[\X^A(\s),\X^B(\s')]=  2i\lambda^2 \left[ \pi \omega^{AB}-\eta^{AB}\theta(\sigma-\s')\right],
\label{e:msnoncom}
\ee
where $\theta(\s)$ is the staircase distribution, i.e., a solution of $\theta'(\s)=2\pi\delta(\s)$; it is odd and  quasi-periodic with period   $2\pi$.
This intrinsic non-commutativity is tied to the fact that the algebra of vertex operators does not contain any cocycles, and that
vertex operators are found as representations of the above Heisenberg-Weyl algebra.
We remark that upon setting of the symplectic background to zero, and the integration of the dual variables, one recovers the
familiar Polyakov action for the bosonic string propagating in a background spacetime.
The critical dimension of the metastring is still the critical central change $c$ (``critical spacetime dimensions'') of the bosonic string, that is, $c=26$.

Finally, a brief note on the habitat of metastring theory\cite{Freidel:2013zga, Freidel:2014qna, Freidel:2015pka, Freidel:2015uug, Freidel:2016pls, Freidel:2017xsi, Freidel:2017wst, Freidel:2017nhg, Freidel:2018apz, Freidel:2019jor}, the modular spacetime and its extensification:
The space of all commuting subalgebras of the Heisenberg-Weyl algebra defines the modular space. The modular space
can be understood as a self-dual lattice of phase space lifted to the Heisenberg-Weyl algebra. It is equipped with what we call
Born geometry: the symplectic geometry of phase space, the doubly orthogonal geometry associated with the self-dual lattice
of phase space, and the doubly metric geometry associated with the choice of vacuum in this modular formulation of quantum theory,'
in terms of modular variables (originally suggested by Aharonov and collaborators in the context of quantum foundations\cite{Aharonov:2005uc}).
These are precisely the backgrounds of the above action for the metastring. Note that the metastring propagates in modular spacetime.
Note also that the bosonic metastring does not have the tachyon interpretation for the leading vertex operator describing the plane waves
in the above chiral phase space formulation of the metastring (the tachyon interpretation is essentially tied to 
the classical spacetime interpretation of the string target space). The basic ``atoms'' of modular spacetime are the fully ``compactified''
$25+1$ dimensional bosonic string. Because of the modular interpretation there are no issues with closed time-like loops in this context.
One builds spacetime backgrounds out of these atomic building blocks in the process called extensification, described in 
\cite{Freidel:2016pls}. From this point of view, one starts from zero dimensions and builds upwards to 4 spacetime dimensions 
(the 10 dimensional superstring being a very particular, and not a generic extensification).
We will use the modular spacetime nature of the metastring background as well as the concept of extensification
in our argument regarding the metastring resolution of the cosmological constant problem in what follows.

What happens if we ``extensify'' the modular cell,
which is of unit Planckian area, into a cell of area $N$ in Planck units\cite{Freidel:2016pls}? We may
consider changing the lattice in different ways that change the size of its fundamental
cell. This is a non-trivial process, as it changes the structure of the theory, and in particular
modifies the Hilbert space. More precisely, while a change of the lattice shape $L \to L'$
that does not modify its size can be implemented unitarily, a change that modifies its size
leads to inequivalent Hilbert spaces $H_L$ and $H_{L'}$. 

We stress that there are really two ways in which we can extend the cell\cite{Freidel:2016pls}: either we extensify modular space along the 
space directions or the dual space directions.
We also distinguish between the operation of extensification which makes the size of space bigger than the size of
dual space, while the second (called ``coarsening operation'') increases the size of the cell homogeneously in both directions
(space and dual space). We can extend the lattice by keeping the fundamental scales (the fundamental size of space and
the fundamental size of dual space) fixed, or we can keep the lattice fixed while rescaling the fundamental scale of
space but keeping the fundamental size of dual space fixed. The coarsening limit corresponds to a rescaling of
both the fundamental scale of space and the fundamental scale of dual space.
From this perspective both extensification and coarsening correspond to
expanding towards a larger phase space cell.
What distinguishes the two procedures is that during extensification we change the shape of the cell, which
is elongated along the space direction, while during coarsening, we do not.

Finally, each phase cell has one unit of flux. For multiflux states, we can  distribute the fluxes along the
phase cells of the extensified space. First we can map the extensified cell with one flux into the unit cell with many fluxes,
where the Hilbert space of an extensified cell with a unit flux is embedded onto the Hilbert space of a unit cell with many fluxes.
We can also embed the unit cell with many fluxes into many unit cells each with a unit flux, which amounts to an embedding of
a Hilbert space of the unit cell with many fluxes to a 
tensor product of Hilbert spaces of unit cells with unit flux, in a coherent fashion. 

Thus, modular space has properties which are far more intricate
than those of the usual classical space. There exists a possibility of realizing states associated
with larger space as multi-flux states. These identifications in principle blur the line between
what is pure geometry and what can be considered as matter. This unification of matter and
geometry contained in modular space is one of the most fascinating aspects of the construction presented in\cite{Freidel:2016pls}.

\subsection{Vacuum Energy of Metaparticles and Metastrings}
\label{s:VEmeta}
We now comment on the metaparticle and metastring analogues of Polchinski's calculation.

As explained in the original paper\cite{Freidel:2018apz}
we confine our attention to the $x,\tp$ polarization, because of the commutative nature of this pair of variables, which in the quantum theory means that we are considering a specific set of transition amplitudes. 
(Here we redefine $\frac{N}{2} \define e$ and $\tilde{N}\define \te$ in order to match the notation from our original paper\cite{Freidel:2018apz}.)
Since we have two constraints, the quantum states will be further labelled by a pair of  evolution parameters, which we call $\ell,\tilde\ell$. The transition amplitude in the $x,\tp$ polarization has a canonical interpretation
\beq
\begin{aligned}
 K(x_f,\tp_f,\ell_f,\tilde{\ell}_f; x_i,\tp_i,\ell_i,\tilde{\ell}_i)
 &=\langle x_f,\tp_f;\ell_f,\tilde{\ell}_f| x_i,\tp_i;\ell_i,\tilde{\ell}_i\rangle\\
 &=\langle x_f,\tp_f|e^{-i(\ell_f-\ell_i) \widehat\cH-i (\tilde{\ell}_f-\tilde{\ell}_i)\widehat\cD}| x_i,\tp_i\rangle.
\end{aligned}
\label{MetaParticlePI}
\eeq
As we will see later in more detail,
causality imposes that 
$\ell=\ell_f-\ell_i>0$ while we will not restrict the sign of $ \tilde{\ell}=\tilde\ell_f-\tilde\ell_i$.
The relativistic propagator, denoted $G$, is obtained by 
integrating $K$ over $\ell$ and $\tilde{\ell}$.
This comes about when we introduce a parameter
$\tau$ parametrizing the world-line and express the Hamiltonian parameters in terms of the frame fields $(e,\te)$ 
\be
 \ell=\int_{\cal C} |e|(\tau),\qquad \tilde{\ell}=\int_{\cal C} \te(\tau).
\ee 
where ${\cal C}:\tau\to (x,\tx,p,\tp, e,\te)(\tau)$ is a path in phase space, with boundary conditions $(x,\tp)(\tau_{i})=(x_{i},\tp_{i})$ and
$(x,\tp)(\tau_{f})=(x_{f},\tp_{f})$, and the modulus on $e$ follows from the causality requirement, as explained in detail in\cite{Freidel:2018apz}.

As in the usual derivation of the particle path integral, we use factorization repeatedly to find a continuum expression  for $G$. 
Allowing for arbitrary parametrizations of the world-line yields the integration over $e$ and $\te$, as is explained in detail in Appendix~A
of\cite{Freidel:2018apz}. This in turn means that the propagator $G$ is obtained 
by integrating over all Lagrange parameters and is independent of the coordinate times $\tau_i,\tau_f$. Thus we find
\begin{equation}
 G(x_f,\tp_f; x_i,\tp_i)
 =\int [\rd e\,\rd\te] \int_{x_{i},\tp_{i}}^{x_{f},\tp_{f}} [\rd^dx\,\rd^d\tp]
   \int [\rd^dp\,\rd^d\tx ]~
    e^{i\int_{\cal C}(p\cdot\rd x-\tx\cdot\rd{\tp}
        +\pi\alpha' p\cdot\rd{\tp}-|e|{\cH}-\tilde e{\cD})},
\end{equation}
The derivation of this path integral proceeds without incident because the operators $\hat{x},\hat{\tp}$ commute. 
Computing it, we find after gauge fixing, that the propagator in $x,\tp$ space is
\beqn
G(x_f,\tp; x_i,\tp_i)
\sim
\delta^{(d)}(\tp-\tp_{i})
\int \frac{\rd^dp}{(2\pi)^d} \int \rd\ell\, \rd\tilde\ell~ 
e^{-i\ell\cH -i\tilde\ell \cD}\,e^{ip\cdot (x_f-x_i)},
\eeqn
Integrating over  $\ell\in (0,\infty)$ and $\tilde\ell\in (-\infty,\infty)$ then yields the metaparticle propagator
\beqn\label{doubletramp}
G(x,\tp; 0,\tp_i)
\sim
\delta^{(d)}(\tp-\tp_{i})
\int \frac{\rd^dp}{(2\pi)^d}~\frac{e^{ip\cdot x} }{p^2+\tp^2+m^2-i\varepsilon}~
     \delta(p\cdot\tp-\mu).
 \label{e:mGreen}
\eeqn
There are two differences compared to an ordinary relativistic particle propagator:
 One is the $\delta$-function of the $\cD$ constraint, and
 the other is the presence of $\tp$ in the denominator. 
We note that the propagator is invariant under the change $\mu\to -\mu$, if we simultaneously change $\tp\to-\tp$. Consequently, we will without loss of generality assume that $\mu>0$.

We have been cavalier about the $dl$ and $d\tilde{l}$ measures. In order to understand that question let us recall how Polyakov
derives the $\frac{\rd l}{2l}$ term in the measure for the ordinary particle\cite{rAP-GFS} (also found in Polchinski's calculation). 
For a circular path, one has the translation symmetry $\tau \to \tau+a$ and this must be excluded from the measure.
This is obvious from the action for the ordinary particle.
Polyakov implements this symmetry via the following relation 
$
\int_0^l \rd a \int_0^l \frac{\rd \tau}{l} \delta(f(\tau) -a) =1.
$
So, Polyakov ends up with the measure
$
\frac{\rd l}{l} 
$
which is Polchinski's answer as well $\frac{\rd l}{2l}$, up to a factor of $2$ which really takes into account the reverse paths\label{p:Joe's2}.
(This is essentially the Faddeev-Popov procedure.)
A different justification for this measure was given in Green, Schwarz and Witten, where the above expression
for the one loop partition function and the vacuum energy was given from the field theory point of view, and the measure results from
the regularization of the Schwinger parametrization of that field theoretic expression for vacuum energy\cite{rGSW1,rGSW2}.

The metaparticle can be viewed as a rigid string, that is a string without string oscillators (just zero modes).
The measure for metaparticle should thus be the same one as the measure of the torus for the one loop partition function
of the string that we reviewed following Polchinski.
In order to do that we need to rescale the dimensionful $\tilde{l}$ by the size of the string $l_s$ (or equivalently, $\alpha'=l_s^2$).
Let us call $\tilde{l}/l_s \define \theta$. Then the correct measure for the metaparticle Schwinger parameters is
the torus measure of the string (with the additional modular invariance requirements which lead to the fundamental domain
of integration as in Polchinski's calculation).

Thus, in order to evaluate the metaparticle vacuum energy density we set $x=0$ and ${\tp}_i=0$ in the above formula 
for the metaparticle propagator and use the Schwinger parametrization of the metaparticle propagator in order
to get the following expression as the correct analogue of Polchinski's result for the vacuum energy density (the above 
expression for $Z_{S^1} (m^2)$ divided by the spacetime volume factor, $V_D$, as well as, in our case, the volume of the dual spacetime)
\begin{equation}
V_{\tp}\, Z_{mp} (m^2, \mu) = \Big (\int_{-\pi}^{\pi} \frac{\rd \theta}{2 \pi}\, e^{i(p{\cdot}\tp - \mu) l_s^2 \theta/2} \Big )
\int \frac{\rd^D {\tp}\,\rd^D p}{(2 \pi)^{2D}} 
\int_0^{\infty} \frac{\rd l}{2l}\, e^{i(p^2 + \tp^2+m^2) l/2} .
\end{equation}
{Note that this expression has to be divided by the volume $V_{\tp}$ of the $\tp$ space in order to give the correct
units for the vacuum energy density.}
Unlike Polchinski's formula it is a Lorentzian expression, and it reduces
to Polchinski's in the singular limit $\mu \to 0$, $\tp \to 0$.

Given the homogeneity of the measure over the metaparticle Schwinger parameters
the above expression for the metaparticle one loop partition function (divided by the volume $V_{\tp}$ of the $\tp$ space, in order to
get the correct units for the vacuum energy density) becomes
(here we only concentrate on the scaling with the momentum space volume)
\begin{equation}
Z_{mp} (m^2, \mu) = a_D  
\int \frac{\rd^D p}{(2 \pi)^D},
\end{equation}
where $a_D$ is a factor that does not depend on the momentum scale in $D$ spacetime dimensions.
(For simplicity we will omit the dependence on $m^2$.)
Let the cutoffs in the $p$ and $\tp$ spaces be denoted by $\Lambda$ and $\tilde{\Lambda}$.
The cutoffs $\Lambda$ and  $\tilde{\Lambda}$ are naturally implied by the fundamental non-commutativity of the metaparticle.
(In what follows we will use the diffeomorphism constraint in order to emphasize the possible
relation between the cosmological constant problem and the gauge hierarchy problem, to be discussed at the end of
this section.
We will also connect to the seesaw formulae derived in our toy model from Section~\ref{s:dSdefo}.)
Then, first, we have from the diffeomorphism constraint that, in principle 
\begin{equation}
\Lambda \tilde{\Lambda} = \mu_{\text{vac}}
\end{equation}
and second, that 
\begin{equation}
Z_{mp} (m^2, \mu) = a_D 
\int \frac{\rd^D p}{(2 \pi)^D} = a_D \Lambda^D  = a_D {\Big(\frac{\mu_{\text{vac}}}{\tilde{\Lambda}}\Big)}^D .
\end{equation}
Here $\mu_{\text{vac}}$ denotes the vacuum energy $\mu$ parameter. (Note the $\mu$ parameters are different for different particles,
and different physical situations.) This seesaw formula is natural in the sense of 't\,Hooft because the increase of the
$\tilde{\Lambda}$ leads to the lowering of the vacuum energy.

Given the intrinsic non-commutativity of the metaparticle, the spacetime polarization is not fixed.
Still one can pick a particular, ``chiral'' spacetime polarization in the above ``doubled'' formulation.
In particular, we could have picked the dual, $\tx, p$ polarization, because of the commutativity of that pair of variables.
Then we would have obtained by replacing $p$ by $\tp$
\begin{equation}
Z_{mp} (m^2 , \mu) = a_D {\Big(\frac{\mu_{\text{vac}}}{{\Lambda}}\Big)}^D  .
\end{equation}
Thus, in either case we get the following ``seesaw'' formula for the observed vacuum energy density $\rho_0$
 which incorporates the fundamental UV/IR relation of the metaparticle
\begin{equation}
\rho_0 = a_D  {\Big(\frac{\mu_{\text{vac}}}{\tilde{\Lambda}}\Big)}^D 
\end{equation}
which is what is apparently observed if $\mu_{\text{vac}}$ is the $\text{TeV}^{2}$ scale, $\tilde{\Lambda}$ is the Planck scale, and $D\,{=}\,4$.

The main point here is that the big spacetime is balanced by a Planckian dual spacetime, and vice versa.
This fits the picture advertised in\cite{Freidel:2021wpl}. And that is the first important point regarding the cosmological constant problem.
The metastring should make the whole calculation both UV and IR finite; see below.
In the following subsections we will explain the justification for the use of the Planck scale for $\tilde{\Lambda}$ and the TeV scale for $\mu_{\text{vac}}$ (the vacuum scale associated with the
Higgs part of the matter sector.)

Before doing that, we note that the metastring analogue of Polchinski's expression for the canonical string written as an infinite sum of
particle excitations is  (after writing $m^2 = \frac{2}{\alpha'} (N_L + N_R -2)$
and $\mu=  \frac{2}{\alpha'} (N_L - N_R )$, where $N_L$ and $N_R$ denote left and right string oscillators) after summing
over all oscillators 
\begin{equation}
V_{\tp} \sum_{L, R} Z_{mp} (m^2, \mu) \define V_{\tp}\, Z_{ms}
\end{equation}
so that we get
\begin{equation}
V_{\tp}\, Z_{ms}= \sum_{L, R} \Big (\int_{-\pi}^{ \pi} \frac{\rd \theta}{2 \pi}\,
e^{i[p{\cdot}\tp - (N_L - N_R)] l_s^2 \theta/2} \Big )
\int \frac{\rd^D {\tp}\, \rd^D p}{(2 \pi)^{2D}} 
\int_0^{\infty} \frac{\rd l}{2l}\, e^{i[p^2 + \tp^2+(N_L + N_R)] l/2} .
\end{equation}

This expression should in principle demonstrate an explicit
relation between the UV and IR finite behavior of the metastring, which is expected on general
grounds in metastring theory (as opposed to the UV finite and IR divergent behavior
of the canonical string). 
We will comment on this issue in what follows.

On the other hand, we have once again
\begin{equation}
Z_{ms} =b_D \int \frac{\rd^D p}{(2 \pi)^D} = b_D \Lambda^D= b_D {\Big(\frac{\mu_{\text{vac}}}{{\Lambda}}\Big)}^D,
\end{equation}
where $b_D$ is another purely numerical factor in $D$ spacetime dimensions. Therefore
the metastring vacuum energy scales as  
\be
\rho_{ms} = b_D \Lambda^D = b_D  {\Big(\frac{\mu_{\text{vac}}}{\tilde{\Lambda}}\Big)}^D.
\ee
This seesaw behavior (again technically natural in the sense of 't\,Hooft) comes about explicitly from the momentum integral in the partition function $Z_{ms}$, as in the case of metaparticles.
Note that both the metastring vacuum energy is UV finite as in the case of Polchinski's original computation in bosonic string theory.

We expect that the metastring should resolve the IR divergence that is due to the tachyon of the canonical bosonic string. The metastring should
display a self-dual fixed point and the UV/IR correspondence, and thus UV finiteness should imply IR finiteness, and thus the absence of
the tachyon, as stated in the original papers on metastrings and modular spacetime
\cite{Freidel:2013zga, Freidel:2014qna, Freidel:2015pka, Freidel:2015uug, Freidel:2016pls, Freidel:2017xsi, Freidel:2017wst, Freidel:2017nhg}.
In what follows we concentrate on the cosmological constant problem.

\subsection{The Cosmological Constant Problem}
\label{s:resolveCC}
Given these results we now come to our main point: the above vacuum energy of the metastring can be naturally
small thus clarifying the cosmological constant problem.
This is the central result of the recent work by Freidel, Kowalski-Glikman, Leigh and Minic\cite{Freidel:2022ryr}. As already emphasized, in that paper a UV/IR feedback for the vacuum energy was explored
as arising by defining a notion of quantum (modular) space-time vacua, and a regularized microscopic count of vacuum states 
was defined via a process that called modular regularization and that was used to compute the vacuum energy. The calculation of vacuum energy was then tied to a ground state degeneracy and by equating this microscopic ground state degeneracy with macroscopic gravitational entropy, or holography, a prediction for the vacuum energy 
was obtained that was consistent with observations.
Also, the concept of contextuality (one of the hallmarks of quantum theory) emerged as a central guiding idea --- as opposed to statistical, anthropic, reasoning\cite{Weinberg:1987dv,Polchinski:2006gy}.

In what follows we will not repeat this argument, but we will emphasize its essential features in the context of metastring theory.
The reason for the clarification of the cosmological constant problem in this context
involves a combination of the modular spacetime background nature of the metastring, the phase
space properties of the metastring, and the holographic bound on the number of degrees of freedom
of any quantum theory of gravity, including the metastring.

{\em First:} The metastring lives in modular spacetime (as explained in \SS\;\ref{s:MP+MS}). In that context one builds a spacetime
background out of modular spacetime ``blocks'' which are of the Planck size $l_P$. 
(In general, the fundamental scale is the string scale $\lambda$, that we take to be of the order of the Planck scale $l_P$.)
This procedure, named extensification, should be contrasted to the usual compactification procedure.

Now we imagine that one has 4 spacetime dimensions extensified and all the other dimensions of the order of the Planck scale (or string scale).
Similarly for the dual directions. In the energy momentum space we also extensify 4 energy momentum dimensions (conjugate to
the extensified spacetime dimensions). The dual momenta $\tilde{\Lambda}$ are also assumed to be of the Planck size.

One might ask: why 4 spacetime dimensions? The natural answer is that cosmologically, as the spacetime extensifies from the
Planck phase, the string (having a 2 dimensional worldsheet) becomes ``critical'' in 4 dimensions (``$2{+}2\,{=}\,4$'') as suggested
long ago by Brandenberger and Vafa in their paper on string gas cosmology\cite{Brandenberger:1988aj} (which also
agrees with our recent work\cite{Freidel:2021wpl}).

{\em Second:} Given this setup, we have the following very simple but profound formula (that is not true in effective field theory).
The phase space volume of the combined 4d extensified spacetime and 4d extensified energy-momentum space is given the number of degrees of freedom $N$.
Let $l$ denote the length of the extensified spacetime dimension and $\Lambda$ the length of the extensified energy momentum space dimension.
Then (without paying attention to the numerical factor, which we denote by $c_4$, in 4 spacetime dimensions, in order to make the main point as clearly as possible) we have that
\be
\int \rd^4 x \int \rd^4 p =c_4\, l^4 \Lambda^4 = N.
\ee
Essentially, this relation represents a UV/IR mixing, but of a very precise kind. {Note that this UV/IR mixing
is not possible in effective field theory.}
Given the fact that our theory (metastrings) contains gravity, we know that the cosmological constant as defined by
Einstein's equations (that also follow from the conformal invariance of the metastring)
is given in terms of the vacuum energy density $\rho$ (as discussed precisely in the previous section for the case
of the metastring, that is scaling
as $\Lambda^4$ in 4 spacetime dimensions) and the Newton constant $G_N=l_P^2$
\be
\lambda_{cc} = 8 \pi \rho\, l_P^2 = 8 \pi b_4\, \Lambda^4\, l_P^2.
\ee
which can be rewritten in the following suggestive way (with $b$ being the numerical factor from the previous section)
\be
\lambda_{cc} = 8 \pi \frac{b_4}{c_4}\, N \Big(\frac{l_P^2}{l^2}\Big) \frac{1}{l^2}.
\ee
Note that we have neglected numerical terms that can be extracted from the precise form of the
metastring one loop (toroidal) partition function defined in the previous section.

{\em Third:} Now, we have to remember that the metastring that exists in this extensified spacetime gravitates in such a way
that the semiclassical gravity conditions regarding the holographic scaling of entropy $S$ of de Sitter space (associated with
the cosmological horizon of 4 dimensional de Sitter space) or the number of degrees of freedom $N$, hold.
In other words, apart from the vacuum energy density, we can look at the holographic formula for entropy
(up to another numerical factor $d_4$, in 4 spacetime dimensions)
\be
N \define S = \frac{d_4\, l^2}{l_P^2} \to N \frac{l_P^2}{d_4\, l^2} = 1.
\ee
Here we are invoking the fact that Einstein-Hilbert action can be derived by the metastring and that in the
semiclassical limit, by the usual computation a la Gibbons-Hawking\cite{Gibbons:1993uz}, we get that the entropy is given by the
co-dimension one volume of spacetime --- that is area of the cosmological horizon for a space with a positive
cosmological constant. Note that the metastring, via the low energy action found in
context of a doubled formulation of the world sheet theory of the closed string, does imply a positive 
cosmological constant. This is discussed in\cite{Berglund:2020qcu} which builds
on our previous metastring papers and it can be seen more explicitly as follows: 
To lowest (zeroth) order of the expansion in the non-commutative parameter $\lambda$
the low energy effective action (in 4d) takes the form:
\be
S_{4d} = - \iint\! \sqrt{-g(x)} \sqrt{-\tilde{g}(\tx)}~ [R(x) + \tilde{R}(\tx)],
\label{e:TsSd}
\ee
a result which first was obtained almost three decades ago 
by assuming that $[\hat x^a,\htx_b]=0$\cite{Tseytlin:1990va}. 
In this leading limit, the $\tx$-integration in the first term defines the gravitational constant $G_N$, and in the second term produces a {\it positive} cosmological constant constant $\L>0$. 
In particular, we are lead to the following low energy effective action valid at long distances of the observed accelerated universe 
(focusing on the relevant $3{+}1$-dimensional 
spacetime $X$, of the ${+}\,{-}\,{-}\,{-}$ signature):
\be
S_{\text{eff}} =\frac{-1}{8 \pi G}\int_X \sqrt{-g}~
  \big( \lambda_{cc}+ {\textstyle\frac{1}2} R + O(R^2)
\big),
 \label{e:Seff}
\ee
with $\lambda_{cc}$ the positive cosmological constant (corresponding to the observed vacuum energy scale of $10^{-3}$\,eV) and
the $O(R^2)$ denote higher order corrections. The inclusion of the Gibbons-Hawking boundary term\cite{Gibbons:1993uz}, leads in the evaluation
of the Euclidean partition function defined by this effective action, whose logarithm gives the free energy and thus, after taking the
derivative of the free energy with respect to the euclidean temperature, the above entropy of the de Sitter cosmological horizon, or,
equivalently, our $N$.

Therefore we get our first prediction, that the cosmological constant in the context of metastrings
can be naturally related to a large (cosmological) scale $l$ of spacetime
\be
\lambda_{cc} = 8 \pi \Big(\frac{b_4 d_4}{c_4}\Big)\frac{1}{l^2},
\ee
where $b_4, c_4, d_4$ are purely numerical, $O(1)$ factors in 4 spacetime dimensions.
The most important observation here is that for large $l$ (the observed Hubble scale) we get the observed small
and positive
cosmological constant.
This result is also
radiatively stable and technically natural\cite{Freidel:2022ryr}, as it depends only on the IR (and not UV) scale.

Notice that the usual effective field theory argument (which is carried over in the canonical discussion of string theory,
as in our review of Polchinski's treatment of this problem) assumes that effectively $N\,{=}\,1$, and thus the
scale of the background spacetime and the energy momentum scale are linked which leads to a Planckian universe
with a huge cosmological constant and the cosmological constant problem.

It is the combination of the phase space nature of the metastring, together with the fact that the metastring lives in
extensified spacetime built out of Planckian (modular) ``blocks'', that via holography gives us naturally small
cosmological constant.
Note that we get that the associated size of the energy momentum space is given by a mixing of the cosmological size $l$
and the Planck size $l_P$ (up to numerical factors $d$ and $c$)
\be
\lambda_{cc} =  8 \pi \Big(\frac{b_4 d_4}{c_4}\Big) \frac{1}{l^2} \define 8 \pi b_4 \Lambda^4 l_P^2 \to \Lambda = \Big(\frac{d_4}{c_4}\Big)^{\!1/4} \frac{1}{\sqrt{l\,l_P}}.
\ee
This expression taken together with the Planckian size of the dual momenta $\tilde{\Lambda} \define l_P^{-1}$ gives us the
immediate scale $\mu_{\text{vac}}$ associated with the vacuum energy (and thus, related to the Higgs scale in the realistic
situation) which is much lower than the Planck scale, and it is order of $1$\,TeV
\be
\Lambda \tilde{\Lambda} = \mu_{\text{vac}} \to \mu_{\text{vac}}
 =\Big(\frac{d_4}{c_4}\Big)^{\!1/4} \frac{1}{\sqrt{l\,l_P}\,l_P}.
\ee
This is a stable scale by our construction and it offers a new perspective on the hierarchy problem.
We see that in the context of the metastring approach to quantum gravity, the cosmological constant problem
and the hierarchy problem are naturally related.
We will comment more on this relation at the end of this section.

We note that the whole discussion presented above for 4 spacetime dimensions can be formally extended to
any $D{+}1$ spacetime dimensions.
We just summarize the relevant formulae (by setting all numerical factors to be of order one, for simplicity)
\be
\begin{aligned}
   l^{D+1} \Lambda^{D+1} &\sim N,&\quad
  N \frac{l_P^{D-1}}{l^{D-1}} &\sim 1,&\quad
  \lambda_{cc} &\sim \frac{1}{l^2},\\*[-1mm]
  \Lambda^{D+1} &\sim \frac{1}{{l^2 l_P^{D-1}}},&\quad
  \tilde{\Lambda} &\sim l_P^{-1},&\quad
  \mu_{\text{vac}} &\sim \Lambda \tilde{\Lambda}.
\end{aligned}\vspace*{-1mm}
 \label{e:4}
\ee

Finally, one might ask ``why is $l$ large?'' This is equivalent to asking ``why is $N$ large?''
The reason for this is the same as Schr\"{o}dinger's answer to the question ``why are atoms small?'',
presented in his famous book ``What is Life?''\cite{rES}.
The answer is that if a large object is built of atoms then in order for that object to have a well defined size,
the number of atoms $n$ should be large, because the statistical fluctuation scales as $\frac{1}{\sqrt{n}}$.
Similarly, the number of modular spacetime ``blocks'' (spacetime atoms) $N$ should be large so that
the extensified spacetime should be stable in size, given the statistical fluctuation $\frac{1}{\sqrt{N}}$.
Note that numerically $N \sim 10^{124}$, which gives the following number of modular cells per each direction (say, $x$ direction, in 4 spacetime dimensions), $N_x \sim 10^{31}$. This number is not
that far off from the Avogadro number used in the
canonical atomistic reasoning.

Also, in this regard we could use the Bekenstein-Hawking formula for the entropy of the cosmological horizon associated
with de Sitter space (that was also an inspiration for\cite{Cohen:1998zx}) and the statistical
reasoning based on the Boltzmann-Einstein formula for entropy in order to argue that the small vacuum energy is entropically preferred
(as was argued by Horava and Minic in\cite{Horava:2000tb}), and the probability distribution $P \define \exp(S)$, where
$S$ is the above entropy of de Sitter space, given by the area of the cosmological horizon, is of the Baum-Hawking form\cite{Gibbons:1993uz},
$dP \sim \exp(c M^2_P/\lambda_{cc}) d\lambda_{cc}$, where $P$ is the probability,
$\lambda_{cc}$ is the cosmological constant in 4 spacetime dimensions and $M_P$ is the Planck energy scale in 4 spacetime dimensions. 
This gives us a large probability for small $\lambda_{cc}$ and thus, a large probability for large $l$ or large $N$.
Furthermore, when combined with the above seesaw formula, this gives us a statistically preferred
value for the $\mu_{\text{vac}}$ which is of the order given by the matter (or Higgs) sector, which, 
as we already emphasized, sheds light on the hierarchy problem
($\mu_{\text{vac}}$ being exponentially small compared to the Planck energy scale).

Finally we note the following:
The size of the extensified spacetime can be defined as 
$l = \frac{l}{l_P} l_P \define m l_P$ where $m$ denotes the number of Planckian ``units''.
Similarly, according to the above argument,
the size of energy-momentum space (in $3+1$ spacetime dimensions) $\Lambda = \frac{1}{\sqrt{l\, l_P}}$
can be defined as $\Lambda = \frac{n}{l}$ thus leading to large number of ``units'':
\be
m \define \frac{l}{l_P}, \quad n \define \Big(\frac{l}{l_P}\Big)^{\!1/2}
\ee
Note that while spacetime units are Planckian of size $l_P$, the energy momentum space is given in units of $\frac{1}{l}$,
where $l$ is the size of the entire spacetime. This, once again is not the usual reasoning based on effective field theory,
according to which, the Planckian size of a spacetime unit leads to the inverse Planckian size of the energy-momentum space unit.
Note that it is the ratio of the Hubble scale and the Planck scale that provides for the hierarchy needed
to resolve the cosmological constant problem
\cite{Freidel:2022ryr}.

Finally, a comment about the infrared finiteness of the metastring partition function.
Polchinski makes the point in his textbook that the vacuum energy of the bosonic string is 
UV finite (the essential feature of string theory) but IR divergent, because of the tachyon.
In particular, the partition function of the 26 dimensional bosonic string that Polchinski writes looks like
\be
2 \pi V_{26} \int \frac{\rd \tau^2}{\tau_2} (4 \pi^2 \alpha' \tau_2)^{-13}
              \sum_i\, e^{-\pi \alpha' m_i^2 \tau_2}
\ee
where the ground state is tachyonic with the tachyon mass squared being $- 24/l_s^2$ (and $\alpha' \define l_s^2$).
The tachyonic contribution diverges. That is why traditionally invokes supersymmetry (SuSy) and then
compactifications followed by SuSy breaking, upon which the cosmological constant problem returns as
it does in any effective field theory formulation.

However, the modular cells of the metastring can be considered as a fully compactified (or dimensionally
reduced to zero dimensions) bosonic string in 26 dimensions.
It is well known that the mass squared of the tachyon is proportional to $-(D{-}2)/l_s^2$, and thus
the tachyon is {\it massive and stable} in zero dimensions. Thus the partition function is
IR finite for each modular cell and any finite number of modular cells. The IR divergence 
appears only in the thermodynamic limit of the infinite extensification, that is, in the classical
spacetime limit.

\subsection{Comments on the Vacuum Energy and Holography}
\label{s:VE+h}
Here we want to rewrite our central formulas~\eqref{e:4},
 {\it{\bfseries a}})~$l^4 \Lambda^4 = N$,
 {\it{\bfseries b}})~$\rho_{\text{vac.}} \define \Lambda^4=N/l^4$,
 {\it{\bfseries c}})~$N= l^2/l_P^2$ and
 {\it{\bfseries d}\,})~$\lambda_{cc} \define \rho_{\text{vac.}} l_P^2 = 1/l^2$,
 in a more precise way.

In the limit of a classical spacetime extensification, we want to
write, instead of $l^4$ the precise formula for a (regularized) volume of spacetime
\be
l^4 \to \int_l \sqrt{-g}\,\rd^4 x
\ee
where $g$ is the (bulk) classical spacetime metric in 4d.
So, the quantization of the momentum space volume ($\Lambda^4$) should amount to
\be
\Lambda^4 \to \frac{N}{\int_l \sqrt{-g}\,\rd^4 x}
\ee
 
According to 
\cite{Freidel:2022ryr}
the first non-trivial feature of an extensified modular spacetime is that the (covariant) phase space volume is
quantized in the units of $N$ (the number of modular cell fluxes). The second non-trivial claim is that
$N$ is given by the holographic bound.
More precisely, in the limit of a classically extensified spacetime, we have
the Gibbons-Hawking term (needed to make the classical action additive), 
evaluated at a boundary of spacetime, given an induced metric $h$ on that boundary,
and the second fundamental form $K$ of the boundary
\be
\frac{1}{8 \pi G} \int \sqrt{-h}\,\rd^3 x~K
\ee
The Gibbons-Hawking term is related to the area of the boundary of spacetime
(assuming that the boundary could be chosen to be the (time) t-axis times a sphere of
radius $l$)
\be
 \int_l \sqrt{-h}\,\rd^3 x = 4 \pi l^2 \int\rd t~\big(1 + O(1/l) \big)
\ee
as follows
\be
\frac{1}{8 \pi G} \int\sqrt{-h}\,\rd^3 x~K
 = \frac{1}{8 \pi G} \frac{\partial}{\partial n} \int \sqrt{-h}\,\rd^3 x
\ee
where $ \frac{\partial}{\partial n} $ is the relevant normal derivative.

So, more precisely, our holographic ``quantization'' condition should involve
the regularized area of the boundary of the classically extensified spacetime
\be
4 G N =  4 \pi l^2 = \frac{ \int_l \sqrt{-h}\,\rd^3 x}{\int\rd t~\big(1{+}O(1/l)\big)}
\ee
Thus the exact formula for the cosmological constant reads as follows
\be
\lambda_{cc} = 8 \pi G  \frac{N}{\int_l \sqrt{-g}\,\rd^4 x}
\label{e:lcc}
\ee
or after using holographic ``quantization''
\be
\lambda_{cc} = \frac{8 \pi^2 l^2 }{\int_l \sqrt{-g}\,\rd^4 x} 
\ee
Thus, given the measured values for $\lambda_{cc}$, $G$ and the volume of the observed spacetime $V_4$, we have
\be
N = \frac{\lambda_{cc} V_4}{8 \pi G}
\ee
In particular,
\be
V_4 = a l^4
\ee
with $a$ being a very specific constant.
Then we have that (after using $N= \pi l^2/G$)
\be
N = \Big(\frac{8 \pi^2}{a}\Big) \frac{1}{\lambda_{cc} G}
\ee
It would be interesting to test these relations empirically. We will make a
few comments about this possibility at the end of this subsection.

Finally, let us rewrite the above formula~\eqref{e:lcc} for the cosmological constant as
\be
\lambda_{cc} \int_l \sqrt{-g}\,\rd^4 x = 8 \pi G N = 2 \pi N (4G)
\ee
which reminds us of the Bohr-Sommerfeld quantization, 
where $4G$ or $4 l_P^2$ is the effective ``Planck constant of quantum gravity''.
Now, we know that in the Einstein Hilbert action, there is precisely such a term that involves the cosmological constant
\be
\frac{1}{8 \pi G} \lambda_{cc} \int_l \sqrt{-g}\,\rd^4 x 
\ee
Thus, we have a 
quantization of this part of the classical action.
Note that this makes sense given the fact that $\lambda_{cc}$ can be interpreted as ``energy''
(the Wheeler-DeWitt equation with the cosmological constant reads $H \Psi = \lambda_{cc} \Psi$). 
Note that the conjugate ``time'' variable is the spacetime volume $\int_l \sqrt{-g} d^4 x$.
However, if one term in the action is quantized then all terms presumably follow the same quantization.
Hence, the curvature term should be quantized as well
\be
\frac{1}{2} \int_l R  \sqrt{-g}\,\rd^4 x = 2 \pi N (4G)
\ee
where the $1/2$ terms takes into account the usual normalization of $1/16 \pi G)$ in front of the curvature term.
Then the boundary Gibbons Hawking term, which makes the Einstein Hilbert action additive,
should be quantized as well. Thus
\be
 \int_l K  \sqrt{-h}\, d^3 x = 2 \pi N (4G)
\ee
Therefore, if we take the boundary of spacetime to be a sphere times the time axis, the
area of the boundary should be quantized as well, and if we took the Euclidean time to be simply $2 \pi$, this leads us to the holographic relation
\be
 \mathrm{Area}(l) \define 4 \pi l^2 = 4 G N
\ee
which is precisely what we have already used in our argument.
So, from this viewpoint, the holographic argument is not needed as an independent input --- it is already contained
in the quantization condition of phase space, that follows from modular polarization of quantum theory and
the process of extensification.
Notice that the quantization condition involves $4G$ as the effective ``Planck constant of quantum gravity'',
which also makes sense, given the appearance of precisely this factor in the Bekenstein-Hawking area law\cite{Gibbons:1993uz}.
Also the factor of $2 \pi$ makes sense as the length of the Euclidean time.
(Note that the above formula is related to the well known entropy formula for black holes, provided
$N$ is replaced by $S$ --- the black hole entropy. It is interesting to note that in Matrix theory $S \sim N$\cite{Banks:1996vh},
where $N$ is the number of partons ($D0$-branes) that make a black-hole bound state.)

In some sense, the above new quantization procedure (invisible in effective field theory (EFT), as discussed in the next subsection) 
that leads to a small positive cosmological constant
and thus ensures that a big spacetime is stable, is analogous to what 
Bohr-Sommerfeld quantization does for the stability of the Rutherford-Bohr atom.
The Bohr-Sommerfeld quantization follows from fundamental commutators, and thus 
we might expect that the vacuum energy, the volume of spacetime and curvature are turned into
operatorial quantities with appropriate (Heisenberg-like) commutators that lead to the above
quantization formulae. 
It would be 
interesting 
to understand if
that follows from modular quantization.

Finally, the quantization of area has been postulated by Bekenstein and Mukhanov
for black hole horizons in their 1995 paper\cite{Bekenstein:1995ju}.
Proposals to
detect this quantization of the black hole horizon area via ``echoes'' in gravitational waves
using LIGO-Virgo Collaboration
have been discussed
in the literature
(see, for example, \cite{Foit:2016uxn} and more recently, \cite{Coates:2021dlg}). 
In our context, the discovery of such ``echoes'' could be interpreted as empirical evidence for modular spacetime.
Here we want to mention the recent critical discussion of echoes of gravitational waves as presented in\cite{Guo:2022umn}, in which the original literature on this subject is listed (such as\cite{Wang:2019rcf}). The conclusion of
this critical essay is: ``Thus if echoes are actually
detected, then we would face a profound change in our understanding of physics.''
In our case, the new quantum number $N$ is very large
($N \sim 10^{124}$), and thus we are really discussing the quantum chaos phase of the quantum black hole physics.
Thus the gravitational wave signals could be expected to display (apart from echoes) certain robust signatures of quantum chaos, such as
scars --- that is, a well known phenomenon where the eigenstates of a classically chaotic quantum system have enhanced 
probability density around the paths of unstable classical periodic orbits. This could be a distinguishing (``smoking gun'')
signature in our context.

\subsection {Double Scaling Limits and EFT}
\label{s:DS+EFT}
Here we discuss how the extensification procedure in modular space reproduces in a 
limit of infinite number of fluxes $N$ and the infinite extension of spacetime $l$ the results
of effective field theory (EFT).

Let us repeat the crucial formulae~\eqref{e:4} of the argument presented in 
\cite{Freidel:2022ryr}:
 {\it{\bfseries a}})~$l^4 \Lambda^4 = N$,
 {\it{\bfseries b}})~$\rho_{\text{vac.}} \define \Lambda^4=N/l^4$,
 {\it{\bfseries c}})~$N= l^2/l_P^2$ and
 {\it{\bfseries d}\,})~$\lambda_{cc} \define \rho_{\text{vac.}} l_P^2 = 1/l^2$.
Now, let us define the precise $N \to \infty$ and $l \to \infty$ limit in which  EFT results are reproduced, so 
that our procedure of separating the IR scale $l$ from the $UV$ scale $l_P$ and the number of modular cells $N$ could be understood
as a precise regularization of the usual (wrong) EFT result.

First, let us introduce the scale $l_{\Lambda}$ which is defined in the following double scaling limit
\be
N \to \infty, \quad l \to \infty, \quad \frac{N}{l^4} \define 1/l_{\Lambda}^4 = \textit{finite}.
\ee
Then let us also take $l \to \infty$ and $l_P \to 0$ so that the ``holographic ratio'' that defines N is infinite
\be
l \to \infty, \quad l_P \to 0, \quad  l^2/l_P^2 \define N \to \infty
\ee
Then from these two equations we have that our finite scale $l_{\Lambda}$ is given as
\be
N \define l^4/l_{\Lambda}^4 \define  l^2/l_P^2, \quad N \to \infty, \quad l \to \infty,  \quad l_P \to 0, \quad l_{\Lambda} \to \textit{finite},
\ee
where the finiteness of $l_{\Lambda}$ is assured by the following double scaling limit that follows from the above two
definitions of $N$ (the phase space/extensification definition and the holographic definition)
\be
 l \to \infty, \quad l_P \to 0, \quad l_{\Lambda} \define \sqrt{l\,l_P} = \textit{finite}.
\ee
Note that for {\it finite but small (as compared to some other characteristic length)} $l_{\Lambda}$ we recover the usual EFT result for the vacuum energy (which is very large
and tied to the energy-momentum cutoff)
\be
\rho_{\text{vac}} \define \Lambda^4 =1/l_{\Lambda}^4
\ee
However, because we have arranged that $l_P \to 0$ (the non-gravitating limit) the cosmological constant vanishes
($ \lambda_{cc} \define \rho_{\text{vac}} l_P^2 = l_P^2/l_{\Lambda}^4$) by definition.
(As already mentioned, the use of cutoffs in the computation of the cosmological
constant in the context of effective field theory has been recently
carefully discussed by Donoghue in\cite{Donoghue:2020hoh}.)

Our construction is recovered in the limit of finite $N$, $l$ and $l_P$, in which case we recover our original formulae~\eqref{e:4}, {\it\bfseries a}, {\it\bfseries b}, {\it\bfseries c}, and {\it\bfseries d}.
Obviously the EFT is completely ignorant of $N$, which is crucial in the argument for the clarification of the cosmological constant 
problem as presented in the recent work by Freidel, Kowalski-Glikman, Leigh and Minic\cite{Freidel:2022ryr}.

\subsection{Comments on the Effective Field Theory Breakdown}
\label{s:EFTbreaks}
As we have mentioned the usual effective field theory argument assumes that effectively $\Lambda \sim 1/l_P$ (even though
the EFT limit can be understood as a double scaling limit, as discussed in the previous subsection), and thus the
scale of the background spacetime and the energy momentum scale are linked which leads to a Planckian universe
with a huge cosmological constant and the cosmological constant problem. Many authors have
recently attempted the discussion of the breakdown of effective field theory in the context of quantum gravity;
see, for example, \cite{Cohen:1998zx, Dvali:2010jz, Banks:2019oiz, Brandenberger:2021zib, Brandenberger:2021kjo, Castellano:2021mmx, Berglund:2022qcc}. Nevertheless, we consider the separation of various scales
$\Lambda$, $\tilde{\Lambda}$ and $\mu$ as a unique feature of metastring theory. This, as far as we know, has not been pointed
out before in the literature.
(Here we recall the suggestion of Veltman, mentioned in his 1999 Nobel prize lecture, that one should work in phase space in order to 
address the cosmological constant problem --- and that is 
a crucial feature of the argument presented in 
\cite{Freidel:2022ryr}, as summarized in this section.)

In particular, we can connect our work to Cohen, Kaplan and Nelson\cite{Cohen:1998zx}
because these authors discuss a constraint on the validity of effective field theory in the limit of a large volume.
(That would correspond to a very small, Planckian, volume of the dual space, which does not appear in their discussion.)
In their well known work, Cohen, Kaplan and Nelson\cite{Cohen:1998zx} say that ``the length $L$, which acts as an IR cutoff,
cannot be chosen independently of the UV cutoff, and scales as $\Lambda^{-3}$,'' where $\Lambda$ is the UV cutoff.
Note that we have established that there is a kind of seesaw relation between UV in ``our'' space and IR in the dual one, 
and so the natural question is: how can this provide us with any relation between UV and IR in ``our'' spacetime?
We claim that this is a matter of what kind of spacetime interpretation we adopt. For example, 
if we think of our $\Lambda$ as the spacetime IR 
scale (whose fourth power gives the observed vacuum energy density, or the cosmological constant) and the $\tilde{\Lambda}$
as the spacetime UV (Planck) scale, then they are not independent: Their product is given by $\mu_{\text{vac}}$, which is naturally given by the Higgs scale --- since the minimum of the Higgs potential {\em\/defines\/} the vacuum. This interpretation agrees with\cite{Cohen:1998zx}.

Here we also note, as an aside, that the above seesaw relation is characteristic of the relation between the quantum (Compton) scale and
the gravitational (Schwarzschild) scale for a given mass $M$.
The Compton scale is $L_Q = \hbar/(M c)$ and the Schwarzschild scale in $3{+}1$ dimensions is $L_{GR} = 2M G_N/c^2$
and thus their product is $ L_Q L_{GR} = 2 \hbar G_N/c^3$, which is essentially the square of the Planck scale $L_Q L_{GR} \sim L_P^2$.
It is interesting that in the above ``seesaw`` formula for the cosmological constant\cite{Berglund:2020qcu} the product of the IR and the UV scale is
given by an intermediate (Higgs) scale that determines the observed vacuum of the matter degrees of freedom, which is much lower
than the Planck scale, indicating a possible relation between the hierarchy problem and the cosmological constant problem\cite{Berglund:2020qcu}, to which we will return below.

To summarize: in this section we have discussed  the vacuum energy problem in the context of the 
metastring approach to quantum gravity (as well as its metaparticle limit). 
Our central observation is that the solution of 
the vacuum energy problem is naturally offered by
an effective, chirally doubled and non-commutative structure of quantum spacetime, the habitat for metastrings and 
metaparticles. A generically small cosmological constant
is realized in string theory via the interplay between the phase space structure of the metastring and
the holographic principle of quantum gravity. The same set-up implies a generic seesaw formula
between the vacuum energy scale and the Planck scale and it leads to the natural resolution of the
hierarchy problem in string theory. Our discussion is intrinsically stringy and we have pointed out quite explicitly where the usual reasoning based 
on effective field theory breaks down, as seen in our more general approach to quantum gravity based on metastring theory and modular spacetime.
Finally, we point out that our set-up has a predictive power when applied to cosmology, and in particular the
equation of state for the dark energy, as already emphasized in\cite{Jejjala:2022eid}.

Coming back to the problem of de Sitter space in string theory, this new understanding of
the string vacuum and its vacuum energy, as well as its implication for the cosmological
constant problem, sheds light on the meaning of de Sitter entropy as counting the degeneracy of
the quantum vacuum, defined in the modular (quantum) polarization of quantum theory,
together with the process of extensification, that results in a classical spacetime background.

Let us close with a remark about the coincidence problem: Why is the universe in the 
accelerating phase now, and why is dark energy dominant and the dark matter is of the order of the  visible matter? Here the fact that dark and visible matter could be correlated in the metaparticle 
formulation (as explained in the next section) does address the second observed feature regarding the similar abundances of
the invisible (dark) and visible matter degrees of freedom. Similarly, the necessary input that 
holography (and IR condition) provided in the choice of modular vacuum which is determined by the
cosmological constant, does shed light on the fist observed feature --- that we are in the
cosmological phase dominated by the cosmological constant. One would be tempted to say that
the interplay of the IR and UV physics
makes our discussion only ``weakly anthropic'' when it comes to the coincidence problem. 
However we think that this is not a correct interpretation and that anthropics (even of the weak kind)
should be replaced by quantum contextuality, which appears as a fundamental feature of the modular polarization
in quantum theory.

\subsection{Metastring, the Cosmological Constant and the Higgs Mass}
\label{s:MS,CC,Higgs}
In this final subsection we address the following issue:
So far, our argument for the vacuum energy was based on the properties of 
modular spacetime (including the concept of extensification) and holography.
The metastring really enters as an example of a theory of quantum gravity consistent with
modular spacetime, but not in any other fundamental way.
Can string theory enter in a fundamental way and perhaps illuminate the
other hierarchy problem related to the Higgs sector of the Standard Model?
Does the new insight into the vacuum of string theory illuminates the question
of the Higgs mass (and perhaps masses of other elementary particles)?
We will make some preliminary observations about these fascinating questions in what follows.

The metastring is endowed with two Heisenberg algebras\cite{Freidel:2013zga, Freidel:2014qna, Freidel:2015pka, Freidel:2015uug, Freidel:2016pls, Freidel:2017xsi, Freidel:2017wst, Freidel:2017nhg, Freidel:2018apz}.
Apart from the non-zero ``phase space'' commutator familiar from quantum mechanics 
$[x, p] = i$ (with $\hbar =1$) that we have used in the above, there is the same type of commutator 
for the dual ``phase space'' (and that one is taken care of by assuming the Planckian size for
the dual spacetime and the Planckian energy scale for the dual momentum space).
However, there is also the uniquely metastring commutator between the
spacetime and dual spacetime coordinates $[x, \tx]= i l_s^2$, which involves
the string length $l_s$. In the following we insist on the compatibility of these two
Heisenberg algebras.

Let us apply the same type of reasoning 
presented in 
\cite{Freidel:2022ryr}
by combining
the modular space associated
with this second Heisenberg algebra (together with the concept of extensification) and holography,
and let us find out the consequences of that line of reasoning.

First we have that (in $4$ extensified spacetime dimensions) with $N$ fluxes in the modular
space of $x, \tx$, where the extensified length of $x$ is now $l_{\Lambda}$, the length scale
associated with the vacuum energy (already determined in our previous discussion to be the geometric mean $l_{\Lambda} = \sqrt{l\, l_P}$
of the observed size of spacetime $l \sim 10^{27}$m and the Planck length $l_P \sim 10^{-35}$m, giving $l_{\Lambda} \sim 10^{-4}m$), 
and the length of $\tx$ is $\tilde{l}$
\be
l_{\Lambda}\,\tilde{l} = l_s^2 N_L^{~1/4}
 = l_s^2 \Big(\frac{l_{\Lambda}^2}{l_P^2}\Big)^{\!1/4} = l_s^2  \Big(\frac{l_{\Lambda}}{l_P}\Big)^{\!1/2}
\ee
where as before the number of fluxes is given by the holographic bound in the dual space $N_L =  \frac{l_{\Lambda}^2}{l_P^2}$.
Note that here we have crucially used the vacuum energy length scale $l_{\Lambda}$, because we are concerned with the
properties of the vacuum of string theory.
What we mean is that the natural ``holographic'' bound in the dual space is set by the size of 
the energy momentum space that reproduces the observed vacuum energy. Thus $N_L \sim l_{\Lambda}^2/l_P^2$.

Note that the string length $l_s$ and the Planck length $l_P$ (related to $G$) are related via the string coupling $g_s$
\be
g_s l_s = l_P, \quad M_s = g_s M_P
\ee
where $M_P$ is the Planck energy scale and $M_s$ the string energy scale.
Taking this conversion between $l_P$ that features in the holographic bound (and thus, the number of fluxes, $N_L$)
and the string length $l_s$ that features in the second Heisenberg algebra of the metastring, we get
\be
\tilde {l} = \frac{l_P}{g_s^2} \Big(\frac{l_P}{l_{\Lambda}}\Big)^{\!1/2}
\ee

Now, let us demand that the extend of the dual spacetime is essentially $l_P$, and so $\tilde{l} \define l_P$.
(This was an implicit demand in our reasoning about the vacuum energy, but it turned
out that it did not enter that discussion. However, this demand is consistent with the
assumption that the size of the ``phase space'' defined by the dual spacetime and dual momenta is Planckian.
Thus, $\tilde{l} = l_P$  is a consistent assumption.)

Then the above formula gives us the value of the string coupling $g_s$ in terms of the 
size of the vacuum energy length scale $l_{\Lambda}$ and the Planck length $l_P$:
\be
g_s = \Big(\frac{l_P}{l_{\Lambda}}\Big)^{\!1/4} \define \Big(\frac{M_\L}{M_P}\Big)^{\!1/4}
      \to g_s^2 = \Big(\frac{M_\L}{M_P}\Big)^{\!1/2}
\ee
where $M_\L$ is the energy scale associated with the vacuum energy.
Note that this is a very small value, which would justify our one-loop computation of
the metastring partition function (the two and higher loops would be suppressed by this small
string coupling). Also, this would suggest a much lower scale for metastrings which would
get close to the scales of accelerator-based particle physics.

Naturally this leads us to the gauge hierarchy problem and the mass of the observed Higgs.
We will use the results regarding the stringy computation of
the Higgs mass presented in Ref.\cite{Abel:2021tyt}.
In that paper Abel and Dienes  claim to ``establish a fully string-theoretic framework for calculating one-loop Higgs masses directly from first principles in 
perturbative closed string theories. This framework makes no assumptions other than worldsheet modular invariance and is therefore applicable 
to all closed strings, regardless of the specific string construction utilized. This framework can also be employed even when spacetime supersymmetry 
is broken (and even when this breaking occurs at the Planck scale), and can be utilized for all scalar Higgs fields, regardless of the particular gauge 
symmetries they break. This therefore includes the Higgs field responsible for electroweak symmetry breaking in the Standard Model.'' Notably, using this
framework, Abel and Dienes demonstrate ``that a gravitational modular anomaly generically relates the Higgs mass to the one-loop cosmological constant,
thereby yielding a string-theoretic connection between the two fundamental quantities which are known to suffer from hierarchy problems in the absence of spacetime supersymmetry''\cite{Abel:2021tyt}.

Thus, let us look at the formula of Abel and Dienes\cite[Eq.\;(2.57)]{Abel:2021tyt} that relates
the Higgs mass to the cosmological constant
\be
m_H^2 = \frac{\xi \Lambda^4}{M_P^2} - \frac{g_s^2 M_s^2}{8\pi^2} \langle\cX\rangle,
\label{e:AbelDienes}
\ee
where $\xi$ being defined in\cite[Eq.\;(2.48)]{Abel:2021tyt} as a particular $T$-matrix trace, and $\cX$ is a suitably normalized insertion in the second moment of the partition function\cite[Eq.\;(2.33)]{Abel:2021tyt}.
Let us examine the first term. By looking at it
one might say immediately that it does not fit the observed 
Higgs mass $m_H \sim 125$\,GeV) if one plugs in the new stringy result 
for the vacuum energy (cosmological constant)
reviewed in the previous subsection. Thus this term seems practically irrelevant.
The second term, after we take the square root and take into account the absolute value of this term, gives
\be
m_H \sim g_s M_s \sqrt{\frac{\langle\cX\rangle}{8 \pi^2} }
 = g_s^2 M_P \sqrt{\frac{\langle\cX\rangle}{8 \pi^2} }
\ee
Now let us use the derived consequence of the second Heisenberg algebra of the metastring (the fundamental
non-commutativity of the spacetime and dual spacetime coordinates) and holography in the observed 4 dimensional
spacetime, namely that the string coupling is $ g_s^2 = (\frac{M_\L}{M_P})^{1/2}$.

This implies the following seesaw formula for the Higgs mass in string theory (in which the cosmological
constant is also small and positive, as in our previous argument regarding the vacuum energy in metastring theory)
\be
m_H \sim \sqrt{M_\L M_P }  \sqrt{\frac{\langle\cX\rangle}{8 \pi^2} }
\ee
where $M_\L$ is energy scale associated with the vacuum energy ($10^{-3}$\,eV) and
$M_P$ is the Planck energy ($10^{19}$\,GeV).
The geometric mean of these two energy scales gives us a TeV, and thus the observed Higgs mass
can be obtained if $\sqrt{\frac{\langle\cX\rangle}{8 \pi^2} } \sim 10^{-2}$, of if $\langle\cX\rangle \sim 1$ --- consistent with the defining equations\cite[Eq.\;(2.27) and\;(2.33)]{Abel:2021tyt}.

Note that the above argument relies on a subtle combination of
holography in $x$-space with the second
Heisenberg algebra between $x$ and
$\tx$. In this argument $\tx$ is still of
the Planck size as we assumed in the argument
regarding the cosmological constant. The factor of
$N_L$ really comes from holography in $x$-space 
but taking into account the effective length scale associated
with vacuum energy $l_{\Lambda}$ 
and the power $N_L^{~1/4}$ is consistent with
what we said about phase space in our argument
regarding the cosmological constant.
(One may say that the UV/IR is manifested
though the relation between the dual
space and momentum space via the 
string length. So the two Heisenberg 
algebras, one between $x$ and $p$ and the
other between $x$ and $\tx$ are mutually consistent.)
Also the relation between holography 
in $x$-space and the non-commutativity 
between $x$ and $\tx$ leads to a unique 
relation between the Planck and string
scales as well as the vacuum energy scale.
(And, as consequence of that, we get a very 
specific statement about the Higgs mass.)

Finally, the string scale turns out to correspond 
roughly to the geometric mean of the Higgs and 
the Planck scales, which numerically corresponds 
to the GZK scale found in high energy cosmic rays. 
(The QCD axions also have a very similar scale, 
according to the most recent bounds.)
This can be seen by plugging $g_s = (M_{\Lambda}/M_P)^{1/4}$ into 
$M_s = g_s M_P$ and using the above formula for
the Higgs mass $m_H \sim \sqrt{M_{\Lambda} M_P }  \sqrt{\frac{\langle\cX\rangle}{8 \pi^2} }$
in order to obtain (approximately)
\be
M_s \sim \sqrt{m_H M_P}
\ee
which does yield to a pretty low scale of around $10^{11}$\,GeV.

The above argument intuitively makes sense because, in string theory, the
one-loop expression for the vacuum energy is related to the one-loop expectation value for the
dilaton, which is in turn related to the string coupling, and this in turn sets the relation
between the string and Planck scales. However, in order to understand that precisely, we
need to properly implement modular quantization of the metastring.
Also, we note that the above seesaw formula is consistent with the 
seesaw-like form of the diffeomorphism constraint in metastring theory.
We will comment on this seesaw relation (a feature of our toy model as well) in Section~\ref{s:Lessons}.

This argument taken together
with the above formula that relates the Higgs mass and the
vacuum energy in string theory (evaluated in the modular spacetime/metastring formulation
of string theory which combines the phase space nature with the holographic
nature of the string vacuum) does lead to a new, and completely stringy view 
of the hierarchy problem. Note finally, that the whole set-up of Abel and Dienes is
perfectly suitable for the modular spacetime formulation of string theory (the metastring).
It would be interesting to generalize the formalism of Abel and Dienes for the case of fermion masses
and check whether ``seesaw-like'' formulae appear as well;
see\cite{Bordes:2014eaa,rBJ-MM}) as well as\cite{Faraggi:1996pa}.

\section{Defining dS Spacetime via Holography} 
\label{s:Define}
Now we review de Sitter holography (and how it contrasts to the AdS/CFT correspondence) 
in the light of the preceding section devoted to the problem of vacuum energy.
We start our discussion with a review of
of the original proposal for the dS/CFT correspondence\cite{Hull:1998vg, Strominger:2001pn, Balasubramanian:2001rb, Witten:2001kn, Balasubramanian:2001nb, Balasubramanian:2002zh},
and we also note various arguments against dS/CFT\cite{Dyson:2002nt}.
We also note papers on the explicit realization of dS/CFT in the context of high spin theories/Vasiliev's gravity,
which can be understood as a limit of the tensionless string\cite{Anninos:2011ui}.
We will summarize the main issues here and
review the proposal\cite{Balasubramanian:2001rb, Balasubramanian:2001nb, Balasubramanian:2002zh} that
quantum gravity/string theory in large dS spaces is defined as two entangled and non-canonical CFTs that live on
the two ``boundaries'': the infinite past $I^{-}$ and infinite future $I^{+}$.
In particular, we relate this proposal to our discussion from Section~\ref{s:VEST} of the vacuum energy and the cosmological constant problem
in the context of the metastring formulation of string theory in modular spacetime.

\subsection{AdS Holography vs.\ dS Holography}

Another approach 
to understanding de~Sitter space in string theory is to start with de~Sitter spacetime (as implied by the cosmological observations) and try to define it in terms of a dual, holographic description, that is in principle
consistent with a broad definition of the space of string theory in terms of the abstract space of D-dimensional conformal field theories and the renormalization group flows on this space (extending the old view of 
string theory in terms of the abstract space of 2d conformal field theories and the corresponding 2d RG flows).
This approach, dubbed as 
dS/CFT duality\cite{Hull:1998vg, Strominger:2001pn, Balasubramanian:2001rb, Witten:2001kn, Balasubramanian:2001nb, Balasubramanian:2002zh} 
is obviously inspired by its better known progenitor, the AdS/CFT\cite{Maldacena:1997re, Gubser:1998bc, Witten:1998qj} duality.
Note that the emphasis here is on the CFT side of duality, so the fact that AdS is realized in the geometric (spacetime) formulation of string theory is not central.

One starts from the Lorentzian $\SO(D{+}1,1)$ isometries of the Euclidean $\AdS_{D+1}$
\be
\rd s^2 = \rd r^2 + e^{A\,r} \rd s^2_{\text{flat}}, \quad (\AdS)
\ee
and Lorentzian $\dS_{D+1}$ spaces (see Section~\ref{s:Prelim})
\be
\rd s^2 = - \rd t^2 + e^{H\,t} \rd s^2_{\text{flat}}, \quad (\dS)
\ee
(written in the corresponding flat patches, with $A \sim \sqrt{|\Lambda_{\AdS}|}$ and $H \sim \sqrt{\Lambda_{\dS}}$).
That is, both spaces, Lorentzian de~Sitter (\dS) ($\Lambda >0$) and Euclidean Anti-de-Sitter (\AdS) ($\Lambda < 0$) 
are hyperboloids
\be
-x_0^2 +x_1^2 +\dots+x_{D+1}^2 = l^2, \quad |\Lambda| = \frac{D(D{-}1)}{2l^2},
\ee
embedded in flat space with signature $(1, D{+}1)$.

Given this geometric fact, there exists a non-local map from the Lorentzian dS space to the corresponding Euclidean AdS space, which respects the corresponding bulk isometries and
thus the boundary conformal structures.
In particular\cite{Balasubramanian:2002zh},
\be
\psi(Y) = \int \rd X~ K(Y,X)\, \phi(X),
\ee
where $\rd X$ denotes the invariant measure of dS space, and the kernel $K(X,Y)$ commutes
with the $\SO(D{+}1,1)$ actions.
Under this map, the two boundaries of dS are mapped onto the single boundary of Euclidean AdS.
However, this does not imply that the dual of dS is a single conformal field theory, as is the case of Euclidean AdS.
Due to a non-local nature of the map, if one knows a field in the Euclidean AdS, one cannot reconstruct the integrals
of the corresponding field on dS along null geodesics. This is the reason for the claim that the dual of dS involves
two entangled conformal field theories\cite{Balasubramanian:2001rb, Balasubramanian:2001nb, Balasubramanian:2002zh}.

More precisely, as discussed in\cite{Balasubramanian:2002zh},
\be
 K(gX, gY) = K(X,Y)
\ee
where $ g \in \SO(D{+}1,1)$. Using suitable $g$, we can always achieve that $X_i = 0$ for $i < d$, and $X_{D+1} = 1$. 
Call such a  point $E$. Then it is sufficient to know $K(E, Y )$, because by acting with the group
we can recover the rest of $K$ from this. The point E is preserved by a $\SO(D, 1)$ subgroup.
This can be used to write $Y$ as
\be
Y (\xi) \define (Y_0, \cdots\!, Y_{D+1}) = (\sqrt{1 + \xi^2}, 0, \cdots\!, 0, \xi).
\ee
Therefore, the kernel is completely determined by a function of a single variable,
\be
K(E, Y(\xi)) \define K(\xi)
\ee
We can write the kernel explicitly, once we are given $K(\xi)$. 
To describe that result, we start with the description of Lorentzian de Sitter spacetime as a hyperboloid immersed in the flat Minkowski space. de Sitter spacetime is described by
\be
P(X,X) = 1
\ee
where we have set the cosmological constant to one and where
\be
P(X, Y ) = -X_0Y_0 + X_1Y_1 + \cdots + X_{D+1}Y_{D+1}
\ee
Note that $P(gX, gY ) = P(X, Y )$, and also $P(E, Y (\xi)) = \xi$. Thus one can show that\cite{Balasubramanian:2002zh}
\be
K(X,Y) = \int \rd \xi~ \delta\big(P(X,Y) - \xi \big)\, K(\xi)
\ee
satisfies all the required properties. In particular, if $K(\xi) = \delta(\xi)$ we have the Radon transform
known from the studies of tomography\cite{Gelfand:1958aa}.

In particular, as discussed in\cite{Balasubramanian:2002zh}, given a point $X$ of Lorentzian 3 dimensional de Sitter spacetime 
defined by $P(X, X) =1$, and $U$ a point on conformal boundary of de Sitter spacetime specified by $P(U,U)=0$, then
the explicit non-local map to Euclidean 3 dimensional Anti-de Sitter spacetime (and a point $Y$ on it) is given by 
\be
\psi(Y, U) = \int \rd X~ \delta\big(P(X,Y)\big)\, |P(X,U)|^{-1-i\rho/2}
\ee
where $i \rho$ is related to the conformal dimension of the dual operator in the AdS/CFT correspondence.

Given this non-local transform from the Euclidean AdS space to the Lorentzian dS space, one is tempted to
define dS/CFT correspondence in analogy with the well known AdS/CFT case.
The problem is that these two spaces are radically different and thus one has to be extremely careful in
such a holographic definition of dS spacetime in string theory.

In the context of the AdS/CFT dictionary we have a well defined background in the supergravity limit.
This is precisely the issue for the case of dS spacetime, as reviewed in Section~\ref{s:Find}, even though we will argue
that there exists a realization of dS spacetime in string theory if one does not follow the traditional route, as discussed in the previous section and in what follows.
However, even without a supergravity solution, one should expect to be able to
construct, or {\em\/define\/} a dS spacetime in string theory via the holographic dictionary.

As discussed in\cite{Balasubramanian:2002zh}, the issue is that one CFT will not suffice.
Even if one starts on the well defined AdS side and applies the above non-local map in order to get to the dS side, there is an issue with the causal structure which becomes degenerate under such a non-local map. In order to properly take into account the causal structure of dS (and, in particular, the presence of the cosmological horizon and the related dS entropy --- the reasons for the ``cosmological breaking of supersymmetry''), one should consider two entangled CFTs and thus realize the finite entropy of dS space
as an entanglement entropy. We will sharpen this message in what follows.

\subsection{de Sitter Holography and Entangled CFTs}

The main message of\cite{Balasubramanian:2001rb, Balasubramanian:2001nb, Balasubramanian:2002zh} is
that de~Sitter holography involves non-canonical CFTs involving non-canonical conjugations of the generators of conformal transformations, and in particular, the Virasoro generator conjugation is $L_n^{\dagger} = - L_{-n}$. 
Also, the holographic dual of dS (in global coordinates) requires two entangled non-canonical CFTs so that the thermal properties of the vacuum de~Sitter spacetime are recovered\cite{Arias:2019pzy,Arenas-Henriquez:2022pyh}.
Finally, the holographic dictionary that relates the generators of correlations functions in two entangled
theories to the bulk action in the Lorentzian space, requires a non-canonical Hilbert space structure of
the dual Hilbert space. This is the crucial difference in understanding of holographic descriptions in the
AdS and dS cases, and points to the crucial issues in realizing dS space in string theory.

\noindent
{\it A very useful toy example:} The above picture does have a very nice toy example.
Consider $\AdS_3$ and $\dS_3$ from the viewpoint of the 3d Chern-Simons (CS) theory\cite{Witten:1989ip}. 
The main point here is that the two solutions $\AdS_3$ and $\dS_3$, come equipped with a canonical and non-canonical inner product of the
corresponding Hilbert spaces. Such non-canonical Hilbert space structures represent the counterpart of the non-canonical conformal field theory
involved in the holographic description of de Siter space.

The reason for the unusual inner product follows from the relation between CS theory in the
3d bulk and the WZW theory on the boundary.
The CS action relevant for the positive $\Lambda$ case
\be
S = c [I_{CS} (A) - I_{CS} (\bar{A})],
\ee
where, $c$ is a constant (that includes the complex WZW level), and $I_{CS}$ is the usual CS action
\be
I_{CS} (A) = \int \rd^3 x~ \Tr\Big(A \wedge dA + \frac{2}{3} A^3\Big).
\ee
Here the gauge field $A$ takes values in $\SL(2,\IC)$, $\bar{A}$ is the complex conjugate of A, 
and the Vielbein and the spin connections are given as
\be
2 i e = A - \bar{A}, \quad 2 \omega = A + \bar{A}.
\ee
The main point here is that, in the negative $\Lambda$ case (AdS) one has independent $\SL(2,\IR)$ theories, and in the 
positive $\Lambda$ case, one has a different hermiticity condition that involves an exchange of left and right movers in
the WZW theory. This is precisely what one finds by looking at a free scalar in dS space\cite{Balasubramanian:2001rb, Balasubramanian:2001nb, Balasubramanian:2002zh}.

We remark that the question of observables in the context of dS holography is still debated\cite{Chandrasekaran:2022cip}. However,  the
above definition of de~Sitter holography can be connected to metastring theory, modular spacetime and modular polarization
in quantum theory, and thus the questions of observables should be framed in that context. We will comment on that
question in the next subsection.
 The main point of the holographic approach to de~Sitter spacetime in string theory
is that the relevant dual boundary CFT (holographically dual to bulk de~Sitter space)
has a non-canonical structure
\cite{Balasubramanian:2001rb, Balasubramanian:2001nb, Balasubramanian:2002zh},
which is manifested with a non-canonical inner product on the associated Hilbert space.\footnote{See also the recent discussion of dS space based on von Neumann algebras\cite{Chandrasekaran:2022cip}, as well as the old work by Connes and Rovelli about time, the modular group and von Neumann algebras\cite{Connes:1994hv}.
This is particularly appropriate in our discussion of dS holography based on the insights gained from the metastring which is intrinsically non-commutative and lives in modular spacetime backgrounds.}
This non-canonical structure is hard to account for in the context of compactifications (especially the ones that lead
to spaces compatible with spacetime interpretation in string theory, such as Anti-de~Sitter space). This is also one of the main obstacles to
finding de~Sitter spacetime via string compactifications in the spacetime formulation of string theory.
In what follows we will tie the non-canonical CFT to dual spacetime, and formulate de Sitter holography,
for a large dS spacetime (with a naturally small cosmological constant), using the insights based on the concepts of modular spacetime and metastrings, which we have utilized
to give a clarification of the vacuum energy problem in string theory in the previous section.

\subsection{de Sitter Holography, Metastrings and Modular Spacetime }
In this section we want to relate above intuitions about dS holography to metastrings and modular spacetime.
First we propose the following intuitive picture: The ``wave function'' of a large dS spacetime, viewed as an entangled state of two CFTs, one at
infinite past and one at infinite future, can be modeled as a BCS wave function for superconductivity, where the
vacuum state is given by the Euclidean, Hartle-Hawking vacuum. This picture incorporates the well known
view of an eternal AdS black hole as an entangled state in two boundary CFTs
\cite{Maldacena:2001kr}.
We will place this proposal in
the context of the metastring formulation of string theory in modular spacetime that we used in our
discussion of vacuum energy problem in string theory.

Let us first discuss this analogy between dS holography and the BCS theory of superconductivity.
In the BCS theory of superconductivity, the superconducting phase is a state with broken symmetry (at low temperature)
and the metal phase is an unbroken, normal phase, with restored symmetry (at high temperature).
Phenomenologically, the current is directly related to the vector potential (as in the London equation) --- this is
just the conserved stationary current of the Abelian Higgs model, describing the simplest symmetry breaking (Higgs) mechanism: $\vec\jmath = c \vec{A}$.
The entanglement in the BCS set-up involves pairs of quasiparticles (quasi-electrons) found around the Fermi surface, with
opposite momenta (Cooper pairs). The pairing happens dynamically via an electron-phonon interaction.
The normal phase is a Fermi liquid, with a well defined Fermi surface and free, dressed, quasi-electrons around it.
(An insightful presentation 
of the effective field theory of the Fermi surface is given in Polchinski's 1992 TASI lectures\cite{Polchinski:1992ed}).
The BCS theory proposes a variational wave-function that describes the ground state of
such pairs, which minimizes the energy of the BCS interacting Hamiltonian. Then one gets the gap equation, for the
mass gap around the Fermi surface. This formula is essentially non-perturbative.
(Polchinski also discussed the effective field theory approach to superconductivity in his 1992 TASI lectures.)

We would like to propose an analogue of this famous picture for the two entangled CFTs of dS holography, one living in the infinite past and the other in the infinite future.
The analogue of the superconducting phase is dS space. The analogue of the normal phase is Euclidean AdS space.
The analogue of the London equation is the relation between the energy-momentum tensor and the metric, which
involves the cosmological constant: $T_{ab} = \Lambda g_{ab}$. The analogue of the BCS wave-function is the entangled wave-function of dS universe. The analogue of the gap is the cosmological constant, and the gap equation is 
replaced by a non-perturbative formula for the cosmological constant (found in our cosmic-string-like toy model of de Sitter space
in string theory
as well as in our discussion of the vacuum energy problem in metastring theory).

Let us recall some basic relations from the BCS theory, following the presentation from Feynman's book\cite[Chapter~10]{rRF-StatMech} on superconductivity.
The BCS wave-function reads as follows:
\be
\Psi_{BCS} = \prod_k \psi_k
\ee
where 
\be
\psi_k = (u_k + v_k a_{k}^{\dagger} a_{-k}^{\dagger}) \phi_0
\ee
where $\phi_0$ is the ground state described by the Fermi liquid (of free, dressed quasi-electrons around the Fermi surface) and $|u_k|^2 + |v_k|^2 =1$. These variational parameters are found by minimizing
the BCS Hamiltonian, given by the quadratic and quartic terms:
\be
H_{BCS} = \sum_k (e_k a_k^{\dagger} a_k + e_{-k} a_{-k}^{\dagger} a_{-k })
+ \sum_{k, k'} V_{k, k'} a_{k'}^{\dagger}a_{-k'}^{\dagger}a_k a_{-k }.
\ee
The minimization gives the gap equation:
\be
E_k = \sqrt{e_k^2 +\Delta_k^2},
\ee
where the gap is defined by the equation for the variational parameters upon minimizing the BCS Hamiltonian,
\be
\Delta_k = - \sum_{k'} V_{k k'} u_{k'} v_{k'},
\ee
and where also
\be
  u_k^2 = \frac{1}{2}\Big(1- \frac{e_k}{E_k}\Big),\quad
  v_k^2 = \frac{1}{2}\Big(1+ \frac{e_k}{E_k}\Big).
\ee
One thereby obtains the gap equation:
\be
\Delta_k = - \sum_{k'} V_{k k'} \frac{\Delta_{k'}}{2\sqrt{e_{k'}^2 +\Delta_{k'}^2}}.
\ee
For an isotropic situation, we have constant interaction $V$ with the range of $e_c$ above and below
the Fermi surface, and zero otherwise. This produces the following fairly simple gap equation for $\Delta$:
\be
1= -V \sum_k \frac{1}{2\sqrt{e_{k}^2 +\Delta^2}},
\ee
which gives (upon the replacement of the sum by an integral) the following non-perturbative formula (characteristic of dimensional transmutation in the purely exponential regime) for the gap $\Delta$:
\be
1 = \frac{|V|}{2}\int_{-e_c}^{e_c} \frac{D(e)\,\rd e}{\sqrt{e^2 +\Delta^2}}
\sim |V| D(0)\int_{0}^{e_c} \frac{\rd e}{\sqrt{e^2 +\Delta^2}},
\ee
where $D(e)$ is the density of states. Thus, finally,
\be
\Delta = \frac{e_c}{\sinh(\frac{1}{|V|D(0)})}
\ee
where $D(0)$ is the density of state at the Fermi surface. $V$ is positive for the electron-phonon interaction
which leads to the formation of Cooper pairs.
All of this can be generalized to finite temperatures 
and successfully compared to the
actual experiment.

Now we propose analogous equations for dS holography, concentrating on $\dS_3$ and two entangled 2d CFT duals.
We will replace the creation and annihilation operators $a$ by their Virasoro counterparts (in momentum space) $L_n$
(where we have to sum over all $n$ in our case), with non-canonical conjugation properties that we have already discussed.
Also, the gap will become the cosmological constant.
The reason why the cosmological constant is the analogue of the gap is contained in the well known formula
for acceleration in dS space\cite{Deser:1997ri}
\be
a_{\text{dS}} = \sqrt{a^2 +a_0^2}
\ee
where $a_0$ is the natural acceleration $c H_0$ and $H_0$ is the Hubble constant in 
dS space, set by the cosmological constant.
By the Unruh-Hawking formula
\cite{Gibbons:1993uz},
this acceleration is related to the Unruh-Hawking temperature, which is essentially of the units of energy, and which makes the analogy with the corresponding equation found in the context of superconductivity reasonable.
Note that this formula also works in AdS space, where AdS is consistent with supersymmetry (SuSy) and so corresponds to the normal phase with unbroken SuSy,
whereas dS can be understood as the broken phase, wherein SuSy is
cosmologically
broken.
Thus the gap in our dS case should be 
related to the dS acceleration, and thus, the cosmological constant. The SuSy breaking scale could be related to an effective low string scale of $10^{11}$ GeV discussed in the previous section.

This analogy leads to the following wave-function for a large dS universe
\be
\Psi_{\text{dS}} = \prod_{k} \psi_{k}
\ee
where 
\be
\psi_k = \prod_n (u_{k,n} + v_{k,n} L_{k, n}^{\dagger} L_{-k, -n}^{\dagger}) \phi_0
\ee
where the positive momentum label refers to the future infinity and negative momentum label to past infinity.
Note that we have to use the momentum space version of the Virasoro generators.
The dS analogue of the BCS Hamiltonian is
\be
\begin{aligned}
H_{\text{dS}} &= \sum_{k,n} (e_{k,n} L_{k,n}^{\dagger} L_{k,n} + e_{-k,-n} L_{-k, -n}^{\dagger} L_{-k, -n })\\*[-2mm]
 &\qquad
+ \sum_{k, k'; n, n'} V_{k, k';n,n'} L_{k', n'}^{\dagger}L_{-k', -n'}^{\dagger}L_{k, n} L_{-k, -n }.
\end{aligned}
\ee
This is one of our crucial points. In order to describe a large dS spacetime in terms of two entangled CFTs with non-canonical conjugation properties and (in general, dynamical) Hilbert space structure, one needs to invoke a second quantized, and in principle, non-perturbative formulation. Such a formulation is quite natural from the point of view of the metastring, and we will comment on that in the remainder of this section.

Given this model Hamiltonian for dS holography, we follow the steps already presented in the BCS theory. The final gap equation gives essentially a non-perturbative formula for the cosmological constant,
involving an exponential of the interaction $V$ and the density of states in the AdS CFT, $D_{AdS}$.
\be
M_{\Lambda} = {M_c}_{\text{AdS}}\exp\Big(-\frac{1}{|V|D_{\text{AdS}}}\Big)
\ee
Here ${M_c}_{\text{AdS}}$ is some critical energy scale, with naturally Planckian dimensions, associated
with the AdS ground state.
{Notice that precisely such an exponential hierarchy emerges in our toy model, as reviewed in Section~\ref{s:dSdefo}. 
This is essentially the seesaw formula which includes the exponential hierarchy. This formula is also consistent with our discussion
of vacuum energy that combines the phase-space/modular spacetime features of the metastring with its holographic features\cite{Freidel:2022ryr}.}
We remark that $V$ should be positive 
as in the BCS theory.

A couple of comments are in order:
The London equation captures the diamagnetic character of superconductivity --- the expulsion of
the magnetic field lines. On the dS side, this would amount to an ``effective antigravity'', which is exactly what the
positive cosmological constant describes: the equation of state is $p = -\rho$.
Similarly, the London equation implies the vanishing of resistance in the superconducting phase.
On the dS side, the analogue of resistance is the Newton constant (because it relates the field to the source in Einstein's equations), so the weakness of gravity is captured by
this fact, which translates into widely separated scales involving the characteristic matter scale and the natural Planck scale.
Also, one can  work in the two-fluid picture 
of normal and superconducting electron fluids. It is tempting to associate this to the two-spacetime pictures,
of normal and dual spacetime as offered by
the metastring formulation of general string theory reviewed in Section~\ref{s:VEST}.

The whole description of the analogy of dS holography and superconductivity is quantum, gravitational 
(gravity would be needed for providing an attractive and thus positive interaction $V$) and, in principle, non-perturbative.
So, from this point of view a non-perturbative formulation of string theory (already outlined in the previous section and also discussed later in the review)
in a holographic limit of large dS spacetimes does lead to dS spaces with a naturally small cosmological constant,
which is also related to an exponential hierarchy between the Planck scale and the observed characteristic matter scales.
This non-perturbative formula should be also indicative of a non-perturbative (cosmological) breaking of supersymmetry that
underlies this holographic dS description.

Note that the original proposal by Balasubramanian, de Boer and Minic (BdBM) \cite{Balasubramanian:2001rb, Balasubramanian:2001nb, Balasubramanian:2002zh}, was mainly concerned with
$\dS_3$, and as such, it has involved two entangled 2d CFTs with Virasoro generators. This is really analogous
to Maldacena's treatment of black holes in $\AdS_3$
\cite{Maldacena:2001kr}. However, the BdBM proposal can be extended to the
realistic case of $\dS_4$, in which case the holographic definition involves two entangled 3d CFTs. The natural
candidate here is the Gustavsson-Bagger-Lambert theory
\cite{Gustavsson:2007vu,Bagger:2006sk,Bagger:2007jr},
which appears in the non-perturbative formulation
of $n$-interacting membranes, of the ABJM model
\cite{Aharony:2008ug}, which is the logical 3d analogue of the $N=4$ conformal SYM in 4d. However, these theories should be quantized in a generic modular polarization\cite{Freidel:2022ryr}.

Finally, in view of the new understanding of the string vacuum presented in the previous section, such a holographic formulation of de Sitter space
(holography being one of the central features of the argument presented in Section~\ref{s:VEST})
should involve both the spacetime and dual spacetime, as well as the corresponding momenta, in such a way that in the early universe the dual degrees of freedom dominate, whereas in the late universe the visible degrees of freedom are dominant, as pointed out in\cite{Freidel:2021wpl}.
Thus de Sitter space should be given as a BCS-like entangled state between
two CFTs, one of which, in the infinite past, is formulated in terms of dual degrees of freedom, and
the one in the infinite future, in terms of visible degrees of freedom. 
The discussion presented in\cite{Freidel:2021wpl} was applied to metaparticles (the zero modes of the metastring) in an expanding universe
(such as de Sitter spacetime) and it indeed revealed that the dual degrees of freedom dominate in the early and young
universe (when the visible sector is frozen), whereas in a large and old universe the dual degrees of
freedom are frozen and the visible degrees of freedom dominate
\cite{Freidel:2021wpl}.
However, the same picture should be true for
the metastring in a dS background.\footnote{The same picture in principle applies to black holes in de Sitter spacetime, suggesting that the interplay of visible and dual degrees of freedom could be crucial for understanding the quantum dynamics of black holes, including the so-called information paradox. By extrapolation to zero cosmological constant, the same picture would apply to black holes in asymptotically flat spacetimes.}

We stress that this proposal fits into our new understanding of de Sitter entropy 
as counting the degeneracy of the quantum vacuum of string theory\cite{Freidel:2021wpl}. This implies that the above Virasoro
generators have to annihilate a degenerate modular vacuum\cite{Freidel:2016pls}. That means that the dS entropy counts the
degeneracy of two entangled theories with degenerate vacua, where the temperature of dS spacetime
emerges after one of the theories (in the infinite future) is traced over. Note that the dual degrees of freedom
are ``frozen'' according to the general behavior of metaparticles in a cosmological background associated with
an expanding universe, 
as the universe gets older and bigger, and thus, one could trace over the dual degrees of freedom, in an effective spacetime description.
This, in turn, is consistent with how the effective cosmological constant emerges from 
a low energy description of the metastring framework, which will be discussed in the next section.
Also, the new quantum number $N$ introduced in Section~\ref{s:VEST},
 which is crucial for understanding how a small positive cosmological constant is realized in string theory,
is the consequence of modular polarization.
Therefore, the dual CFTs have to be quantized in the modular polarization in which this 
new quantum number is apparent. Similarly, dS observables should be defined in the modular formulation of quantum theory, given the fact that the Schr\"{o}dinger 
polarization is generically singular
from the modular spacetime point 
of view\cite{Freidel:2016pls}.

One of the most important new points here is that it was very difficult in the past to understand the appearance of CFTs with non-canonical conjugations for their Virasoro generators. Now however, such non-canonical CFTs are associated with the metastring formulation of string theory, and thus the physical picture, including the reason for the natural appearance of a small positive cosmological constant and a de Sitter background in string theory, is logically  complete, at least conceptually.

In conclusion, given the new insight about the vacuum energy in string theory as presented in Section~\ref{s:VEST}, 
one can approach the problem of de Sitter space in string theory in the context in which the vacuum energy problem has
been understood, that is, in the framework of modular spacetime, and in particular, in the metastring formulation
of string theory.
However, even if we have a new insight on the vacuum energy problem and de Sitter space in the context of metastring theory
(a duality symmetric and intrinsically non-commutative formulation of string theory) we need to understand
how the ``real world'' emerges in such a de Sitter background with a naturally small cosmological constant.

\section{de~Sitter Spacetime, String Theory and Reality}
\label{s:Lessons}
In section~\ref{s:VEST} we have presented a new view on the cosmological problem in string theory, and we have argued
that the new insight regarding the vacuum energy and gravitational entropy should be formulated
in a $T$-duality-covariant, chiral, phase-space like formulation of string theory --- the metastring formulation.
The holographic definition of de~Sitter space, based on non-canonical dual CFTs, and the new insights
about the vacuum energy, also implies that string theory ought to be reformulated in a generalized-geometric fashion in terms of abstract CFT data.
 Such a description has been recently developed in\cite{Freidel:2013zga, Freidel:2014qna, Freidel:2015pka, Freidel:2015uug, Freidel:2016pls, Freidel:2017xsi, Freidel:2017wst, Freidel:2017nhg}, showing that string theory reaches to the foundations of quantum theory in terms of a manifestly non-local formulation of quantum mechanics and quantum field theory. This new approach is based on a quantum spacetime the geometry of which captures the essential quantum non-locality of any quantum theory and is encoded by:
 ({\bf1})~a mutually compatible symplectic structure $\omega_{AB}$,
 ({\bf2})~the symmetric polarization metric $\eta_{AB}$ and
 ({\bf3})~the doubled symmetric spacetime metric $H_{AB}$.
It turns out\cite{Freidel:2018tkj,Freidel:2017yuv} that this chirally doubled, phase-space like non-commutative stringy target spacetime incorporates ``double field theory''\cite{Tseytlin:1990nb,Tseytlin:1990va,Tseytlin:1990hn,Siegel:1993xq,Siegel:1993th,Alvarez:2000bh,Alvarez:2000bi,Hull:2004in,Hull:2009mi,Coimbra:2011nw,Vaisman:2012ke,Vaisman:2012px,Aldazabal:2013sca,Hohm:2013vpa,Blumenhagen:2014gva,Hassler:2016srl} as well as the ``generalized (complex, K{\"a}hler) geometry''\cite{Gualtieri:2004wh,Grana:2008yw,Coimbra:2011nw,Koerber:2010bx,Gualtieri:2007ng,Hull:2012dy,Gualtieri:2014ux,Candelas:2016usb,K_k_nyesi_2018,Candelas:2018lib}. Continuing our brief review in \SS\;\ref{s:MP+MS}, we now continue highlighting the immediately relevant aspects of this still novel structure, seeking out its implications to dark energy, dark matter and their effect on the hierarchy of scales.

The emphasis of this section is as follows: even if we are able to obtain de~Sitter spacetime with a naturally small cosmological
constant, such a universe has to contain the observed visible matter, described by the Standard Model, as well
as the dark matter sector needed for cosmological structure formation. In this section, after outlining the general formulation
of string theory and quantum gravity based on the metastring approach, we discuss the origin of dark energy in such a
formulation, as well as the origin of the dark matter sector (in the form of the zero modes of the metastring --- the metaparticles),
as well as outline a new picture on the emergence of the Standard Model in string theory.
This picture completes, at least conceptually, our proposal for a string theory of the observed world.

\subsection{General String Theory and Quantum Gravity} 
\label{s:GS}
Strings have both left- and right-movers, both of which should in general be taken into account, and treated on the same footing --- which is frequently not the case: 
 In the usual spacetime interpretation, the sum of the left and right zero modes, $x^a{:=}\frc1{\sqrt2}(x^a_{\sss L}{+}x^a_{\sss R})$, are identified as {\em\/the\/} spacetime coordinates; the dual (winding) coordinates, $\tx^a{:=}\frc1{\sqrt2}(x^a_{\sss L}{-}x^a_{\sss R})$ are, in typical textbook treatments, regarded as an afterthought if not outright ignored.\footnote{This coordinate doubling turns up, and most importantly, in the study of so-called {\em\/twisted\/} sectors in Landau-Ginzburg orbifold description of string compactification\cite{Vafa:1989xc,Intriligator:1990ua}.}
 However, by examining the simplest example of the canonical free string compactified on a circle, in an intrinsically $T$-duality covariant formulation of the Polyakov string, it was proven that these variables do not commute pair-wise\cite{Freidel:2017wst,Freidel:2017nhg}:
\begin{equation}
   [\hat{x}^a,\hat{\tx}_b]=2\pi i\l^2\d^a{}_b.
 \label{e:Qxtxcomm}
\end{equation}
 The full $x$-spacetime covariance is maintained in this description, 
and the string tension is given by the ratio of the fundamental length\footnote{This {\em\/fundamental,} {\em\/primordial\/} length-scale, $\l$, should not be confused with the $3{+}1$-dimensional Planck length, $(M_P^{\sss(4)})^{-1}\!\sim\!10^{-35}$m: For example, the exponential hierarchy~\eqref{GN} in the \BHM\ models implies that $\l\sim (M_P^{\sss(10)})^{-1}\lesssim 10^{-19}$m.} and energy scales $\alpha' = \lambda/{\epsilon}$, and~\eqref{e:Qxtxcomm} clearly corresponds to the last of equations~\eqref{xtxcomm} if we set $\e=1/\l$.
Also, this fundamental non-locality is derived by examining the algebra of vertex operators in the 2d CFT of a free string compactified on a circle\cite{Freidel:2013zga, Freidel:2014qna, Freidel:2015pka, Freidel:2015uug, Freidel:2016pls, Freidel:2017xsi, Freidel:2017wst, Freidel:2017nhg}.

\paragraph{Born Triple:}
As summarized in Section~\ref{s:MP+MS}, one of the central features of this refinement is that the zero-mode variables
 $\X^{A}\define (x^a/\lambda ,\tx_a/\lambda )^{T}$ span a phase-space like non-commutative, chirally doubled spacetime. Here, $\tx_a\,{:=}\,\delta_{ab}\,\tx^b$ uses the neutral metric $\eta_{AB}$, and the Hilbert structure (i.e., the inner product) is captured by $\omega_{AB}$, as specified in~\eqref{etaH0}.
These structures are usually ignored in the canonical spacetime interpretation of string theory.
 String theory (as a theory of quantum gravity) thus becomes that quantum theory whose intrinsic (Born) geometry of quantum non-locality is made dynamical.
For more information about Born geometry encoded by the triple $(\omega_{AB},\,\eta_{AB},\,H_{AB})$, see\cite{Freidel:2017yuv, Freidel:2018tkj}.

It is important to discuss the question of the physical meaning of $(\hat{x}^a, \hat{\tx}_b)$.
Following the usual logic that connects the non-trivial commutators to the mutual indeterminacy of the non-commuting operators, we can understand the 
meaning of $(\hat{x}^a, \hat{\tx}_b)$, as follows: the $\hat{x}^a$ coordinate can be associated with short-distance (UV) ``spacetime'' and the $\hat{\tx}_b$ with the long-stance (IR) ``spacetime'', 
such that their corresponding indeterminacies are necessarily complementary, i.e.: $ \Delta {x}^a \Delta {\tx}_b \sim \lambda^2 \delta^a_b$, as follows from~\eqref{e:Qxtxcomm}.
 Furthermore, local effective fields $\phi(x)$ should be replaced with bi-local fields $\phi(x, \tx)$, leading to intrinsically non-commutative field theories.
 In turn, such theories generically display mixing between the UV and IR physics\cite{Douglas:2001ba, Szabo:2001kg, Grosse:2004yu}, and in order to define such theories in the continuum one has to appeal to a double-scale renormalization group (RG) and the self-dual fixed points. Therefore, the effective field theory scale does not have
to be the UV scale, but some scale that is a geometric mean between the UV and IR scales (which should be compared to the discussion in section 4).

Besides having a choice in the Hilbert space structure (that is, the inner product), this seems to be what is called for by the holographic formulation of de~Sitter space: an intrinsic CFT formulation with the UV and IR cut-offs both explicitly taken into account.
One of the issues associated with de~Sitter holography was always that the parameter that defines the size of de~Sitter spacetime is the observed cosmological constant, which is one of the 
explicit observables sensitive both to the UV and the IR physics. This appears as a parameter in the holographic formulation, which is not in principle tied to understanding its origin and naturalness.
However, the generalized-geometric formulation offers in principle a new outlook on this fundamental problem.

This new approach to string theory suggest the natural unified theory of strings to be the bosonic
string reduced on a $25{+}1$-dimensional torus (but interpreted from a generalized-geometric, quantum spacetime-based point of view,
so that the usual puzzles associated with possible closed timelike curves etc.\ are avoided), 
and then ``extensified'' (this was the new concept used in\cite{Freidel:2013zga, Freidel:2014qna, Freidel:2015pka, Freidel:2015uug, Freidel:2016pls, Freidel:2017xsi, Freidel:2017wst, Freidel:2017nhg}). As suggested in these papers, 
the strongly coupled version of this fundamentally bosonic and non-commutative
construction would then provide the proper definitions
of the conjectured supersymmetric non-perturbative limits, that is, $M$- and $F$-theory.

This generalized geometric formulation of string theory originates from the
{\it chiral} worldsheet (metastring) description~\eqref{e:MSAction}:
\be
S^{\text{ch}}_{\text{str}}=\frac{1}{4\pi}\int_{\Sigma}\rd^2\sigma~
 \Big[\pa_{\tau}{\X}^{A} \big(\eta_{AB}(\X)+\w_{AB}(\X)\big)
- \pa_\s\X^A H_{\!AB}(\X)\Big] \pa_\s\X^B, 
\label{e:MSA}
\ee 
where $\Sigma$ is the worldsheet, and $\X^A$, ($A=1,\cdots\!,26$, for the critical bosonic string) combine the sum ($x^a$) and the difference ($\tx_a$) of the left- and right-moving {\em\/chiral bosons\/} on the string.\footnote{This formulation also leads to a natural proposal for a non-perturbative string theory which will be discussed
in a latter part of this section. Here we concentrate on the perturbative metastring formulation.}
The mutually compatible dynamical fields
are
the antisymmetric symplectic structure $\w_{AB}$,
the symmetric polarization metric $\eta_{AB}$ and
the doubled symmetric metric $H_{\!AB}$, respectively, defining the so-called Born geometry\cite{Freidel:2013zga}.
For orientation and omitting torsion, $B_{ab}$, the fundamental triple,
 $(\eta_{AB},\,H_{AB},\, \w_{AB})$ defined in~\eqref{etaH0}, specifies respectively the neutral metric, the doubled spacetime metric and the symplectic structure of the {\em\/flat\/} metastring chirally doubled target $\X^A$-space. These easily define the following key tensors:
\begin{subequations}
 \label{e:IJK}
\begin{alignat}9
 (I&:=\w^{-1}H)^A{}_B&
   &=\begin{pmatrix}0&-h^{ab}\\h_{ab}&~~0\end{pmatrix},&\qquad I^2&=-1;\\
  (J&:=\eta^{-1}H)^A{}_B&
    &=\begin{pmatrix}0&h^{ab}\\h_{ab}&0\end{pmatrix},&\qquad J^2&=+1;\\
 (K&:=\eta^{-1}\w)^A{}_B&
   &=\begin{pmatrix}-\d^a{}_b&0\\~~0&\d_a{}^b\end{pmatrix},&\qquad K^2&=+1.
\end{alignat}
\begin{equation}
  [I,J]=2K,\quad
  [K,I]=2J,\quad
  [J,K]=-2I;\qquad
  \{J,I\}=\{I,K\}=\{K,J\}=0.
\end{equation}
\end{subequations}
These provide the $\X^A$-space with a simultaneous {\em\/complex\/} ($I$),  {\em\/chiral\/} ($J$) and {\em\/real\/} ($K$) structure.\footnote{Alternatively, $\inv2\{iI,iJ,iK\}$ generate the $\mathfrak{so}(2,1)$ Lie algebra. In turn, the ordered triple $\{I,iJ,iK\}$ of mutually anticommuting complex structures satisfy the quaternionic product algebra, $e_\a{\cdot}e_\b=\e_{\a\b}{}^\g\,e_\g$.} These structures straightforwardly generalize to curved $\X^A$-spacetime models, and the inclusion of the stringy $B_{ab}$-field even provides for torsion; see~\eqref{Bcommrel}, below.

Quantization renders the doubled ``phase-space'' operators $\hat{\X}^A=(\hx^a/\l, \htx_a/\l)$ inherently non-commutative, inducing\cite{Freidel:2013zga}
\be
[ \hat{\X}^A, \hat{\X}^B] = i \w^{AB}.
\label{e:CnCR}
\ee
In components and for constant non-zero $\w^{AB}$, as in~\eqref{etaH0},~\eqref{e:CnCR} is equivalent to:
\be
[\hx^a,\htx_b]=2\pi i\l^2 \delta^a_b,\qquad
[\hx^a,\hx^b]=0=[\htx_a,\htx_b],
\label{e:CnCR1}
\ee
where $\l$ denotes the fundamental length scale, such as the Planck scale --- in the original, $9{+}1$-dimensional setting. Then, $\e=1/\l$ is the corresponding fundamental energy scale
and the string tension is 
$\a' = \l/{\e}=\l^2$.
Note that the Hamiltonian and diffeomorphism constrains,
\begin{alignat}9
 \pa_\s\X^A H_{AB} \pa_\s\X^B
 &=\l^{-2}\big[ g_{ab}(x,\tx)\,(\pa_\s x^a)(\pa_\s x^b)
               +g^{ab}(x,\tx)\,(\pa_\s\tx_a)(\pa_\s\tx_b)\big]=0,\label{e:HamC}\\
 \pa_\s\X^A \eta_{AB} \pa_\s\X^B
 &=2\l^{-2}\big[ (\pa_\s x^a)(\pa_\s\tx_a) \big]=0,\label{e:DifC}
\end{alignat}
respectively, are treated on the equal footing.
In particular, the usual spacetime interpretation of the zero mode sector of string theory\cite{Polchinski:1998rq} is
tied to the solution of the diffeomorphism constraint~\eqref{e:DifC} by level matching.

Note that in the $B$-background the commutators read
 \begin{equation}\label{Bcommrel}
[\hat{x}^a,\hat{x}^b]=0,\qquad 
[\hat{x}^a,\hat{\tx}_b]=2\pi i\lambda^2 \delta^a{}_b,\qquad 
[\hat{\tx}_a,\hat{\tx}_b]=-4\pi i\lambda^2 B_{ab}.
\end{equation}
This provides a new interpretation to the antisymmetric $B$-field, as essentially stemming from the non-trivial symplectic structure of general string theory, after an $O(D,D)$ transformation.
 Similarly, there is a conjugate background (called $\beta$-background) characterized by
the commutation relations:
 \begin{equation}\label{betacommrel}
[\hat{x}^a,\hat{x}^b]=4\pi i\lambda^2 \beta^{ab},\qquad 
[\hat{x}^a,\hat{\tx}_b]=2\pi i\lambda^2 \delta^a{}_b,\qquad 
[\hat{\tx}_a,\hat{\tx}_b]=0,
\end{equation}
which reveal genuine non-commutativity of spacetime in string theory. Indeed, the direct non-commutativity in the observable spacetime as seen in~\eqref{betacommrel} is exactly of the kind that has been discussed in ``non-commutative quantum field theory,'' and which exhibits a priori measurable violations of Lorentz symmetry\cite{Carroll:2001ws}. It is thus worthwhile noting that such studies have a ``home'' within the metastring/metaparticle framework, and are related to the $B$-background via the~\eqref{betacommrel} vs.~\eqref{Bcommrel} gauge choice of sorts.

In this more general and generically non-commutative chiral formulation, the spacetime interpretation 
is replaced by a modular spacetime (or quantum spacetime)
realization (also found in the context of quantum foundations)\cite{Freidel:2013zga}.
Also, all effective fields must be regarded a priori as 
bi-local $\phi(x, \tx)$ and non-commutative, 
subject to~\eqref{e:CnCR1}, and therefore inherently non-local in 
the conventional $x^a$-spacetime\cite{Freidel:2013zga}. Such non-commutative field theories\cite{Douglas:2001ba} 
generically display a mixing between the ultraviolet (UV) and infrared (IR) physics
with continuum limits defined via a double-scale renormalization group (RG) and the 
self-dual fixed points\cite{Douglas:2001ba,Freidel:2013zga}. 
This has profound implications for the generic physics of string theory, and in particular the problems
of dark energy, dark matter and the separation of scales. Thus, an effective field theory limit based on
the traditional spacetime interpretation of string theory is simply not generic, but an exceptional and singular limit of this more general,
non-commutative chiral formulation.

The underlying reason for the phase-space like non-commutativity in the general formulation of string theory may also be seen as follows:
 In the canonical presentation of string theory, say, compactified on a
circle, one assumes that the coordinates and their duals commute, and one consequently finds that the algebra of vertex operators contains cocycles. Such cocycles are simply ubiquitous in string theory.
 The closer analysis of the underlying symplectic structure finds that these cocycles are simply place-holders for a non-commutative formulation\cite{Freidel:2017wst,Freidel:2017nhg}:
The coordinates and their duals do not commute~\eqref{e:CnCR1}, but form a Weyl-Heisenberg-algebra, and the vertex operators form a representation this Weyl-Heisenberg algebra --- now without cocycles.
 The usual spacetime formulation of strings is therefore
simply a singular limit of this generic non-commutative and doubled, phase-space like, formulation. In essence, standard textbook presentation of string theory expands around a commutative limit of this intrinsically non-commutative formulation.
 This singular limit is very similar to the so-called Seiberg-Witten limit of non-commutative field theory\cite{Seiberg:1999vs}, where the physical observables must be identified with great care.
 Analogously, the physical observables in this general, non-commutative closed string theory must be identified as representations of the underlying non-commutative algebra.

Similarly, backgrounds like the $B$- and the $\beta$-background, respectively~\eqref{Bcommrel} and~\eqref{betacommrel}, are found to be part of the
symplectic structure of the non-commutative phase-space like formulation, and that the dilaton naturally comes from the 
the volume preserving transformations of this underlying phase space. The geometry of this formulation is
Born geometry which unifies the symplectic and doubly orthogonal transformation with the natural double metric.
This geometry turns out to be the underlying geometry of quantum theory in its modular spacetime polarization, in which
the symplectic structure captures the discrete phase space of the underlying spacetime and its dual, the orthogonal
structure is the symmetry of this doubled spacetime, and the doubled metric structure is related to the choice of
quantum vacuum in this generic modular polarization. 

This implementation of Born geometry is the consequence of the compatibility of
the fundamental length/time (non-locality) with Lorentz symmetry (causality) which are the deep reasons behind
quantum theory. String theory (in its general formulation) is a very particular quantum theory in which this
Born geometry is dynamical. Thus the road from quantum theory to string theory, viewed as a theory of quantum gravity,
is analogous to the road taken from special to general relativity. Finally, the natural habitat of string theory is
modular spacetime, as a precise form of quantum spacetime. The observed classical spacetime is only an
emergent feature (extensification) of this quantum spacetime. Also, instead of the procedure of quantization of classical formulations, we should properly start from an intrinsic quantum formulation (modular polarization, given by a complete 
set of unitary operators of translation and dual translation), and then look for various consistent classical limits.

We emphasize the trade-off between 
a stable non-commutative (symplectic) bosonic picture of general string theory (and thus {\em\/not\/} the traditional bosonic picture 
with instabilities) and the traditional stable supersymmetric picture. In some sense we have 
a trade-off between the symplectic and complex structures (reminiscent of mirror symmetry).

\paragraph{Born Dynamics:}
In view of the preceding discussion regarding the generic formulation of string theory and its relation to the problem of
dark energy and dark matter, in the remainder of this section we discuss a proposal for a non-perturbative formulation of string theory (and its $M$-theory and $F$-theory avatars). Indeed, the metastring offers such a new view on the fundamental question of a non-perturbative formulation of quantum gravity\cite{Freidel:2013zga}
by noting the following:
In the metastring formulation the target space is found to be a modular space (quantum spacetime), but 
the same can be also said of the world-sheet.
If the world-sheet is made modular in the metastring formulation, by doubling of $\tau$ and $\sigma$, so
that  $\X (\tau, \sigma)$ can  be
in general viewed as an infinite dimensional matrix (the matrix indices coming from the Fourier components of the
doubles of $\tau$ and $\sigma$), then the corresponding metastring action becomes
\be
\int \Tr \big[ \partial_{\tau} \X^A \partial_{\sigma} \X^B (\omega_{AB} + \eta_{AB}) - 
\partial_{\sigma} \X^A H_{AB} \partial_{\sigma} \X^B \big]\; \rd \tau\, \rd \sigma ,
\ee
where the trace is over the matrix indices.
Then we could associate the natural partonic degrees of freedom with matrix entries.
We arrive at a {\it non-perturbative quantum gravity} by replacing the sigma derivative with a
commutator involving one extra $\X^{26}$ (with $A=0,1,2,\cdots\!,25$). (Note that the idea
that the canonical world-sheet of string theory
might become non-commutative in a deeper, non-perturbative formulation, was suggested in\cite{Atick:1988si}.)
\be
\partial_{\sigma} \X^A \to [\X^{26}, \X^A] .
\ee
This dictionary suggests the following fully interactive and non-perturbative formulation of metastring theory
in terms of a ($M$-theory-like) matrix model form of the above metastring action (with $a,b,c=0,1,2,\cdots\!,25, 26$ )
\be
\int \Tr\big[ \partial_{\tau} \X^a [\X^b, \X^c] \eta_{abc}  - 
H_{ac} [\X^a, \X^b] [\X^c, \X^d] H_{bd}\big]\; \rd \tau ,
\ee
where the first term is of a Chern-Simons form and the second of the Yang-Mills form, and $\eta_{abc}$ contains both 
$\omega_{AB}$ and $\eta_{AB}$.
This is then the non-perturbative ``gravitization of the quantum''\cite{Freidel:2013zga}.
We remark that in this non-perturbative matrix theory like formulation of the metastring (and quantum gravity),
the matrices emerge from the modular world-sheet, and the fundamental commutator from the
Poisson bracket with respect to the dual world sheet coordinates (of the modular/quantum world sheet) --- that is,
quantum gravity ``quantizes'' itself, and thus quantum mechanics originates in quantum gravity.
(However, this formulation should be distinguished from Penrose's ``gravitization of the quantum'' and gravity-induced ``collapse of the wave function''\cite{Penrose:2014nha}.
Also note some similarity of the metastring formulation, in its intrinsic non-commutative form, to
the most recent proposal by Penrose regarding ``palatial'' twistor theory\cite{Penrose:2015lla}.\footnote{Note that the bosonic string can be rewritten in terms of twistors\cite{Deguchi:2010gg}.
This formulation can be generalized to include not only the
original algebra of twistors, but also the ``palatial twistor algebra''
in which the same doubled Heisenberg algebra of metastring theory appears.
This formulation might be also useful in elucidating the nature of the
algebraic structure that replaces supersymmetry in the general non-commutative
phase-space like formulation of string theory. Note that twistors can be viewed as
complex phase-space coordinates for massless degrees of freedom, and that
they are, in some sense, also naturally related to supersymmetry (but without supersymmetry).
In turn, this relation should be tied to the relevance of the phase space structure in the context of Born geometry\cite{Freidel:2018tkj}, and especially the uniqueness of the
connection in Born geometry.})

We also note that the essential points regarding the phase space, non-commutative formulation of
string theory have been already suggested in the well known work by Atick and Witten in 1988\cite{Atick:1988si}.

In thinking about non-perturbative matrix model formulations of string theory it is natural to invoke the IIB matrix model\cite{Ishibashi:1996xs}, based on D-instantons as well as the matrix model of 
$M$-theory\cite{Banks:1996vh}, based on D0-branes. However, these matrix models lack very important covariant
properties associated with $F$-theory and $M$-theory. In our proposal we can do better.
Given our new viewpoint we can suggest a 
new covariant
non-commutative matrix model formulation of $F$-theory, by also writing in the large $N$ limit 
\begin{equation}
\pa_{\tau}\X^C = [\X, \X^C], 
\end{equation}
in terms of commutators of two (one for $\pa_\s\X^C$ and one for $\pa_{\tau}\X^C$) extra $N\,{\times}\,N$ matrix valued chiral $\X$'s.
Notice that, in general, we do not need an overall trace, and so the action can be viewed as a matrix, rendering the entire non-perturbative formulation of $F$-theory as
purely quantum in the sense of the original matrix formulation of quantum mechanics 
by Born-Jordan and Born-Heisenberg-Jordan\cite{vdW-history} 
\begin{equation}
 \S_{\text{ncF}}\,{=}\,\frac{1}{4\pi} 
[{\X}^{a},  {\X}^{b}] [{\X}^{c},  {\X}^{d}]  f_{abcd},
\end{equation}
where instead of $26$ bosonic $\X$ matrices one would have $28$, with supersymmetry\footnote{More precisely, this should be a ``re-bosonized'' preimage of supersymmetry in line with the sketch~\eqref{e:HS}--\eqref{e:mSuSy2} of the heterotic string model.} emerging in 10($+$2) dimensions from this underlying
bosonic formulation. Thus, in this formulation one can realize the $SL(2,Z)$ symmetry of IIB string theory.
In this non-commutative matrix model formulation of $F$-theory, in general,
$f_{abcs} ({\X})$ is a dynamic background, to be determined from the matrix analogue of the vanishing of
the relevant beta function.
By $T$-duality, the new covariant $M$-theory matrix model reads as 
\begin{equation}
 \S_{\text{ncM}}\,{=}\,\frac{1}{4\pi}  
\int_{\tau} 
\big(\pa_{\tau} \X^i [{\X}^{j},  {\X}^{k}] g_{ijk} - [{\X}^{i},  {\X}^{j}][{\X}^{k},  {\X}^{l}] h_{ijkl}\big),
\end{equation}
with 27 bosonic $\X$ matrices, with supersymmetry emerging in 11 dimensions.
Once again, the backgrounds $g_{ijk} ({\X})$, and $h_{ijkl} ({\X})$ are fully dynamicals, and they
should be determined by a matrix-analogue of the Renormalization Group (RG) equation, and the vanishing of the
corresponding beta function. 

{Note the possibility of fundamental time asymmetry in this formulation.
This is due to the cubic nature of the first term, and it is rooted in the fundamentally chiral nature of the 
worldsheed description of metastring theory. Such fundamental time asymmetry must be important for
the fundamental cosmological problems, such as the fine tuning of the initial state, as well as the problem of
matter-antimatter asymmetry.}

{The fundamental equations of the metastring are self-dual, and using the large-$N$ prescription 
$\pa_{\tau}\X^C = [\X, \X^C]$, we have the general form of Nahm's equations\cite{Nahm:1981nb,Hitchin:1983ay,Donaldson:1984tm}, which in turn lead
to the fundamental equations of Nambu quantum theory\cite{Minic:2002pd,Minic:2020zjb}.
This would be a concrete implementation of the idea of ``gravitizing the quantum'' that is naturally
implied by this proposal for a non-perturbative formulation of quantum gravity in the guise of string theory.
The chiral formulation of the metastring has to be reflected in the choice of matrices --- skew-symmetric or
anti-Hermitian matrix. This should be an important point, especially in connection with the covariant formulation
of Matrix theory or $M$-theory in 11d.}

Note that in this approach
holography\cite{tHooft:1993dmi,Susskind:1994vu} (such as AdS/CFT\cite{Maldacena:1997re, Gubser:1998bc, Witten:1998qj}, which can be viewed as a ``quantum Jarzynski equality on the space of geometrized RG flows''\cite{Minic:2010pw,Gray:2013rv}) is emergent in a particular semi-classical ``extensification''
of quantum spacetime, in which the dual spacetime degrees of freedom are also completely decoupled.
 The relevant information about
 $\w_{AB},\eta_{AB}$ and $H_{AB}$ is now contained in the new dynamical backgrounds $f_{abcd}$ in $F$-theory, and $g_{ijk}$ and $h_{ijkl}$ in $M$-theory.
This proposal offers a new formulation of covariant Matrix theory in the $M$-theory limit\cite{Minic:1999js,Awata:1999dz,Minic:2002pd,Minic:2003nx},
which is essentially a partonic formulation --- strings emerge from partonic constituents in a certain limit. This new matrix formulation is fundamentally bosonic and thus it is reminiscent of bosonic $M$-theory\cite{Horowitz:2000gn}.
The relevant backgrounds $g_{ijk}$ and $h_{ijkl} $ should be determined by the matrix RG equations.
Also, there are lessons here for the new concept of ``gravitization of quantum theory'' as well as the idea that (observer dependent) dynamical Hilbert spaces or 2-Hilbert spaces (here represented by matrices) are fundamentally needed
in quantum gravity\cite{Bunk:2018qvk,Balasubramanian:2002wy,Smolin:1995vq,Smolin:1995vq,Crane:1995qj,Minic:1998nu,Banks:2001yp}.
This matrix like formulation should be understood as a general non-perturbative formulation of 
string theory. In this partonic (quantum spacetime) formulation closed strings (as well as branes) are collective excitations, in turn constructed from the product of open string fields. Similarly, our toy model can be understood as a collective excitation in this more
fundamental ``partonic'' formulation. The observed classical spacetime emerges as an 
``extensification''\cite{Freidel:2013zga}, 
in a particular limit, out of the basic building blocks of quantum spacetime.
Their remnants can be found in the low energy bi-local quantum fields, with bi-local (metaparticle) quanta, which
were a motivation for our discussion of dark matter in string theory.

Finally, it is an old realization that the 10d superstring can be found as a solution of the bosonic string theory\cite{Casher:1985ra}.
This is precisely what we have in our proposed non-perturbative formulation.
(Such a bosonic formulation is also endowed with higher mathematical symmetries, as already observed
in\cite{Moore:1993zc}.)
Supersymmetry (in the guise of $M$- and $F$-theory) is emergent from our non-perturbative and seemingly entirely
bosonic formulation in a similar fashion.
This might allow us to go around some obvious problems raised by the apparent falsification of supersymmetry at
the observable LHC energies.

A comment regarding our toy model: Given the fact that our toy model can be understood as a non-holomorphic 
deformation of D7-branes in IIB string theory, we can imagine that D7-branes emerge as 
bound states of ``monads'' (partons of our non-perturbative formulation of IIB string theory)
in the supersymmetric limit. Thus, in the general, non-supersymmetric case, such initial D7-brane bound states
get deformed into our cosmic-brane co-dimension two solutions in six dimensions, times a stable K3.
Our toy model can be also understood by starting from a stringy cosmic string,
which is supersymmetric, after supersymmetry is broken in a very particular way. From the point of view
of string theory, the world sheet description can be understood as an effective string theory formulation
that goes back to Polchinski and Strominger\cite{Polchinski:1991ax}. The leading order quadratic terms in that formulation, can be rewritten
following the metastring prescription, that is in a chiral, doubled and intrinsically non-commutative form.
Thus our \BHM\ model can be understood as an effective string theory, accompanied by effective Born geometry,
and therefore, can be related directly to the above discussion of general string theory.

\subsection{General String Theory and Dark Energy}
\label{s:DarkE}
In our recent papers\cite{Berglund:2019ctg,Berglund:2019yjq} 
we have argued that the generalized geometric formulation of string theory discussed above provides for an effective description of dark energy that is consistent with a de~Sitter spacetime. This is due to the theory's chirally and non-commutatively~\eqref{e:CnCR1} doubled realization of the target space 
and the stringy effective action on the doubled non-commutative~\eqref{e:CnCR1} spacetime $(x^a,\tx_a)$ 
(where we also include the matter Lagrangian $L_m$) 
\be
  S_{\text{eff}}^{\textit{nc}}
  =\iint \Tr \sqrt{g(x,\tx)}~ \big[R(x,\tx) +L_m(x,\tx) +\dots\big],
\label{e:ncEH}
\ee
where the ellipses denote higher-order curvature terms induced by string theory (and where the
corresponding Planck lengths are set to one). This result can
be understood as a generalization of the famous calculation by Friedan\cite{rF79b}.
Dropping the matter Lagrangian for now, to lowest (zeroth) order of the expansion in the non-commutative parameter $\l$
of $S_{\text{eff}}^{\textit{nc}}$ takes the form~\eqref{e:TsSd} in $d$ observed spacetime dimensions.
This is a striking result which first was obtained almost three decades ago by Tseytlin\cite{Tseytlin:1990nb,Tseytlin:1990va,Tseytlin:1990hn},  effectively neglecting the intrinsic non-commutativity of general string theory
and thus $\w_{AB}$ in~\eqref{e:CnCR}, and  
by assuming that $[\hat x,\htx]=0$\cite{Tseytlin:1990nb,Tseytlin:1990va,Tseytlin:1990hn}.
In general the observed spacetime and its dual are correlated, which should be important for
phenomenology, but to first order in the fundamental length we will treat them as independent sectors. 

In particular, in the leading limit, the $\tx$-integration in the first term of~\eqref{e:TsSd} defines the gravitational constant $G_N$, and in the second term produces a {\it positive} cosmological constant $\L>0$. Hence, the weakness of gravity is determined by the size of the canonically conjugate dual space, while the smallness of the cosmological constant is given by its curvature.  In particular we have:
\be
 \bar{S} 
= \frac{\int_X\! \sqrt{-g(x)}\, \big[R(x) +\dots\,\big]}{ \int_X\! \sqrt{-g(x)}}+\dots
 \label{e:TsLb11}
\ee
which leads to a seesaw-like formula for the cosmological constant, discussed below.
Note that if we include the matter sector explicitly we can argue that the dual part of the matter sector 
appears as dark matter, which is in turn both sensitive to dark energy\cite{Ho:2010ca,Ho:2011xc,Ho:2012ar,Edmonds:2013hba,Edmonds:2016tio,Ng:2016qvh,Edmonds:2017zhg,Edmonds:2017fce,Edmonds:2020iji,Edmonds:2020tug} 
and also dynamically correlated with the
visible matter. One outstanding feature of this approach is the unity of the description of the entire dark energy and matter sector, induced and determined by the properties of the dual spacetime, as predicted by this general, non-commutatively phase-space doubled formulation of string theory. We emphasize that our emergent cosmological constant is naturally small and radiatively stable,
given the argument presented in section~\ref{s:VEST}.

Note also that our $G_N$ could vary by varying the volume of the dual spacetime.
This can be connected to the efforts to
resolve the $H_0$ tension by effectively varying $G_N$\cite{Ballesteros:2020sik} and references therein. For a recent review of the $H_0$ tension
see\cite{DiValentino:2021izs}.

\subsection{Dark Sector, String Theory and the Hierarchy of Scales}
\label{s:DS+Hierachy}
In this subsection we stress that the observed hierarchy of scales, already discussed in
Section~\ref{s:VEST}, can be understood from the point of view of effective spacetime description of the
metastring formulation, and that it also nicely meshes with the general features of our toy model,
discussed in Section~\ref{s:dSdefo}.

In general string theory (which is chiral and non-commutative),
all effective fields must be regarded a priori as 
bi-local $\phi(x, \tx)$\cite{Freidel:2013zga}, subject to~\eqref{e:CnCR1}.
Moreover, the fields are doubled as well, and thus for every  $\phi(x, \tx)$ there exists 
a dual $\tilde{\phi}(x, \tx)$, as implied by the background field approach, discussed
in the next subsection. These fields are in general correlated, due to their non-commutativity.
However, to first order in the fundamental length, we can treat them as independent.
Thus, in general,
to lowest (zeroth) order of the expansion in the non-commutative parameter $\l$
of $S_{\text{eff}}^{\textit{nc}}$ takes the following form (that also includes the matter sector and its dual),
which generalizes equation \eqref{e:TsSd}
:
\be
\iint\! \sqrt{{g(x)}\,{\tilde{g}(\tx)}}~ 
 \big[R(x) + \tilde{R}(\tx)+ L_m (A(x, \tx)) + \tilde{L}_{dm}   ( \tilde{A}(x, \tx)) \big]
\label{e:TsSd1}
\ee
Here the
corresponding Planck lengths are set to one and
the $A$ fields denote the usual Standard Model fields, and the $\tilde{A}$ are their duals, as predicted by the general formulation of 
quantum theory that is sensitive to the minimal length (the non-commutative parameter $\l$\cite{Freidel:2013zga}).
In the following subsection we will elaborate more explicitly on the dual matter degrees of freedom.
Right now we are concerned with the generalization of the discussion summarized in the previous section.

Therefore, after integrating over the dual spacetime, and after taking into account $T$-duality, equation 
\eqref{e:TsLb11} now reads
\be
 \bar{S'} 
= \frac{\int_X\! \sqrt{-g(x)}\, \big[R(x) +L_m(x) + \tilde{L}_{dm} (x)\big]}{ \int_X\! \sqrt{-g(x)}}+\dots
 \label{e:TsLb1}
\ee
The main point here is that the dual sector (as already indicated in the previous section) should be interpreted as the dark matter sector,
which is correlated to the visible sector via the dark energy sector, as discussed in\cite{Ho:2010ca,Ho:2011xc,Ho:2012ar,Edmonds:2013hba,Edmonds:2016tio,Ng:2016qvh,Edmonds:2017zhg,Edmonds:2017fce,Edmonds:2020iji,Edmonds:2020tug}.
Once again, we stress the unity of the description of the entire dark sector based on 
the properties of the dual spacetime, as predicted by the generic formulation of string theory (viewed as 
a quantum theory with a dynamical Born geometry)\cite{Freidel:2013zga}.

Let us turn off the dynamical part of gravity and consider the hierarchy of scales in this formulation.
First we have
\be
S_0= - \iint\! \sqrt{{g(x)}{\tilde{g}(\tx)}}~\big[ L_m (A(x, \tx)) + \tilde{L}_{dm}   ( \tilde{A}(x, \tx)) \big],
\label{e:TsSd2}
\ee
which leads to the following non-extensive action to lowest order in $\L$
\be
 \bar{S_0} 
= \frac{\int_X\! \sqrt{-g(x)}\, \big[L_m(x) + \tilde{L}_{dm} (x)\big]}{ \int_X\! \sqrt{-g(x)}}+\dots
 \label{e:TsLb2}
\ee
after integrating over the dual spacetime.
This in turn implies a seesaw formula which involves the matter scale from the matter (and dark matter) part of the action
and the scales related to the UV (Planck scale, $M_P$) and the IR (dark energy scale, $M_\L$):
\be
 M_\L   M_P \,{\sim}\,M_H^2 .
 \label{e:MLMPMH}
\ee
where $M_H$ denotes the matter scale (the Higgs scale).
This provides for the consistent relation $\mu \sim M_H^2$, where $M_H \sim 1$\,TeV.

In general, the relation~\eqref{e:MLMPMH} is also implied by the diffeomorphism constraint of the metastring,
which is controlled by the $O(d,d)$ bi-orthogonal metric $\eta$, and which implies, in the limit of zero modes
(and zero momenta)
that $E \tilde{E} = \mu$, where $E$ is the energy scale in the observed spacetime, and $\tilde{E}$ is the energy
scale in the dual spacetime, and $\mu$ is the new parameter, which could be chosen to the Higgs scale (at least for the vacuum).
If we remember that the geometry of the dual spacetime is responsible for the origin of dark energy, then the
dual energy $\tilde{E}$ can be set to the dark energy scale. Then the fundamental energy scale $E$ in the
observed spacetime is the Planck scale. The explicit mixing of the UV and IR scales in a fully covariant formulation, which
is consistent with a properly defined continuum limit,
is a quite unique feature of the metastring.
This feature of the metastring should be contrasted with other approaches to the hierarchy problem which mix the UV and IR scales but which violate covariance, such as\cite{Craig:2019zbn}.
In particular, this discussion meshes nicely with the new view on the cosmological constant problem and its relation to the gauge hierarchy problem, presented in Section~\ref{s:VEST}.

Therefore, we have that the vacuum energy is given as
\begin{equation}
\rho_0 \sim M_\L\!^4 = \Big(\frac{M_H^2}{M_P}\Big)^4
 \label{e:IRUVseesaw}
\end{equation}
and thus $\rho_0 \sim (10^{-3}\,\text{eV})^4$, which is the observed value.
Comparing the seesaw relation~\eqref{e:IRUVseesaw} with~\eqref{e:LongCC} shows that setting ${M_\L}^4=\L_4$, $M_P=M_4$ and $M_H=M_6$ identifies these two relations {\em\/perfectly.} This confirms that this hallmark result was not an artifact of the simplicity of the Section~\ref{s:dSdefo} toy model, and lends support to identifying the Standard Model (Higgs) mass-scale, $M_H$, with the 6d Planck scale, which is in that \BHM\ scenario close to the primordial mass-scale of the underlying string theory.
All of this is consistent with a more detailed picture of the vacuum energy problem given in Section~\ref{s:VEST}.

Also, as highlighted in our previous work, the effective action of the sequester type\cite{Kaloper:2013zca,Kaloper:2014dqa,Padilla:2015aaa}
can be argued to emerge in this discussion
\begin{equation}
 \int_X \sqrt{-g}~\Big[\frac{R}{2G} + s^4 L (s^{-2} g^{ab}) + \frac{\Lambda}{G}\Big]
 + \sigma\Big(\frac{\Lambda}{s^4 \mu_s^4}\Big),
\end{equation}
where $L$ denotes the combined Lagrangian for the matter and
dark matter sectors, $\mu_s$ is a mass scale
and $\sigma(\frac{\Lambda}{s^4 \mu_s^4})$ is a
global interaction that is not integrated over\cite{Kaloper:2013zca,Kaloper:2014dqa,Padilla:2015aaa}. 
This can be provided by our set up: Start with bilocal
fields  $\phi(x, \tx)$\cite{Freidel:2013zga},
and replace the dual labels $\tx$ and also $\l$ (in a coarsest approximation) by the global dynamical scale $s \sim \Delta {\tx} \,{\sim}\,\l^2 \Delta {x}^{-1} $.
Also, normal ordering produces $\sigma$.
This then provides
an effective realization of the sequester mechanism 
in a  non-commutative phase of string theory. 
An important point here is that the low energy effective description of the generic string theory is, to lowest order,
a sequestered effective field theory, and more generally, a non-commutative effective field theory\cite{Douglas:2001ba} of a new kind, which is
defined within a doubled RG, which is covariant with respect to the UV and IR cut-offs, and which is endowed with a
self-dual fixed point\cite{Freidel:2013zga}.

\subsection{Dark Matter and General String Theory}
\label{s:DM+Strings}
In this subsection we discuss the dual matter degrees of freedom and the explicit appearance of dark matter in
the general formulation of string theory\cite{Freidel:2013zga}.
The previous discussion looked at the stringy effective action to lowest order in $\lambda$, neglecting the non-commutative aspects of the generalized geometric formulation of the string worldsheet~\eqref{e:MSA}. In order to see the effect of the leading order correction in $\lambda$ of\eqref{e:TsSd1},
we consider the zero modes of $S^{\text{ch}}_{\text{str}}$. 

The action for the zero modes of generalized string theory,
the metaparticle action, is fixed by the symmetries of the chiral worldsheet in terms of the symplectic form $\omega$, the $O(D,D)$ metric $\eta$ and the double metric $H$ --- the so-called Born geometry. Following the general results about the metastring formulation\cite{Freidel:2013zga} and~\eqref{1}, it takes the form:
\be
S_{mp} = \int {\rm d}\t~( p{\cdot}\dot{x} +\tp{\cdot}\dot{\tx} 
         +\l^2 p{\cdot}\dot{\tp} + N h + \tilde{N} d)
\label{e:SMP}
\ee
where the Hamiltonian constraint, fixed by the double metric,
$h = p^2 + {\tp}^2 + \fm^2$, and the diffeomorphism constraint, fixed by the $O(D,D)$ metric, is
$d = p{\cdot}\tp - \mu$.
These constraints are inherited from the quantization of chiral worldsheet theory. Finally, the symplectic structure fixes
the $ \l^2 p \dot{\tp} $ term.
Note that $\fm^2$ should not be confused with a mass of a particle excitation. In parallel with the usual 
discussion found in introductory chapters of textbooks on quantum field theory one has to understand the representation theory of associated with the symmetries of the underlying Born geometry,
and interpret $\fm^2$ and $\mu$ in terms of the relevant Casimirs in the full representation theory of this
description\cite{Freidel:2021wpl}.
We are not going to pursue this question in what follows, but we alert the reader that $\mu$ is a new
parameter not found in the context of the effective field theory description, while $\fm^2$ can be interpreted
as a particle mass only in a very degenerate limit in which $\tp =0$ and $\mu=0$.

We begin our discussion with a vector background, by shifting the momenta and the dual momenta
by the gauge field and its dual\cite{Freidel:2013zga}, 
\be
p_{a} \to p_{a} + A_{a}(x, \tx) , \qquad {\tp}_{a} \to {\tp}_{a} + {\tilde{A}}_{a}(x, \tx)
\label{e:minC}
\ee
These gauge fields have the usual Abelian gauge symmetries.
Thus in the target space action for the gauge fields we end up with
canonical Maxwellian terms (with the obvious index structure)
plus a characteristic coupling inherited from the symplectic structure
(which is clearly not present in the usual effective field theory approach)
\be
\int_{x,\tx} \big[ F^2 -a\l^2\brb{A}{\tilde{A}} +{\tilde{F}}^2+ F \tilde{F} +\dots\big]
\label{e:SMP2}
\ee
and where the $\l^2\brb{A}{\tilde{A}}$ term is induced directly from the $\l^2p\dot{\tp}$-term in~\eqref{e:SMP} by the ``minimal coupling'' shift~\eqref{e:minC}, and its dimensionful coefficient $a$ will be determined in what follows.
This ``mixing'' term (which is not present in the effective field theory action) may be expressed as:
\begin{alignat}9
 \brb{A}{\tilde{A}}&\define\int_{\t_*}^\t\rd\t'~
  A(x(\t'))\,\frac{\rd\tilde{A}(\tx(\t'))}{\rd\t'},\\
 &=\frc12\big[A(x)\tilde{A}(\tx)\big]_{\t_*}^\t
  +\frc12\int_{\t_*}^\t\rd\t'~
    \Big[A(x(\t'))\frac{\rd\tilde{A}(\tx(\t'))}{\rd\t'}
    -\frac{\rd A(x(\t'))}{\rd\t'}\tilde{A}(\tx(\t'))\Big].
 \label{e:AdotA}
\end{alignat}
The first of these includes a (worldline-local) mixing term, together with its value at some ``reference'' proper time, $\t_*$, the integral of which evaluates in~\eqref{e:SMP2} merely to an additive constant and so is irrelevant. The second part, $[A\dot{\tilde{A}}\,{-}\,\dot{A}\tilde{A}]$, is far more interesting, as it provides a telltale ``angular momentum''-like coupling of $(A,\tilde{A})$-pairs to corresponding ``external/background'' fluxes, represented here by the coefficient ``$a\l^2$.'' This relies on identifying $A,\tilde{A}$ as canonical coordinates in the target-spacetime (classical) field theory. Alternatively, in the underlying worldsheet quantum field theory, the $A,\tilde{A}$ are coefficients of certain quantum states, for which then the $\brb{A}{\tilde{A}}$-term likewise accompanies a Berry-phase like quantity.

The above integral looks very much like the Bohr-Sommerfeld
$\int\!p\,\rd q$ integral, and which, apparently, 
also gets quantized in our case, if we plug this integral
in the whole path integral. 
Bohr-Sommerfeld (semiclassical) quantization has physical consequences: 
quantization of classical energy levels etc. 
So, it is natural to expect something similar in 
our case. First the fields and their duals 
are ``canonically conjugate'' like $q$ and $p$. 
Second, the fields and their duals exist only 
in certain state configurations, at least to first 
order in the appropriate quantization parameter. 
This reasoning goes beyond the rules of effective 
field theory. 

In this sense, the dual particles are conjugate to
the Standard Model particles in analogy with the canonical fact that
momenta $p$ are conjugate to coordinates $q$. 
That is, the dual fields do not commute with Standard Model fields, and 
the two form an effective Heisenberg algebra.
Naturally, that target-space/field-space ``canonical conjugateness'' is 
induced from the domain-space canonical commutation relation between  $x$  and $\tx$.
That is the ``domain-space'' symplectic structure induces a symplectic structure in the target space. 
This is ``obvious'' if one thinks of the target space as being ``generated'' by the fields (as their vevs, 
after a choice of an action and path integral).
Thus the actual canonical conjugateness of ``$p$'' and ``$q$'' induces a conjugateness 
between any given function  $f(p)$  and certain special function  $g(q)$.
In our case, the ``specialness'' of  $\tilde{A}(\tx)$  as related to  $A(x)$  
is that they both stem from the same bi-local function  $A(x,\tx)$.

Note the relevance of the very light dual degrees of freedom in the observed spacetime
in connection with the $H_0$ tension, which could be resolved by changing the
effective number of light dark matter degrees of freedom --- dark radiation.
Such light degrees of freedom could be provided by the dual degrees of freedom that are natural in our discussion.
However, we also note that the vacuum structure of the full metafield description goes
beyond the background field description used in this section\cite{Freidel:2021wpl}.

We note that in the Coulomb gauge (and its dual), the action~\eqref{e:SMP2} becomes:
\be
 \int_{x,\tx} \Big[(\partial A)^2  -a\l^2\brb{A}{\tilde{A}} +(\tilde{\partial}\tilde{A})^2 +\partial A\,\tilde{\partial} \tilde{A} +\dots \Big].
\ee
On the other hand, if we second-quantize the metaparticle action we should end up
with the following structure fixed by Born geometry
\be
\int_{x,\tx} \Big[(\partial A)^2  +(\tilde{\partial} {A})^2
 -a\l^2\brb{A}{\tilde{A}} +(\tilde{\partial} \tilde{A})^2 +({\partial} \tilde{A})^2
 +\partial A\,\tilde{\partial} \tilde{A} +\dots\Big].
\ee
By integrating over the dual spacetime ($\tx$), and by setting, for consistency, $\tp \to 0$ in the
observable $3{+}1$-dimensional $x$-spacetime, we finally obtain (with indices on the respective gauge fields fully restored)\cite{Berglund:2020qcu}
\be
\int_x \Big[(\vd_{[a} A_{b]})^2  -\frac{\l_4\!^2}{L^4}\mathcal{F}^{ab}\brb{A_a}{\tilde{A}_b} 
+({\vd_{[a}} \tilde{A}_{b]})^2 +\dots\Big],\qquad
\tilde{A}_a\define \eta_{ab}\,\tilde{A}^b
\label{e:SMPv}
\ee
The effective length scale $\l_4$ relevant in the observed $3{+}1$-dimensional $x$-spacetime indeed stems from the fundamental length-scale $\l$ from~\eqref{e:CnCR1}, but is rescaled by the restriction to the observable $3{+}1$-dimensions --- which may well involve an exponentially driven hierarchy, such as seen in~\eqref{GN}.
 We have also introduced the size, $L$, of the observed $3{+}1$-dimensional $x$-spacetime, whence its $L^4$ is the scale of its 4-dimensional volume.\footnote{As seen in the toy model discussed in Section~\ref{s:dSdefo} and as will be discussed again below, different cosmological scenarios result in differing estimates for both $\l_4$ and $L$, so here we leave them unspecified.}
Because the ``mixing'' term $\brb{A_a}{\tilde{A}^a}$ is properly normalized on the world-line,
in spacetime it must be multiplied by the volume of spacetime, and so has to be rescaled by
the volume of the observed spacetime, $L^4$.
This term is thereby sensitive both to the UV cutoff $\l_4$ and the IR cutoff $L$, both of which as induced for the observed $3{+}1$-dimensional spacetime. Finally, the dimensionless quantity $\mathcal{F}^{ab}$ encodes the Berry phase-like ``background fluxes.''

In other words, in the $3{+}1$-dimensional observable spacetime, one has both the visible sector and its dual/dark counterpart, coupled via this very particular ``mixing/correlation'' between the two, which is sensitive
to both the UV and IR cutoffs. This correlation is invisible to effective field theory, and it
vanishes as either $\l_4\to 0$  or $L \to \infty$. 
This is consistent with the general set-up of the metastring which has two
cutoffs (UV and IR) and which generically should be defined with self-dual RG fixed points.
Note that this would mean that Standard Model of particle physics (of visible matter) has a dual counterpart 
(a dual Standard Model describing dark matter)
and that these two are correlated, via the fundamental length.
We remark that such a correlation between visible and dark matter involving an IR scale (in this case, the Hubble scale)
 has been observed in astronomical data and has been studied in the context of modified dark matter
in\cite{Ho:2010ca,Ho:2011xc,Ho:2012ar,Edmonds:2013hba,Edmonds:2016tio,Ng:2016qvh,Edmonds:2017zhg,Edmonds:2017fce,Edmonds:2020iji,Edmonds:2020tug}.
However, the ratio $\l_4\!^2/L^4$ should not be considered
to be necessarily trans-Planckian as would be the case if $\l_4$ were the Planck length,
and $L$ the Hubble length. Instead, as already discussed in Section~\ref{s:DS+Hierachy}, these two scales are effective UV and
IR scales in the observed $3{+}1$-dimensional spacetime $\sW^{1,3}$, both of which depend on the details of how this $\sW^{1,3}$ ``sits'' within the total $9{+}1$-dimensional spacetime. This question has already arisen in Section~\ref{s:Class}, and we will return to it again in Sections~\ref{s:Pheno2}.
Still, the appearance of such scales is a direct evidence of the breakdown of
effective field theory.

With the general idea for the vector fields $A_a$ and $\tilde{A}^a$, the corresponding result for scalars $\phi^i$ and their duals $\tilde{\phi}_j$\cite{Berglund:2020qcu} is immediate:
\be
\int_x \Big[\|\partial \phi\|^2 + \|{\partial} \tilde{\phi}\|^2
  - \frac{\l_4\!^2}{L^4}\,\mathfrak{F}_{ij} \brb{\phi^i}{\tilde{\phi}^j} +\dots\Big]
\label{e:SMPb}
\ee
as such terms would be forced in every dimensional reduction framework.
Thus, a Higgs field would have a dual Higgs field and the two would be correlated via the fundamental length.
Note that we could also consider pseudoscalars (axions) $a$ and their duals $\tilde{a}$
\be
\int_x \Big[(\partial a)^2  - \frac{\l_4\!^2}{L^4} \brb{a}{\tilde{a}}
+  ({\partial} \tilde{a})^2 +\dots\Big]
\label{e:SMPb1}
\ee
Once again, an axion field would be correlated to a dual axion field via the fundamental length scale.
We should mention that axions can be viewed as
``boost generators'' (at least for constant profiles) between the observed and dual spacetimes\cite{Freidel:2013zga}.
This can be seen by observing that the constant Kalb-Ramond field can be absorbed into a non-trivial symplectic form
(on its diagonal) after an $O(d,d)$ rotation\cite{Freidel:2013zga}.
Thus the Kalb-Ramond two-form enters into an explicit non-commutativity of the modular spacetime, and
it can be used to rotate between observed and dual spacetime coordinates (as an explicit illustration of
relative, or observer-dependent, locality)\cite{Freidel:2013zga}.
The Kalb-Ramond two-form dualizes into a pseudoscalar in 4d, and thus the 4d axion has the same features.
More generally, for non-constant Kalb-Ramond profiles one ends up with a non-associative structure\cite{Freidel:2013zga}.
Therefore, in general, axions are indicators of non-commutative (when constant) and non-associative (when propagating) structures in modular spacetime.

Similarly, the corresponding action for the fermions (by taking the ``square root'' of the propagating part of the scalar action to accommodate 
the spin-statistics theorem)\cite{Berglund:2020qcu}:
\be
\int_x \Big[ i(\bar{\psi}/\!\!\!\partial \psi)
            +i(\bar{\tilde{\psi}}/\!\!\!\partial\tilde{\psi})
  -\frac{\l_4\!^3}{L^4} \mathfrak{F}_{ij}
    \big(\brb{\bar\psi^i}{\tilde{\psi}^j}+\textit{h.c.}\big) +\dots\Big],
\label{e:SMPf}
\ee
where $\brb{\bar\j^i}{\tilde\j^j}=\frc12(\bar\j^i\tilde\j^j{-}\bar\j^j\tilde\j^i)$ are non-derivative bilinear terms, which couple to the same ``external'' fluxes as the bosons, akin to a worldline super-Zeemann effect\cite{Doran:2008xw}.
 The use of supersymmetry to relate~\eqref{e:SMPb} and~\eqref{e:SMPf} is justified as these are free fields in flat spacetime; it is the omitted interaction terms (including metric and curvature deviations from flat spacetime) that break supersymmetry.
 From the underlying (worldline, or even worldsheet\cite{Freidel:2013zga}) point of view, such terms are induced from generalizing~\eqref{e:SMP} along the by now standard GLSM construction\cite{rPhases,Distler:1993mk}: Including additional, gauged symmetries and degrees of freedom as well as (super)potentials induces the interaction terms indicated by ellipses in~\eqref{e:SMPv}, \eqref{e:SMPb} and~\eqref{e:SMPf}, and is classically equivalent to generalizing~\eqref{e:SMP} into a nonlinear $\sigma$-model. The ``free-field-limit'' terms shown herein however remain unchanged.

The $\l_4\!^3/L^4$ coefficient in the fermionic ``mixing'' term is forced on dimensional grounds, by the canonical 3/2 dimension for the fermion field.
This term is quite explicitly a mass-mixing term, and may well induce a new type of a ``seesaw mechanism,'' perhaps even tunable to lead to naturally small neutrino masses. Also, the inclusion of several mass-scales enables concrete models to incorporate a hierarchy of ``seesaw mechanisms,'' with a more reasonable chance to approach the intricacies of a realistic mass spectrum. This relates back to our discussion at the very end of Section~\ref{s:MS,CC,Higgs} and to the phenomenological studies\cite{Bordes:2014eaa,rBJ-MM,Faraggi:1996pa}.

To summarize, the leading ``kinetic'' parts in the actions~\eqref{e:SMPb} and~\eqref{e:SMPf}, together with the ``mixing terms,'' $\brb{\phi}{\tilde\phi}$ and $\brb{\bar\psi}{\tilde\psi}$, are seen to be natural:
 ({\small\bf1})~by dimensional reduction from~\eqref{e:SMPv} to~\eqref{e:SMPb}, and
 ({\small\bf2})~by supersymmetry from~\eqref{e:SMPb} to~\eqref{e:SMPf}.
In turn, the \BHM\ system\cite{rBHM7} then induces supersymmetry breaking via the interaction terms, represented by the ellipses in~\eqref{e:SMPv}, \eqref{e:SMPb} and~\eqref{e:SMPf}, which is not unlike the Polonyi mechanism\cite{Polonyi:1977pj}; for a recent discussion, see also\cite{Ketov:2018uel,Aldabergenov:2019aag}.

In conclusion of this subsection, we note that the peculiar correlation between the visible and dark sectors, discussed for scalar, pseudoscalar, fermionic
and vector degrees of freedom, can be also found in the gravitational and dual gravitational sectors.
Thus the observed gravity and dark energy are correlated via the scale of non-commutativity.
This might have interesting observable effects for the so-called $H_0$ tension\cite{Verde:2019ivm},
as discussed elsewhere\cite{jejjala2020dynamical}.

\subsection{Phenomenological Implications }
\label{s:Pheno2}
In this section we comment on various phenomenological issues.
The most important generic predictions of the generic string theory\cite{Freidel:2013zga}
is that the geometry of the dual spacetime determines the dark energy sector\cite{Berglund:2019ctg} and that the dual
matter degrees of freedom naturally appear as dark matter degrees of freedom, as discussed in the preceding section.
We note that, quite explicitly, the dark matter sector provides ``sources'' for the visible matter
sector. This follows from the coupling $ \frac{\l_4\!^2}{L^4}\brb{\phi}{\tilde{\phi}}$, 
as predicted by the doubled/non-commutative
set-up. This provides an explicit correlation between the dark matter sector and a visible sector via
the fundamental length. 
Given the seesaw formula for the dark energy which relates the dark energy scale to the fundamental
length, which could be taken to be the Planck energy scale, then the dark matter is also sensitive to the dark energy.
So, the visible matter, dark matter and dark energy are all related.
This is consistent with the observational evidence presented in\cite{Ho:2010ca, Ho:2011xc, Ho:2012ar, Edmonds:2013hba, Edmonds:2016tio, Ng:2016qvh, Edmonds:2017zhg, Edmonds:2017fce, Edmonds:2020iji, Edmonds:2020tug},
already alluded to in this review.

We emphasize that in our discussion of the hierarchy problem
the UV and IR scales are radiatively stable, and so is their product, the Higgs scale.
This new view on the hierarchy problem, that goes beyond the usual tools of effective field theory,
because of the explicit presence of the widely separated UV and IR scales.
The usual suggested approaches to the hierarchy problem: technicolor, SuSy and extra dimensions are
all within the canonical effective field theory.
In the context of string theory, effective field theory (and the approach to the
hierarchy problem via a SuSy effective field theory) can be naturally found via
string compactifications, but in that case one is faced with the issue 
of supersymmetry breaking
(and the fundamental question of ``measures'' on the string 
landscape/swampland\cite{Danielsson:2018ztv, Obied:2018sgi, Andriot:2019wrs, Vafa:2005ui, Palti:2019pca}.)
We claim that these issues are transcended in the general, chirally doubled and non-commutative formulation
of string theory with a fundamentally bosonic and non-commutative formulation, in which spacetime and matter
(as well as supersymmetry at the Planck scale)
can be viewed as emergent phenomena.

Next we comment on the seesaw formula, which mixes UV and IR scales, and the neutrino sector.
Such a seesaw would involve the neutrino and its dual partner --- the dual neutrino. 
Quite generically, the dual sector is the source for the visible sector and the overall effect is
to make the visible sector essentially massive.
If we could argue for such a formula, this would provide a new way of generating neutrino masses. 
Actually, this does provide a curious mixing ``mass'' effect --- even if $\l_4\!^3/L^4$ is very small. To see this, consider the well-known, standard diagonalization of a typical mass-matrix restricted to a fermion and its dual, $\left[\begin{smallmatrix}m&\l_4{}^3/L^4\\\l_4{}^3/L^4&\widetilde{m}\end{smallmatrix}\right]$:
\begin{equation}
  \frc12\Big[(m+\widetilde{m})\pm\sqrt{(m-\widetilde{m})^2+4(\l_4{}^3/L^4)^2}\Big] 
  \approx\Big\{\begin{matrix}m{+}\d m\\\widetilde{m}{-}\d m\end{matrix},\qquad
  \d m\define\frac{\l_4{}^6}{L^8(m{-}\widetilde{m})}+\dots
\end{equation}
What is worth observing is that the mixing contribution in $\d m$ is divided by the difference $(m{-}\widetilde{m})$, so that the mass-gap, $2\d m$, may well be phenomenologically relevant ---  even for a minuscule numerical value of $(\l_4{}^3/L^4)^2$. This clearly involves model-dependent details, and must be left to a case-by-case analysis.

Next, we remark on other phenomenological issues associated with the dual Standard Model.
In the dual QED sector, we should find a dual photon that is correlated to the visible photon, and
that is distinct from the dark photon of effective field theory. The correlation is proportional to
the fundamental length, and it finite even in the limit of zero momenta (deep infrared).
Also, the usual visible photon/dark photon coupling is subdominant to this term that
is inherent in our story. 
Similarly, in the dual of the weak sector of the Standard Model, we have a dual of the visible $Z$.
This dual of $Z$ should be distinguished from the usual $Z'$ by its sensitivity to the fundamental
length and by its correlation to the visible $Z$. 
These type of correlations might be found in correlated events in the accelerators, that are
not products of any standard particle decays.

Also, in the dual QCD we should find interesting phenomenology in the deep infrared, even though that
is a very difficult region to study in QCD. 
In particular, given the new view of the axion field in the
above discussion,
we have a novel and possibly useful viewpoint of the strong CP problem in QCD:
The first observation here is that according to\cite{Freidel:2013zga}, 
the constant Kalb-Ramond field mixes $x$- and $\tx$-spacetimes (it acts as a boost that linearly combines the spacetime and its dual in the context of a larger doubled and non commutative quantum spacetime). Also the commutator of dual spacetime coordinates is given by the constant $B$ Kalb-Ramond 2 form. So, for $B=0$ we get just the observed (4d) spacetime. Also its $H=\rd B$ field strength is trivially zero. But $H$ is dual (in 4d) to 
the axion ($a$), which is also constant. But $B$ is zero (there is no preferred background direction) and so this constant axion may be interpreted as a uniform distribution for the axion (whose constant values can be positive and negative). 
Now, let us assume that our axion $a$ is the QCD axion, which is relevant for the strong CP problem. The CP violating term is proportional to the axion ($a F\wedge F$). So, if we average this term which is linear in the axion over a uniform distribution for this axion we get zero ($\int_{-k}^k\rd a~ a=0$, with $k \to \infty$).
In order to complete this argument we would have to study small fluctuations of the axion field in order to understand the robustness of this new viewpoint on the strong CP problem.

In conclusion of this subsection, let us comment on the problem of inflation in string theory.
It is known that string theory does not lead to natural slow-roll parameters (see\cite{Baumann:2014nda}), and thus, inflation
is somewhat problematic in string theory. On the other hand, it is known that the original Starobinsky inflation,
which can be also viewed as an example of $f(R)$ gravity (with a specific, and renormalizable, $R^2$ term)
does lead to very successful comparison with observation; see\cite{Starobinsky:1980te} and\cite{Mukhanov:1981xt}). As is well known $f(R)$ gravity can 
be rewritten as ordinary Einstein gravity coupled to a scalar field with a particular potential.
Starobinsky's choice of the $R^2$ terms leads to a very specific inflationary potential, the inflaton being that
effective scalar field. Given that phenomenological success,
various physicists have tried to connect Starobinsky inflation to particle physics, and in particular, to the Higgs sector,
the Higgs field being the natural scalar degree of freedom in the standard model. This is the so called
Higgs inflation approach\cite{Bezrukov:2007ep}. However, there are
problems with fine tuning, and the relation between the Higgs inflation and the Starobinsky inflation is
not completely clear\cite{Bezrukov:2007ep}. Now, given our discussion
regarding the dual sector, the Higgs sector has a natural dual, and this dual scalar field (the dual Higgs) 
can provide for a natural inflaton, that leads to an effective Starobinsky-like inflation.
The central reason here is that in the dual sector we have other parameters, and other scales that
could be adjusted to yield the natural connection with the Starobinsky inflation.
Therefore, another phenomenological avenue of our discussion, which also relates to the problem of de~Sitter
space (given the quasi-de-Sitter nature of the inflationary phase), is that the dual Higgs degree
of freedom may serve as a Starobinsky-like inflaton. This has a reasonable chance of resolving the apparent tension between Starobinsky inflation and supersymmetry.

\subsection{Re-Bosonized Heterotic Strings and the Standard Model}
\label{s:nC-hetS}
The chirally doubled $(26{+}26)$-bosonic description of stringy target spacetime\cite{Freidel:2015pka} may be adapted to discussing all stringy models, superstrings included, by reexamining the template of the partial and asymmetric (heterotic) fermionization\cite{Casher:1985ra,Gross:1984dd,Gross:1985fr,Gross:1985rr}; see also\cite{Siegel:1993th}.
 These standard constructions build around treating left- and right-moving degrees of freedom as completely independent and commuting, which is the $\w\!\to\!0$ (and so $\l\!\to\!0$) limit of the inherently non-commutative metastring\cite{Freidel:2015pka,Freidel:2017xsi}. Via the relations $x^a{:=}\frc1{\sqrt2}(x^a_{\sss L}{+}x^a_{\sss R})$ and $\tx_a{:=}\frc1{\sqrt2}(x^a_{\sss L}{-}x^a_{\sss R})$, non-commutativity~\eqref{e:CnCR1} implies that $[x^a_{\sss L},x^b_{\sss R}]=2\pi i\l^2\d^{ab}$.

\paragraph{The Standard Construction:}
In the $\l\to0$ limit, the construction of the heterotic string may be depicted by the following (standard light-cone restricted) diagram:
\begin{equation}
  \vcenter{\hbox{\begin{tikzpicture}[xscale=.9]
     \path[use as bounding box](-.7,.1)--(15.4,-1.1);
      \foreach\x in {0,...,25} \path(\x*.5,0)node{\scriptsize\x};
      \path(-.5,-.5)node{$x_{\sss L}$};
       \foreach\x in {0,...,25} \draw(\x*.5,-.5)circle(.5mm);
       \draw[blue](4.8,-.3)rectangle++(3.9,-.4);
       \draw[blue](8.8,-.3)rectangle++(3.9,-.4);
       \path[blue](13.9,-.5)node{\footnotesize YM: $E_8{\times}E_8$};
      \path(-.5,-1)node{$x_{\sss R}$};
       \foreach\x in {0,...,25} \draw(\x*.5,-1)circle(.5mm);
       \draw[red](4.8,-.8)rectangle++(7.9,-.4);
       \path[red](14.15,-1)node{\footnotesize SuSy: $T_M{+}T^*_M$};
       \draw[green!75!black](.8,-.3)rectangle++(3.9,-.9);
       \path[green!75!black](2.75,-.75)node{\scriptsize$M$};
                 \end{tikzpicture}}}
 \label{e:HS}
\end{equation}
Here the 16 chiral bosons $x_{\sss R}^{10},\cdots\!,x_{\sss R}^{25}$, {\color{red}red-boxed} in~\eqref{e:HS}, are ``fermionized,'' and their dynamics is restricted so as to act as the on-shell $8{+}8$ worldsheet $(0,2)$-superpartners to the $(x^2,\cdots\!,x^9)$, {\color[rgb]{0,.6,0}green-boxed} in~\eqref{e:HS}, in the light-cone gauge:
\begin{equation}
  \Bigg( \genfrac{\{}{\}}{0pt}{}
          {x_{\sss L}^2,\cdots\!,x_{\sss L}^9}{x_{\sss R}^2,\cdots\!,x_{\sss R}^9}
   \!=\! \genfrac{\{}{\}}{0pt}{}
          {x^2,\cdots\!,x^9}{{\color{gray!60!cyan}\tx^2,\cdots\!,\tx^9}} \Bigg)
 \overset{\sss\rm SuSy}\longleftrightarrow
  \Bigg( \genfrac{\{}{\}}{0pt}{}
          {\color{red}x_{\sss R}^{10},\cdots\!,x_{\sss R}^{25}}
          {{\color{red}x_{\sss R}^{10},\cdots\!,x_{\sss R}^{25}}}
                \!\dec{$\sss B$}{$\sss F$}{\leftrightharpoons}\!
          \genfrac{\{}{\}}{0pt}{}
          {\color{red}\j_{\sss R}^2,\cdots\!,\j_{\sss R}^9}
          {{\color{red}\bar\j_{\sss R}^2,\cdots\!,\bar\j_{\sss R}^9}} \Bigg).
 \label{e:mSuSy1}
\end{equation}
The so-formed eight worldsheet $(0,2)$-supermultiplets,
 $(x^\mu|\j_{\sss R}^\mu,\bar\j_{\sss R}^\mu)$ with $\mu\,{=}\,2,\cdots\!,9$
explicitly omit using the corresponding $\tx^\mu$ ({\color{gray!60!cyan}pale blue} in~\eqref{e:mSuSy1}).
 Regarding the target space in turn, the 2-component worldsheet Majorana-Weyl fermions, $\j_{\sss R}^\mu$ ($\mu=2,\cdots\!,9$), and their Dirac conjugates, $\bar\j_{\sss R}^\mu$, span the 8-dimensional spinor of the Lorentz light-cone subgroup $\SO(8)\subset\SO(1,9)$. Identified via $\SO(8)$ triality with $\SO(8)$-vectors, these span the tangent and cotangent vectors at the target spacetime point indicated by the $x^\mu$.
 Although highly nonlocal and nonlinear, the chiral boson to fermion transformation is a 1--1 mapping\cite{Coleman:1974bu,Mandelstam:1975hb}; see also\cite{Witten:1984vr,Harada:1989qp,vonDelft:1998pk,Senechal:1999us}. Correspondingly, the linear worldsheet supersymmetry has a correspondingly nonlinear and nonlocal ``re-bosonized'' preimage:
\begin{equation}
 \big( \underbrace{x^\mu\overset{\sss\text{``SuSy''}}\longleftrightarrow
      (x^{\mu+8}_{\sss R},x^{\mu+16}_{\sss R})}_{M~\leftrightarrow ~T_M+T^*_M\quad}
       \TikZ{\path[use as bounding box](0,0);
              \draw[blue, thick, stealth-stealth](-1,-.1)to[out=-60,in=180]++(1,-.6)
               --node[above=-2pt]{\scriptsize boson/fermion-ization}
               ++(2.5,0)to[out=0,in=-120]++(1,.6);
            }
       \big)
  ~~\dec{$\sss B$}{$\sss F$}{\leftrightharpoons}~~
 \Big( x^\mu\overset{\sss\rm SuSy}\longleftrightarrow
 \big( \j^\mu_{\sss R}[x^{\mu+8}_{\sss R},x^{\mu+16}_{\sss R}],
        \bar\j^\mu_{\sss R}[x^{\mu+8}_{\sss R},x^{\mu+16}_{\sss R}]\big) \Big),
 \label{e:mSuSy2}
\end{equation}
complete with a corresponding preimage of all supersymmetry features. In particular, this identifies/restricts $\{x^{10}_{\sss R},\cdots\!,x^{25}_{\sss R}\}$ as spanning the tangent and cotangent space to the target (sub)spacetime, $M$, coordinatized by $x^\mu$. While the existence of this identification is evident, its details remain to be worked out along\cite{Freidel:2015pka,Freidel:2017xsi}, and compared with the various known supergravity constraints in the standard formalism.
 This meta\-string preimage of supersymmetry thus involves {\em\/solitons,} $\j^\mu_{\sss R}$ and their Dirac conjugates, all of which are represented by exponential functionals\cite{Mandelstam:1975hb}.

On the other hand, the left-moving 16 chiral bosons $x_{\sss L}^{10},\cdots\!,x_{\sss L}^{25}$, {\color{blue}blue-boxed} in~\eqref{e:HS}, are compactified on the $E_8{\times}E_8$ (or $\SO(32)/\ZZ_2$) lattice, and provide for the total of 496 massless YM-gauge degrees of freedom owing to the lattice-specified twisted boundary conditions. These 16 chiral bosons correspond to the simple positive roots of the gauge group, so that group elements are generated by these $x_{\sss L}$'s, and are again represented by exponential functionals.

The chiral boson sketch~\eqref{e:HS}--\eqref{e:mSuSy2} is limited both to the ``transverse'' degrees of freedom in the light-cone gauge, and that the worldsheet $X_{\sss L,R}$'s were all originally assumed to be mutually commuting free fields, with but twisted boundary conditions and the identifications~\eqref{e:HS}--\eqref{e:mSuSy2} required for supersymmetry. As noted in\cite{Gross:1985fr} and explicitly shown in\cite{rCHSW}, the free field dynamics may be generalized to Ricci-flatness, which is consistent with the overall requirements\cite{rFrGaZu86,rBowRaj87,rBowRaj87a,rBowRaj87b,rHHRR-sDiffS1,Pilch:1987eb}.
 In such generalizations, the sketch~\eqref{e:HS}--\eqref{e:mSuSy2} can only be understood to represent {\em\/local\/} coordinates on any given chart on any topologically nontrivial (non-contractible) geometry.

This standard description thus not only omits the $\tx^\mu$ (with $\mu\,{=}\,0,\cdots\!,9$ outside the light-cone gauge) but {\em\/relies\/} on the commutative limit $[x^a,\tx^b]=2\pi i\l^2\d^{ab}\to0$ for all $a\,{=}\,0,\cdots\!,25$.

\paragraph{Born Mirror-Doubling:}
However,  $[x^a,\tx^b]\,{=}\,2\pi i\l^2\d^{ab}\,{\neq}\,0$, and this implies a specific structure to the geometry of the target spacetime.

The overall structure of the metastring target spacetime, $\mathscr{M}$, is chirally doubled into an ``almost symplectic and para-Hermitian manifold''\footnote{Such spaces admit a signature-$(d,d)$ metric $\eta$ and a symplectic structure $\w$ such that $K\!:=\!\eta{\cdot}\w$ is an {\em\/almost product structure,} $K^2=+{\bf 1}$. This $\eta$ need not be flat, and ``almost'' means that $\w$ need not be a closed 2-form. Owing to the almost symplectic structure, the existence of the corresponding flat (Bott) connection guarantees that a foliated space is everywhere locally a product of two half-dimensional affine spaces.} that has a compatible {\em\/foliation\/}: Locally at every point, $x$, this doubled spacetime factorizes, $\mathscr{M}_x\,{=}\,M_x\,{\times}_x\,\Tw{M}_x$, the local coordinates for the two factors being $(x^a,\tx_b)_p$\cite{Freidel:2018tkj}, equipped with a phase-space like structure. For local diffeomorphisms (implemented by the Dorfman generalization of the Lie derivative) to be integrable to finite translations, it is necessary to impose the so-called ``section condition,'' which halves the spacetime akin to the quantum-mechanical restriction of the classical phase space, $(p,q)$, in the coordinate or momentum representation --- or indeed any other $\p\!:=\!(\a p{+}\b q)$-{\em\/polarization,} as familiar from the Geometric Quantization program\cite{rNH-GQ,rNW-GQ}.

It is then a hallmark of this structure that $\p[T\mathscr{M}]=(T{+}T^*)\p[\mathscr{M}]$: the polarization of (the total space of) the tangent bundle on $\mathscr{M}$ is the generalized/doubled tangent bundle on a polarization of $\mathscr{M}$. 
 In local coordinates, elements of $T\mathscr{M}$ in the $\p_x$-polarization are given as $v^a(x,\tx)\vd_a{+}w_a(x,\tx)\rd x^a$, since $\tx$ are locally constant ($\tw\vd^a,\rd\tx_a\!\mapsto\!0$) on $\p_x[\mathscr{M}]=M_x$\cite[Eq.\,(5.1)]{Freidel:2018tkj}. 
 In turn, the same element is $v^a(x,\tx)\rd\tx_a{+}w_a(x,\tx)\tw\vd^a$ in the $\p_\tx$-polarization, since on $\p_\tx[\mathscr{M}]=\Tw{M}_\tx$ the $x$ are locally constant ($\vd_a,\rd x^a\!\mapsto\!0$).
 Thus, swapping the polarization $M_x\leftrightarrow \Tw{M}_\tx$ explicitly identifies $T_{M_x}=T^*_{\Tw{M}_\tx}$ and $T^*_{M_x}=T_{\Tw{M}_\tx}$ --- which is the underlying premise of mirror symmetry. As Calabi-Yau manifolds (compact or not), $M$ and $\Tw{M}$ additionally admit compatible complex structures and K{\"a}hler Ricci-flat metrics. Whether these features are also guaranteed by Born geometry or merely compatible with it is an interesting but for now open question.

The Born geometry of the metastring (chirally and non-commutatively doubled) target spacetime thus explicitly includes mirror symmetry by virtue of having to ``polarize'' on one half of the spacetime, such that polarizing on the complementary half swaps the corresponding tangent and cotangent spaces. This realization of mirror symmetry as $T$-duality seems rather more direct and automatic in the target spacetime than the earlier proposals\cite{Strominger:1996it,Polchinski:1998rr}.

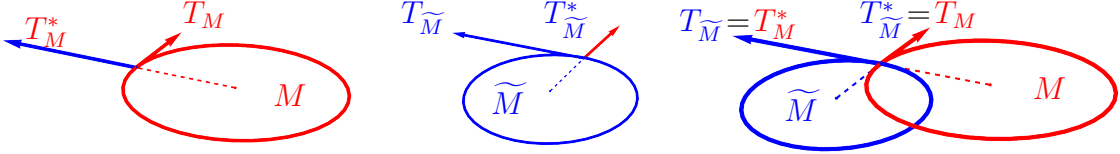
\begin{figure}[htbp]
 \begin{center}
  \begin{picture}(160,15)(-3,7)
   \put(0,-1){\includegraphics[width=50mm]{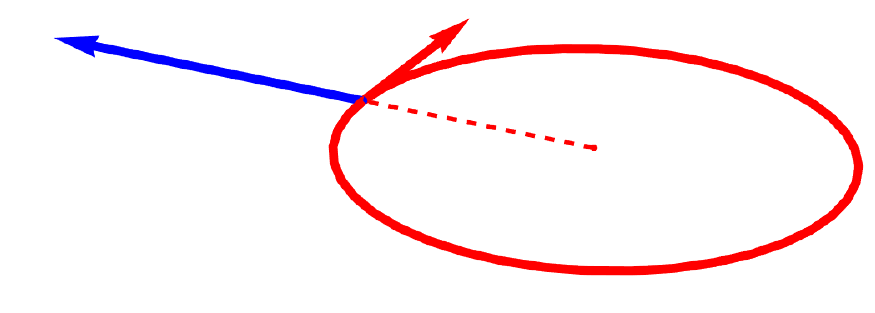}}
    \put(39,6){\color{red}$M$}
    \put(27,17){\color{red}$T_M$}
    \put(6,15){\color{red}$T^*_M$}
   \put(60,0){\includegraphics[width=50mm]{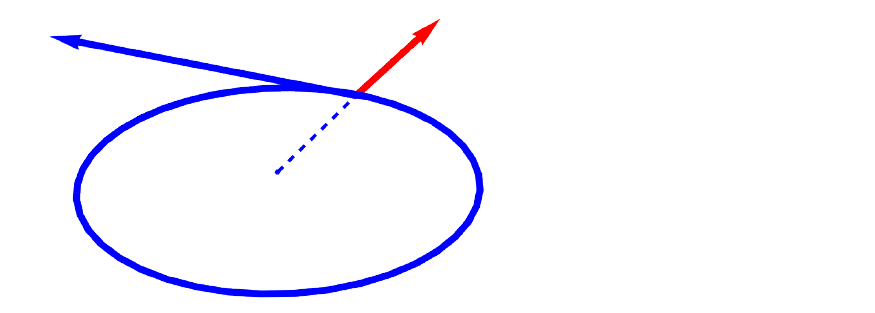}}
    \put(68,5){\color{blue}$\Tw{M}$}
    \put(75,17){\color{blue}$T^*_{\Tw{M}}$}
    \put(56,17){\color{blue}$T_{\Tw{M}}$}
   \put(97,-2){\includegraphics[width=55mm]{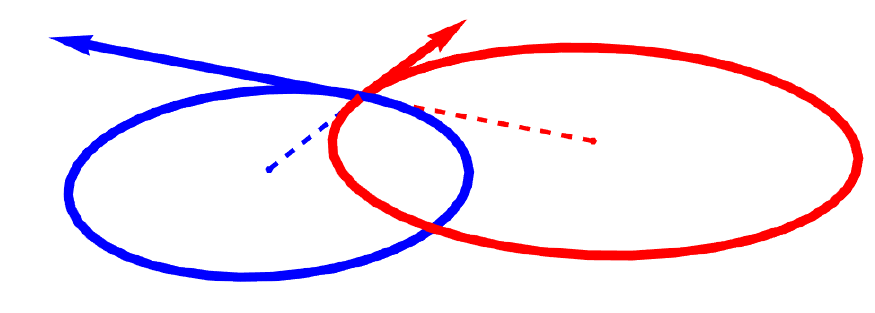}}
    \put(140,7){\color{red}$M$}
    \put(107,4){\color{blue}$\Tw{M}$}
    \put(93,16){${\color{blue}T_{\Tw{M}}}\!=\!{\color{red}T^*_M}$}
    \put(117,17){${\color{blue}T^*_{\Tw{M}}}\!=\!{\color{red}T_M}$}
  \end{picture}
 \end{center}
 \caption{The original $(M,T_M,T^*_M)$, its mirror $(\Tw{M},T_{\Tw{M}},T^*_{\Tw{M}})$, and both depicted together}
 \label{f:M+MM}
\end{figure}

\paragraph{Born Heterotic Strings:}
The foregoing realization may be applied to the heterotic Ansatz~\eqref{e:HS} by swapping the polarization in the factor spanned by the first third of fields, $\{x^a,\,a\,{\leqslant}\,9\}$, while adapting the roles of the remaining 2/3 of fields, $\{x^a_{\sss L,R},\,a\,{\geqslant}\,10\}$ as forced by Born geometry: For example, if say $x^{\mu+8}_{\sss R}$ were identified (by ``re-bosonized'' supersymmetry) to span $T_{M_x}$ to $M_x$ coordinatized by $x^\mu$, then they must span $T^*_{\Tw{M}_\tx}$ to $\Tw{M}_\tx$ coordinatized by $\tx_\mu$.
These relationships are sketched in Figure~\ref{f:M+MM}, which shows that this structure corroborates the argument presented in\cite[\SS\,5.1]{rBHM7}.
 In Calabi-Yau compactifications, the spacetime spanned by $\{x^a,\,a\,{\leqslant}\,9\}$ factors as $M_x\,{=}\,\IR^{1,3}\,{\times}\,\sY^6$, where the polarization swapping is trivial on the $\IR^{1,3}$-factor, but implements the nontrivial mirror symmetry on the $\sY^6$ factor.

Recalling that the original heterotic Ansatz~\eqref{e:HS} assumed the various free fields to be mutually commuting, and to a large extent mutually interchangeable, we consider here the following tripartite heterotic Ansatz,\footnote{In the light-cone gauge, the initial variables $x^0,x^1,\tx^0,\tx^1$ are identified with the worldsheet (``WS'') coordinates, i.e., ``longitudinal.'' In this gauge, the coordinates transversal to the worldsheet are the physical degrees of freedom. In turn, this implies that the worldsheet itself should be coordinatized in this chirally doubled fashion for consistency.} shown already with the Born-mirror identifications:
\begin{equation}
  \vcenter{\hbox{\begin{tikzpicture}
     \path[use as bounding box](-1.2,.5)--(13.2,-1.6);
      \foreach\x in {0,...,25} \path(\x*.5,.4)node{\scriptsize\x};
      \path(-.5,-.5)node{$x$};
       \foreach\x in {0,...,25} \draw(\x*.5,-.5)circle(.5mm);
      \path(-.5,-.95)node{$\tx$};
       \foreach\x in {0,...,25} \draw(\x*.5,-1)circle(.5mm);
       \draw[green!70!black,densely dashed](-.2,-.3)rectangle++(.9,-.9);
        \path[green!70!black](.3,-.1)node{\scriptsize WS};
       \draw[red,densely dashed](.8,-.3)rectangle++(3.9,-.4);
        \path[red](2.75,-.1)node{\scriptsize light-cone coordinates on $M$};
       \draw[red](4.8,-.3)rectangle++(3.9,-.4);
        \path[red](6.75,-.1)node{\scriptsize light-cone
         $T_M={\color{blue}T^*_{\Tw{M}}}$-fiber};
       \draw[red](8.8,-.3)rectangle++(3.9,-.4);
        \path[red](10.75,-.1)node{\scriptsize Simple pos.\ roots of $E_8$};
       \draw[blue,densely dashed](.8,-.8)rectangle++(3.9,-.4);
        \path[blue](2.75,-1.45)node{\scriptsize light-cone coordinates on $\Tw{M}$};
       \draw[blue](4.8,-.8)rectangle++(3.9,-.4);
        \path[blue](6.75,-1.45)node{\scriptsize light-cone
         $T_{\Tw{M}}={\color{red}T^*_M}$-fiber};
       \draw[blue](8.8,-.8)rectangle++(3.9,-.4);
        \path[blue](10.75,-1.45)node{\scriptsize Simple pos.\ roots of $\Tw{E}_8$};
                 \end{tikzpicture}}}
 \label{e:HMS}
\end{equation}
In particular, $\{x^{10},\cdots\!,x^{17}\}$ and $\{\tx^{10},\cdots\!,\tx^{17}\}$ are here identified by ``re-bosonized'' supersymmetry to span the indicated (co)tangent spaces to the chirally doubled space with the $M\!\divideontimes\!\Tw{M}$ foliation.
The {\em\/non-commutative\/} metastring framework and Born geometry then have the following immediate implications:
\begin{enumerate}\vspace*{-8pt}
 \item The metastring metric, $\eta_{AB}|_{10\cdots17}$, reproduces the canonical fiber-wise pairing of the two canonical bundles:
  $\vev{T_M,T^*_M}{=}1{=}\vev{\smash{T^*_{\sss\Tw{M}}},\smash{T_{\sss\Tw{M}}}}$.
  
 \item The metastring metric, $\eta_{AB}|_{18\cdots25}$, implies a dual pairing between the $E_8$'s simple positive roots and those of $\Tw{E}_8$. This implies that the roots of $\Tw{E}_8$ are $\eta$-canonically {\em\/reciprocal\/} to those of $E_8$ --- which identifies $\Tw{E}_8$ as the {\em\/Langlands dual,} $E^\vee_8$\cite{Borel:1979ux}, i.e., the electro-magnetic dual\cite{Kapustin:2006pk,Frenkel:2009ra}! Being simply laced, $E_8^\vee\approx E_8$, and they are physically indistinguishable. The $E_8$-group elements are exponential functionals of the $\{x^{18},\cdots\!,x^{25}\}$, which can be arranged to commute\cite{Freidel:2015pka,Freidel:2017xsi} with the $\Tw{E}_8$-group exponential functionals of the $\{\tx^{18},\cdots\!,\tx^{25}\}$, realizing the standard $E_8\,{\times}\,\Tw{E}_8$ Yang-Mills group, with $[E_8,\Tw{E}_8]=0$ as assumed in the original heterotic Ansatz\cite{Casher:1985ra,Gross:1984dd,Gross:1985fr,Gross:1985rr}.

 \item As indicated in~\eqref{e:HMS}, for $a=2,\cdots\!,9$ identify
 the light-cone gauge ``base'' coordinates $x^a,\tx^a$ 
 and the light-cone gauge ``fibre'' coordinates
 $x^{a+8}\!\mapsto\!\rd x^a\!=\!\tw\vd^a$ and $\tx^{a+8}\!\mapsto\!\vd_a\!=\!\rd\tx_a$,
 and for $a=1,\cdots\!,8$ identify the simple positive roots
 $(x^{a+17},\tx^{a+17})\mapsto(\rho_a,\rho_{\tw a})$.
Then:
\begin{enumerate}\itemsep=-1pt\vspace*{-.5\baselineskip}

 \item All target spacetime fields are a priori bi-local: they {\em\/depend\/} on both
  $x^\mu$ and $\tx_\mu$, with $\mu=0,\cdots\!,9$.

 \item They are {\em\/valued\/} as follows:
  $A_\mu{}^\a$ is an element of $T^*_M{\times}E_8$, i.e., it is an $E_8$-valued 1-form on $M$.
  In turn, $A^{\mu\tw\a}$ is valued in $T_M{\times}\Tw{E}_8$, so that
  $A_\mu^{\tw\a}{:=}H_{\mu\nu}A^{\nu\tw\a}$ is (via $H_{\mu\nu}$-contraction) an $\Tw{E}_8$-valued 1-form on $M$.

\end{enumerate}
 These originally bi-local fields then give rise to both an $E_8$-gauge field {\em\/and\/} its (``dark'') dual field valued in the {\em\/same\/} $E_8$ algebra. Analogously developing the $\Tw{E}_8$-valued gauge fields, we have both the standard $A_\mu{}^\a(x),A_\mu{}^{\tw\a}(x)$, as well as their (``dark'') duals, $\Tw{A}_\mu{}^\a(\tx),\Tw{A}_\mu{}^{\tw\a}(\tx)$.

 \item The metastring non-commutativity correlates, via
  $\frc{\l_4\!^2}{L^4}\int_x\mathfrak{F}^{\mu\nu}\brb{A_\mu^\a(x)}{\Tw{A}_\nu^{\,\tw\b}(\tx)}\eta_{\a\tw\b}$, the standard $E_8$-fields and their dual ``dark'' $\Tw{E}_8$-fields, and of course analogously the standard and ``dark'' fields with $E_8\leftrightarrow\Tw{E}_8$ swapped.
\end{enumerate}

\paragraph{Further Remarks:}
A recent analysis of the conifold transition in the standard formulation of heterotic string theory indicates that ``[effective] field theory alone [\dots] is likely not sufficient to describe the relevant singularities''\cite{Anderson:2022bpo}. In turn, the diagram~\eqref{e:HS} indicates that the usual EFT approach indeed misses the $(\tx^0,{\cdots},\tx^9)$-dependent fields, i.e., the ``dark'' $\tx$-dependent fields --- and then also their curvature (Chern class) contributions. Section~\ref{s:DarkE} already showed that the overall curvature of the $\tx$-spacetime induces a cosmological constant, so too must then the dual, ``dark'' $\Tw{A}_\mu{}^\a(\tx),\Tw{A}_\mu{}^{\tw\a}(\tx)$ induce a previously unaccounted gauge-bundle curvature, correlated via~\eqref{e:SMPv} and other above-enumerated relations.
 Whether this manifestly self-$T$-dual, chirally and non-commutatively doubled formulation affords a resolution to this decades-old puzzle and a physically complete description of conifold transitions for heterotic strings is a tantalizing, open question.
 
Following the discussion in Section~\ref{s:DM+Strings}, the fundamental non-commutativity $[x_{\sss L},x_{\sss R}]\neq0$ also implies the existence of gauge-gravity correlation terms, $\frac{\l_4\!^2}{L^4}\int_{x}\mathfrak{F}^{ab}\brb{A_a}{\Gamma_b}$. (The linear momentum of a charged particle is in curved spacetime modified both by the gauge potential, $A_a$, and by the Christoffel symbol/potential, $\Gamma_a$.)
Such a coupling to background fluxes $\mathfrak{F}^{ab}$ seems closely related to the so-called $\p$-fluxes of\cite{Freidel:2017nhg}, which however explicitly involve the dilaton. Their consistent combination and possible cancellation, defining the critical case in which this assumption of the standard~\eqref{e:HS} heterotic string model\cite{Casher:1985ra,Gross:1984dd,Gross:1985fr,Gross:1985rr} is also met, can only happen when $g_s\!\sim\!O(1)$. Curiously, this coincides with the seemingly unrelated observation of $\vev{g^D_s}_{\sY^2_\perp}\sim O(1)$ in Section~\ref{s:BHM}. More importantly however, this implies that the Born geometry of the metastring formalism, besides being inherently $T$-dual, also involves $S$-duality.

 Although undoubtedly technically demanding, {\em\/there seem to be no insuperable obstacles to\/} reverse-engineer the myriads of standard (super)string models into the metastring framework\cite{Freidel:2015pka,Freidel:2017xsi}. In particular, the Born geometry of this chirally doubled, phase-space like non-commutative stringy target spacetime insures\cite{Freidel:2018tkj,Freidel:2017yuv} that 
 it contains, in the commutative $\w,\l\!\to\!0$ limit, both the formalism of
  ``double field theory''\cite{Tseytlin:1990nb,Tseytlin:1990va,Tseytlin:1990hn,Siegel:1993xq,Siegel:1993th,Alvarez:2000bh,Alvarez:2000bi,Hull:2004in,Hull:2009mi,Coimbra:2011nw,Vaisman:2012ke,Vaisman:2012px,Aldazabal:2013sca,Hohm:2013vpa,Blumenhagen:2014gva,Hassler:2016srl} as well as that of ``generalized (complex, K{\"a}hler) geometry''\cite{Gualtieri:2004wh,Grana:2008yw,Coimbra:2011nw,Koerber:2010bx,Gualtieri:2007ng,Hull:2012dy,Gualtieri:2014ux,Candelas:2016usb,K_k_nyesi_2018,Candelas:2018lib}
 --- and so also their respective results, as well as indications towards leading $\l$-corrections.

\paragraph{Standard model:}
 Finally, we remark on the relation of this new view of the heterotic string and the emergence of the particle physics Standard Model.
 
 The non-commutative geometry of the Standard Model (SM)  \cite{Connes:1994yd,Connes:2008wl}
 allows to place the Higgs field on the same footing with the gauge fields, thus fixing the Higgs sector. Our non-perturbative formulation of string theory allows for 
general non-commutativity and non-associativity, essentially because of the $B$ and $H$ fields in string theory. 
We want to emphasize that the uniqueness of the SM group from the point of view of a unique non-associative quantum theory 
(see G\"{u}rsey and G\"{u}naydin's work\cite{Gunaydin:1973rs,Gunaydin:1974fb}, reviewed in the book by G\"{u}rsey and Tze\cite{Gursey:1996mj}) meshes well with the general non-associativity of our non-perturbative formulation. Also, chiral fermions can arise from the self-dual lattice that gives a fermionic string from the underlying bosonic string formulation. 
In particular, the unique prediction of heterotic string theory which is constructed from bosonic string theory,
the gauge group $E_8$ can be related to
the octonionic geometry (see the review of Baez on octonions and physics\cite{Baez:2001dm}),
which in turn, upon  ``integrating out'' over one octonionic structure (compatible with the double nature
of the non-perturbative formulation of bosonic metastring theory) reduces to octonionic geometry, the Moufang plane\cite{Gursey:1996mj,Baez:2001dm},
which is known to have the projective geometry of $F_4/\SO(9)$, and the unique subroup of $\SO(9)$ compatible with
the Poincare group, being the Standard Model group $\SU(3) \times \SU(2) \times U(1)$ (as observed by G\"{u}rsey and
G\"{u}naydin). Thus the uniqueness of $E_8$, via the non-associative octonionic geometry, could imply the uniqueness
of the Standard Model of Particle Physics. We note that this type of reasoning transcends the domain of the canonical
effective field theory and is complementary to it.

In particular, if the Standard Model gauge group appears from the unique non-associative structure, and
then the non-commutative structure of the non-perturbative formulation forces one into the so-called
non-commutative geometry of the Standard Model as discussed in\cite{Connes:1994yd,Connes:2008wl} and\cite{Chamseddine:2019fjq,Devastato:2019grb} as well as\cite{Aydemir:2013zua,Aydemir:2014ama,Aydemir:2015nfa,Aydemir:2016xtj,Aydemir:2018cbb}
then the question still remains: what physics is responsible for the masses and couplings of elementary particles
and their interactions? One approach that fits the narrative reviewed in this article is
``the Universe as an attractor" approach in which the fundamental interactions vertex of string field theory,
or the above non-perturbative formulation of metastring theory in terms of an abstract, doubled, non-commutative
and non-associative matrix quantum theory, introduces an effective ``horizontal transfer of information''
that has been used in biological physics to argue for the universality of the genetic code, as reviewed in\cite{Argyriadis:2019fwb}.
In that context one uses 
the mechanism of horizontal gene transfer to demonstrate convergence of the genetic code to a near universal solution.
It is tempting to conjecture that a similar mechanism can be used to argue for
a universal solution for 
particle masses and couplings. This also fits the idea of quantum contextuality (as opposed to anthropics) that was already mentioned in our discussion of the vacuum energy problem.
All this taken together with the realization of dark energy as the geometry/gravity of the dual spacetime\cite{Berglund:2019yjq}, as well as dark matter as the metaparticle extension of the Standard Model
presents a complete proposal for how, in principle, string theory, in its metastring formulation
could describe a world with a small and positive cosmological constant together with the visible and
dark matter sectors needed for the observed cosmological and particle physics structures.

\section{Conclusion}
\label{s:Outlook}
In this review we have discussed various aspects of de~Sitter spacetime in string theory: its status as a spacetime solution, 
including a new view on the cosmological constant problem, its (global) holographic definition in terms of two entangled and non-canonical
conformal field theories, and how a realistic de~Sitter universe that includes the observed
visible and dark sectors can be realized in string theory, formulated in an intrinsically non-commutative and generalized-geometric form. In particular, we argue that 
the doubled, non-commutative generalized-geometric formulation of string theory can lead to a positive cosmological constant.
Essentially, the curvature of the dual space is the cosmological constant in the observed spacetime, and the size of the dual space is the gravitational constant in the
same observed spacetime. Also, the three scales associated with intrinsic non-commutativity of string theory, the cosmological constant scale and the Planck scale,
as well as the Higgs scale,
can be arranged, to satisfy a seesaw-like formula. We also discuss various implications for
dark matter, hierarchy problem as well as a new non-perturbative approach to string theory and quantum gravity.
In particular, the vacuum energy of the string balances its phase space nature versus its holographic nature, in order
to provide for a radiatively stable and technically natural cosmological constant.
Also, we have emphasized that these new features of string theory can be implemented in a particular
deformation of cosmic-string like models.

The foregoing indicates that looking for alternatives to de Siter space in string theory, such as in the form of quintessence (another effective field theory description), appears rather unnatural even by the very criteria of effective field theory\cite{Obied:2018sgi, Agrawal:2018own}.
Instead, one should look for cosmological signatures of the intrinsic non-commutativity of string theory and its generalized (Born) geometry. Such cosmological signatures can have both UV and IR guises, and their mixing.
 In particular, one should look for non-local effects at largest possible (Hubble) scale. A preliminary study of such non-local effects associated with a one-parameter vacuum structure of de~Sitter space
was discussed in\cite{Kaloper:2002uj, Danielsson:2002kx, deBoer:2004nd}. Similarly, effective non-commutativity in cosmology was discussed in\cite{Brandenberger:2002nq} and the minimal length in\cite{Brandenberger:1988aj, Nayeri:2005ck, Brandenberger:2006xi}. 
However, such effects should be re-examined from the point of view of the non-commutative description of string theory discussed herein.

From the empirical point of view, in order to test the proposal reviewed in this paper, one should search for metaparticles,
test the crucial relation between the phase space and holographic aspects of the vacuum energy, and look for 
general quantum implications of a dynamical Born geometry of the metastring, such as ``gravitizing the quantum''\cite{Berglund:2022qcc,Berglund:2022skk}.
Clearly, much work remains to be done to advance the ideas described in this review, and we look forward to discussing some of these developments in the future.

\paragraph{Acknowledgments:} 
Our very special thanks go to L.~Freidel, J.~Kowalski-Glikman and R.~G.~Leigh for many years of fruitful collaboration on the central subjects of this review.
We also thank J.~A.~Argyriadis, U.~Aydemir, V.~Balasubramanian, J.~de Boer,
D.~Edmonds, D.~Farrah,
M.~G\"{u}naydin, C.~M.~Ho, P.~Horava, V.~Jejjala, M.~Kavic, 
\hbox{Y.-H.}~He, D.~Mattingly, Y.~J.~Ng, C.~Sun, T.~Takeuchi and C.~H.~Tze for discussions and collaboration.
 We should also like to thank E.~{\'O}.~Colg{\'a}in, G.~B.~De~Luca, F.~Diaz, A.~Hebecker, D.~Junghans, C.~Krishnan and A.~Tomasiello for constructive suggestions about this effort.
 PB thanks
the Simons Center for Geometry
and Physics and the CERN Theory Group, for their hospitality,
and TH is grateful to 
the Department of Physics, University of Maryland, and 
the Physics Department of the University of Novi Sad, Serbia,
for recurring hospitality and resources. 
The work of DM is supported in part by Department of Energy 
(under DOE grant number DE-SC0020262)
and the Julian Schwinger Foundation. DM thanks Perimeter Institute for hospitality and support.

%
\providecommand{\href}[2]{#2}
\begingroup\raggedright
\small\baselineskip=13pt \parskip=0pt plus2pt minus1pt

\endgroup

\end{document}